\documentclass[10pt]{article}%
\usepackage[a4paper,top=2.8cm,bottom=2.8cm,left=3cm,right=3cm]{geometry} %
\usepackage{sgame}
\usepackage{amsmath}%
\usepackage{amsthm}%
\usepackage{amsfonts}%
\usepackage{amssymb}%
\usepackage{mathrsfs}
\usepackage{braket}					
\usepackage{mathtools}
\usepackage{enumitem}
\usepackage{xcolor}%
\usepackage{graphicx}%
\usepackage{bm}
\usepackage{tikz}%
\usetikzlibrary{calc}%
\usetikzlibrary{cd}%
\usepackage{float}%
\usepackage{etoolbox}
\newcommand{\splitatcommas}[1]{%
  \begingroup
  \begingroup\lccode`~=`, \lowercase{\endgroup
    \edef~{\mathchar\the\mathcode`, \penalty0 \noexpand\hspace{0pt plus 1em}}%
  }\mathcode`,="8000 #1%
  \endgroup
}

\usepackage{changepage}

\usepackage{datetime}
\makeatletter
\newcommand\Romanmonth{\@Roman{\month}}
\makeatother
\newdateformat{mydate}{\THEDAY.\Romanmonth.\THEYEAR}
\usepackage{natbib}          
\usepackage[T1]{fontenc}
\usepackage[utf8]{inputenc}
\usepackage{epigraph}
\makeatletter
\patchcmd{\epigraph}{\@epitext{#1}}{\itshape\@epitext{#1}}{}{}
\makeatother
\usepackage{titlesec}
\titleformat{\section}
  {\normalfont\sc\centering}
  {\thesection.}{1em}{}
\titleformat{\subsection}
  {\normalfont\centering}
  {\normalfont\sc\centering\thesubsection}{1em}{}{}
\titleformat{\subsubsection}
  {\normalfont\slshape\centering}
  {\normalfont\sc\centering\thesubsubsection}{1em}{}{}
\makeatletter
\g@addto@macro\@floatboxreset\centering
\makeatother
\makeatletter
\newcommand{\mathleft}{\@fleqntrue}
\newcommand{\mathcenter}{\@fleqnfalse}
\makeatother
\DeclareMathOperator{\marg}{marg}
\DeclareMathOperator{\supp}{supp}

\usepackage{scalerel}

\makeatletter
\def\th@plain{%
  \thm@notefont{}
  \itshape 
}
\def\th@definition{%
  \thm@notefont{}
  \normalfont 
}
\makeatother
\numberwithin{equation}{section}
\usepackage{stmaryrd}

\DeclareMathAlphabet{\mathsfit}{\encodingdefault}{\sfdefault}{m}{sl}
\newcommand{\tens}[1]{\mathsfit{#1}}
\newcommand{\mcal}{\mathcal}
\newcommand{\mscr}{\mathscr}
\newcommand{\mbf}{\mathbf}

\newcommand{\mfrak}{\mathfrak}
\newcommand{\sAlg}{\Sigma}
\newcommand{\Alg}{\mcal{A}}
\newcommand{\CEvents}{\mcal{C}}
\newcommand{\CEventsM}{\mcal{C}_M}
\newcommand{\CEventsT}{\mcal{C}_\Theta}
\newcommand{\bN}{\mathbb{N}}

\newcommand{\la}{\langle}
\newcommand{\ra}{\rangle}

\newcommand{\CPSM}{\Delta^{\CEvents_M}}
\newcommand{\CPST}{\Delta^{\CEvents_\Theta}}
\newcommand{\otbar}[1]{\mkern 1.5mu\overline{\mkern-1.5mu#1\mkern-1.5mu}\mkern 1.5mu}
\newcommand{\id}{\mbf{id}}
\newcommand{\1}{\mbf{1}}
\newcommand{\MSC}{(M, \sAlg_M, \CEvents_M)}
\newcommand{\TSC}{(\Theta, \sAlg_\Theta, \CEvents_\Theta)}
\newcommand{\MS}{(M, \sAlg_M)}
\newcommand{\PowerSet}{\mcal{P}}
\newcommand{\PiSystem}{\mcal{D}}

\newcommand{\Cat}{\bm{\mathsf{C}}}
\newcommand{\SubCat}{\bm{\mathsf{S}}}
\newcommand{\SET}{\bm{\mathsf{Set}}}
\newcommand{\Top}{\bm{\mathsf{Top}}}
\newcommand{\CoAlg}{\bm{\mathsf{CoAlg}}}
\newcommand{\Meas}{\bm{\mathsf{Meas}}}
\newcommand{\MeasM}{\bm{\mathsf{Meas}}_M}

\newcommand{\MeasI}{\bm{\mathsf{Meas}}^{I}}
\newcommand{\Ob}{\mathsf{Ob}}
\newcommand{\op}{\mathsf{op}}
\newcommand{\Coalg}{\mfrak{C}}
\newcommand{\TCoalg}{\mfrak{Z}}
\newcommand{\Functor}{\bm{\Phi}}
\newcommand{\PowerSetF}{\bm{\mcal{P}}}
\newcommand{\Proj}{\mbf{Proj}}
\newcommand{\Cons}{\mbf{K}}
\newcommand{\Id}{\mbf{Id}}
\newcommand{\Ing}{\mbf{G}}

\newcommand{\ProdFM}{\mbf{P}_M}
\newcommand{\ProdFT}{\mbf{P}_\State}
\newcommand{\CPSFM}{\bm{\Delta}^{\bm{\CEventsM}}}
\newcommand{\DeltaC}{\bm{\Delta}^{C}}
\newcommand{\CPSFT}{\bm{\Delta}^{\bm{\CEventsT}}}
\newcommand{\HierF}{\overline{\bm{\mathrm{H}}}}
\newcommand{\TypeF}{{\bm{\mathrm{T}}}}
\newcommand{\TypeFp}{{\overline{\bm{\mathrm{T}}}}}
\newcommand{\Diagram}{{\bm{\mathrm{D}}}}
\newcommand{\cps}{\nu}
\newcommand{\pushf}{\widehat{f}}

\newcommand{\hier}{\chi}
\newcommand{\cone}{h}

\newcommand{\State}{\Theta}

\newcommand{\Car}{\tens{S}}
\newcommand{\car}{\tens{s}}

\newcommand{\trans}{\bm{\kappa}}

\newcommand{\Type}{\tens{T}}
\newcommand{\type}{\tens{t}}
\newcommand{\bel}{\bm{\beta}}

\newcommand{\tmor}{\vartheta}
\newcommand{\cmor}{\bm{\mu}}
\newcommand{\zetap}{\bm{\zeta}_{+}}
\newcommand{\zetat}{\overrightarrow{\bm{\zeta}}}
\newcommand{\Projective}{\tens{P}}

\newcommand{\Terminal}{\tens{Z}}
\newcommand{\belt}{\bm{\zeta}}
\newcommand{\rhot}{\bm{\varphi}}
\newcommand{\terminal}{\tens{z}}
\newcommand{\opchain}{\omega^{\op}}
\newcommand{\Limit}{\tens{L}}
\newcommand{\Coalitions}{\mcal{I}}
\newcommand{\TypeC}{{\bm{\mathrm{C}}}}
\newcommand{\CEventsI}{\mcal{C}_I}
\newcommand{\CPSFC}{\mbf{\Delta}^{\bm{\CEventsI}}}
\newcommand{\ProdFI}{\mbf{P}_I}
\def\SetVertic{\egroup\;\hspace{-0.1cm}\middle|\hspace{-0.1cm}\;\bgroup}
{\catcode`\|=\active
  \xdef\Sets{\protect\expandafter\noexpand\csname Sets \endcsname}
  \expandafter\gdef\csname Sets \endcsname#1%
  	{\hspace{-0.05cm}\left\{\hspace{-0.08cm}%
     		\:{
     			\mathcode`\|32768\let|\SetVertic
     		#1}
     		\:\hspace{-0.08cm}\right\}}
}

{\catcode`\|=\active
  \xdef\Round{\protect\expandafter\noexpand\csname Round \endcsname}
  \expandafter\gdef\csname Round \endcsname#1{\left(%
     \:{
     \mathcode`\|32768\let|\SetVertic
     #1}\:\right)}
}

{\catcode`\|=\active
  \xdef\Rounds{\protect\expandafter\noexpand\csname Rounds \endcsname}
  \expandafter\gdef\csname Rounds \endcsname#1%
  	{\hspace{-0.05cm}\left(\hspace{-0.08cm}%
     		\:{
     			\mathcode`\|32768\let|\SetVertic
     		#1}
     		\:\hspace{-0.08cm}\right)}
}

{\catcode`\|=\active
  \xdef\Square{\protect\expandafter\noexpand\csname Square \endcsname}
  \expandafter\gdef\csname Square \endcsname#1{\left[%
     \:{
     \mathcode`\|32768\let|\SetVertic
     #1}\:\right]}
}

{\catcode`\|=\active
  \xdef\Squares{\protect\expandafter\noexpand\csname Squares \endcsname}
  \expandafter\gdef\csname Squares \endcsname#1%
  	{\hspace{-0.05cm}\left[\hspace{-0.08cm}%
     		\:{
     			\mathcode`\|32768\let|\SetVertic
     		#1}
     		\:\hspace{-0.08cm}\right]}
}

{\catcode`\|=\active
  \xdef\Angle{\protect\expandafter\noexpand\csname Angle \endcsname}
  \expandafter\gdef\csname Angle \endcsname#1{\left\langle%
     \:{
     \mathcode`\|32768\let|\SetVertic
     #1}\:\right\rangle}
}

{\catcode`\|=\active
  \xdef\Angles{\protect\expandafter\noexpand\csname Angles \endcsname}
  \expandafter\gdef\csname Angles \endcsname#1%
  	{\hspace{-0.05cm}\left\langle\hspace{-0.08cm}%
     		\:{
     			\mathcode`\|32768\let|\SetVertic
     		#1}
     		\:\hspace{-0.08cm}\right\rangle}
}
\providecommand{\envert}[2][-1]{
\ensuremath{\mathinner{
\ifthenelse{\equal{#1}{-1}}{ 
\!\left\lvert#2\right\rvert}{}
\ifthenelse{\equal{#1}{0}}{ 
\lvert#2\rvert}{}
\ifthenelse{\equal{#1}{1}}{ 
\!\bigl\lvert#2\bigr\rvert}{}
\ifthenelse{\equal{#1}{2}}{ 
\!\Bigl\lvert#2\Bigr\rvert}{}
\ifthenelse{\equal{#1}{3}}{ 
\!\biggl\lvert#2\biggr\rvert}{}
\ifthenelse{\equal{#1}{4}}{ 
\!\Biggl\lvert#2\Biggr\rvert}{}
}} 
}
\let\abs=\envert

\usepackage{varioref}
\usepackage[pagebackref,hypertexnames=false]{hyperref}
\usepackage[noabbrev,nameinlink]{cleveref}
\definecolor{egtgreen}{RGB}{75, 155, 8}
\definecolor{egtpurple}{RGB}{10, 33, 128}
\definecolor{egtred}{RGB}{103, 25, 15}
\definecolor{burgundy}{rgb}{0.5, 0.0, 0.13}
\definecolor{cyanp}{RGB}{45,129,173}
\hypersetup{colorlinks=true,linkcolor={egtred},citecolor=green!50!black,bookmarksnumbered=true}
\hypersetup{%
pdfcreator={},
  pdftitle={Topology-Free Type Structures with Conditioning Events},%
  pdfauthor={Pf. Guarino},%
  pdfsubject={},%
  pdfproducer={},%
  pdfkeywords={}]%
}
\newcommand{\Mref}[2][cyanp]{%
\hypersetup{linkcolor=#1}%
\Cref{#2}%
\hypersetup{linkcolor=burgundy}%
}
\newcommand{\Sref}[2][burgundy]{%
\hypersetup{linkcolor=#1}%
\Cref{#2}%
\hypersetup{linkcolor=burgundy}%
}
\usepackage{xpatch}\xpatchbibmacro{pageref}{parens}{brackets}{}{}
\makeatletter
\patchcmd{\BR@backref}{\newblock}{\newblock[}{}{}
\patchcmd{\BR@backref}{\par}{]\par}{}{}
\makeatother
\theoremstyle{plain}
\newtheorem{theorem}{Theorem}
\newtheorem{definition}{Definition}[section]
\newtheorem{proposition}[theorem]{Proposition}
\newtheorem{lemma}{Lemma}
\newtheorem{applemma}{Lemma}[section]

\newtheorem{remark}{Remark}[section]
\theoremstyle{definition}
\newtheorem{notation}{Notation}
\theoremstyle{remark}
\makeatletter
\def\thm@space@setup{%
  \thm@preskip=0.5cm 
  \thm@postskip=\thm@preskip 
}
\makeatother
\newtheoremstyle{myclaimstyle} 
    {\topsep}                    
    {\topsep}                    
    {\itshape}                   
    {}                           
    {\itshape}                   
    {.}                          
    {.5em}                       
    {}  

\theoremstyle{myclaimstyle}

\newtheoremstyle{mycasestyle} 
    {3pt}                    
    {3pt}                    
    {}                   
    {}                           
    {}                   
    {:}                          
    {.5em}                       
    {}  

\theoremstyle{mycasestyle}

\newtheoremstyle{named}{}{}{\itshape}{}{\bfseries}{.}{.5em}{\thmnote{#3}}
\theoremstyle{named}

\makeatletter
\newtheoremstyle{case}
  {5pt}
  {5pt}
  {\addtolength{\@totalleftmargin}{0em}
   \addtolength{\linewidth}{-1em}
   \parshape 1 1em \linewidth}
  {}
  {\normalfont}
  {:}
  {.5em}
  {}
\makeatother

\theoremstyle{case}

\AtEndEnvironment{proof}{\setcounter{step}{0}}
\AtEndEnvironment{proof}{\setcounter{claim}{0}}
\AtEndEnvironment{proof}{\setcounter{case}{0}}
\AtEndEnvironment{proof}{\setcounter{Cases}{0}}
\makeatletter
\def\thm@space@setup{%
  \thm@preskip=0.5cm 
  \thm@postskip=\thm@preskip 
}
\makeatother
\usepackage{thmtools}

\declaretheorem[numberwithin=section, style=definition, name=Example, qed=$\diamond$]{example}
\renewcommand\thmcontinues[1]{Continued}
\declaretheoremstyle[
  spaceabove=3pt, spacebelow=6pt,
  headfont=\normalfont\itshape,
  notefont=\mdseries, notebraces={(}{)},
  bodyfont=\normalfont,
  postheadspace=1em
]{innerproof}

\renewcommand\thmcontinues[1]{}

\renewcommand\thmcontinues[1]{}
\makeatletter
\def\@xfootnote[#1]{%
  \protected@xdef\@thefnmark{#1}%
  \@footnotemark\@footnotetext}
\makeatother
\title{\vspace{-1.3cm}Topology-Free Type Structures with Conditioning Events%
\thanks{The present work is a considerable development of \cite{Guarino_2017}, where a sketch of a proof of the result obtained with radically different tools was presented. The main technical bulk of this work has been obtained during an informal visiting period at LUISS Guido Carli back in the early spring of 2018 for which I am extremely grateful to Marco Scarsini. I would like to thank the editor, Nicholas Yannelis, the anonymous handling editor, and two referees for their outstanding reading of the paper and their suggestions that ultimately lead to the present version. Special thanks go to Pierpaolo Battigalli, for his long-term interest in this endeavour, and to Martin Meier, without whose encouragement to pursue the project---quite simply---this manuscript would have never become public. I also would like to thank Michael Greinecker and Gabriel Ziegler for many---most stimulating---conversations on the topic of large interactive structures in general and Marcus Pivato for enlightening discussions on category theory. Finally, I would like to thank the audiences of the EWET 2023 Workshop, the SAET 2023 Conference, and the GRASS XVIII 2024 Workshop. Of course, all errors are mine. I thankfully acknowledge financial support from the Austrian Science Fund (FWF) (P31248-G27), from MIUR under the PRIN 2017 program (grant number 2017K8ANN4), and partial support from the Departmental Strategic Plan (PSD – 2022-2025) of the Department of Economics and Statistics (DIES) of the University of Udine.}%
}
\author{%
Pierfrancesco Guarino\thanks{%
University of Udine (Department of Economics and Statistics -- DIES). %
\textit{E-mails:} %
\texttt{pf.guarino@hotmail.com} \& %
\texttt{pierfrancesco.guarino@uniud.it} %
.}
}
\date{19.X.2024}
\begin{document}

\maketitle

\vspace{-0.5cm}
\begin{center}
FINAL VERSION\\\smallskip
\emph{Accepted for Publication on ``Economic Theory''}\footnote[$\ddag$]{DOI: \texttt{https://doi.org/10.1007/s00199-024-01621-5}}
\end{center}

\vspace{0.1cm}

\begin{abstract}
\noindent Starting without any topological assumption, we establish the existence of the universal type structure in presence of---possibly uncountably many and topologically unrestricted---conditioning events, namely, a type structure that is non-redundant, belief-complete, terminal, and unique up to measurable type isomorphism, by performing a construction in the spirit of the hierarchical one in \cite{Heifetz_Samet_1998}. In particular, we obtain the result by exploiting arguments from category theory and the theory of coalgebras, thus, making explicit the mathematical structure underlying all the constructions of large interactive structures and obtaining the belief-completeness of the structure (unattainable via the standard hierarchical construction \emph{\`a la} \cite{Heifetz_Samet_1998}) as an immediate corollary of known results from these fields. Additionally, we show how our construction, with its lack of topological and cardinality assumptions on the family of conditioning events, can be employed in various game-theoretical contexts.

\bigskip


\end{abstract}

\bigskip

\epigraph{``Before category theory was invented it had been noted in some parts of mathematics that from some privileged objects, often called `universal', one could construct \emph{all} objects of the same kind.''}%
{---\citet[Part 1.IV.8.C, p.151]{Dieudonne_1989}}

\vspace{-0.5cm}

\newpage

\tableofcontents

\newpage

\section{Introduction}
\label{sec:introduction}

\subsection{Motivation \& Results}
\label{subsec:motivation_results}

Type structures are one of the most widely employed tools in theoretical economics. Introduced in \cite{Harsanyi_1967} to handle the technical problems arising from the infinite regress proper of the analysis of games with incomplete information, their usage ranges from applied issues, such as those tackled by mechanism design, to more foundational problems, as exemplified by those works belonging to epistemic game theory.\footnote{Regarding epistemic game theory, see the survey \cite{Dekel_Siniscalchi_2015}, or the two textbooks \cite{Perea_2012} and \cite{Battigalli_et_al_Forthcoming}, completely devoted to the topic.}

In the context of type structures, a particular role is played by those type structures that satisfy what can be deemed `large' properties, which are typically captured via the so-called notion of universality.\footnote{See \Sref{subsec:large_type_structures} for the definition we employ in this work and \Sref{subsubsec:notion_universality} with respect to how the  term is used in the literature.} Intuitively, the Universal Type Structure, where the usage of the definite article comes from the fact that the construct is essentially \emph{unique}, is a particular type structure that contains all the possible types that can arise given a certain domain of uncertainty. Thus, it is immediate to see the importance of this notion in the fields mentioned above, as---for example---pointed out in \citet[Section 3]{Bergemann_Morris_2012}: employing the Universal Type Structure allows the analyst to drop all at once the possible \emph{a priori} assumptions that can be forced upon a game-theoretical analysis by employing smaller type structures, since, in the evocative words of \citet[Section 8, pp.1672--1673]{Aumann_Heifetz_2002},
\begin{quote}
\emph{``The system does not depend on reality; it is a framework, it fits any reality, so to speak, like the frames that one buys in photo shops, which do not depend on who is in the photo and they fit any photo with any subject, as long as the size is right. In brief, there is no substantive information in the system.''}
\end{quote}
In other words, the usage of smaller type structures  forces upon a game-theoretical analysis certain restrictions by  \emph{a fortiori} ruling out some possible beliefs, i.e., contrary to the universal one, there is \emph{``substantive information in the system''} when such a system is a `small' type structure---even if the substantive information is present somewhat implicitly.

Thus, working with the Universal Type Structure is a considerable step further in dropping involuntary---and non-explicit---assumptions that could limit game-theoretical analysis: as such, employing the Universal Type Structure is---for example---a typical (even if admittedly extreme) step when an analyst wants to relax common knowledge assumptions in the spirit of the so-called \emph{Wilson's doctrine} as in \cite{Wilson_1987}. However, there is another way in which potential non-explicit assumptions could arise even when using the Universal Type Structure, namely, when the Universal Type Structure is built from topological assumptions. Indeed, in such a case, it is most natural to ask ourselves if the results we obtain from a given analysis are the consequence of the very topological assumptions needed to built the Universal Type Structure employed. Thus, it is in this context that  topology-free constructions of Universal Type Structures happen to derive their importance: they are constructs that can be used to perform analyses robust to topological details and can provide a conceptual foundation for the topological constructions. 

For the `standard' case appropriate to deal with---for example---static games, the construction of the Universal Type Structure has been obtained in  \cite{Heifetz_Samet_1998}. Rather crucially, this work did not follow the standard procedure based on the construction of coherent infinite hierarchies of beliefs (i.e., infinite hierarchies of beliefs satisfying a coherency requirement across different belief orders) as performed when starting from topological assumptions, in the spirit of \cite{Mertens_Zamir_1985} or \cite{Brandenburger_Dekel_1993}.\footnote{See also \cite{Armbruster_Boge_1979}, \cite{Heifetz_1993}, and \citet[Chapter A.III.1]{Mertens_et_al_2015}.} Indeed, this was a consequence of the striking result obtained in \cite{Heifetz_Samet_1999}, who showed that, when we drop topological assumptions, we lose the equivalence between types (as coherent infinite hierarchies of beliefs that admit an extension to the limit) and coherent infinite hierarchies of beliefs and there are coherent infinite hierarchies of beliefs that are actually \emph{not} types.  As a consequence, the approach taken in \cite{Heifetz_Samet_1998}, based on taking types belonging to `small' type structures as ready-made objects to then collect all of them in a large type structure, had an undesirable side-effect.  That is, the obtained construct had two out of the three properties attached to the idea of universality, namely, terminality (capturing the very idea of collecting all those types in the same large structure) and non-redundancy: the remaining property, namely, belief-completeness, particularly desirable for epistemic endeavours, had to be established in \cite{Meier_2012} by employing tools from infinitary probability logic.\footnote{See \Sref{subsubsec:notion_belief-completeness} for a discussion of the result and the path chosen to establish it.}

In the present work, we build the Universal Type Structure with conditioning events without topological assumptions, thus, extending the work of \cite{Heifetz_Samet_1998} to the presence of conditioning events and the work of 
\cite{Battigalli_Siniscalchi_1999}, that constructed the Universal Type Structure \emph{{\`a} la} \cite{Brandenburger_Dekel_1993} starting from a Polish common domain of uncertainty to deal with the presence of conditioning events, to a topology-free setting. Incidentally, by obtaining our result, we prove true a conjecture made in \citet[Section 2, p.198]{Battigalli_Siniscalchi_1999} regarding the possibility of performing a construction \emph{{\`a} la} \cite{Heifetz_Samet_1998} for type structures with conditioning events. However, whereas to obtain the result we follow the construction performed in \cite{Heifetz_Samet_1998}, at the same time, we actually take a more general approach, relying on tools from category theory and the theory of coalgebras (duly introduced in \Sref{subsec:coalgebras}). In taking this path, built on work by \cite{Viglizzo_2005a},\footnote{See \cite{Lawvere_1962} and \cite{Giry_1982} for the first works addressing measure and probability spaces from a categorical standpoint.} where a coalgebraic analog of \cite{Heifetz_Samet_1998} for the case without conditioning events is derived, we obtain in \Mref{th:main_theorem} all at once the three universality properties of our resulting construct, which is---indeed---non-redundant, belief-complete, and terminal (and unique up to---measurable---type isomorphism).\footnote{See \Sref{subsubsec:notion_universality} regarding the notion of universality we employ and its relation to terminality and the other notions mentioned above.} 

With respect to the tools chosen to obtain our result, it is important to stress one point: there is no `shortcut' to get \emph{all} the relevant universality properties at once in the topology-free case via a construction that does not involve a certain level of `technicalities'. Indeed, as pointed out above, it is the construction in \cite{Meier_2012} that actually established for the first time the universality of the construction in \cite{Heifetz_Samet_1998} by---among other things---proving its belief-completeness by employing tools from infinitary probability logic, with \cite{Viglizzo_2005a} being an alternative based on category theoretical and coalgebraic tools. Thus, in order to establish the universality in the topology-free case, either we choose Scylla or Charybdis, i.e., either we pick tools from infinitary probability logic or we choose the category theoretical and coalgebraic approach:\footnote{With the understanding that the disjunction is used here in an inclusive sense: i.e., it is possible to work with \emph{coalgebraic modal logic}. See \Sref{subsubsec:notion_belief-completeness} for a discussion of this point.} the standard construction of \cite{Heifetz_Samet_1998} does not \emph{automatically} deliver everything we would like to get with belief-completeness being the missing ingredient.\footnote{Where, in particular, what is missing is that the belief functions are \emph{isomorphisms} in the appropriate category under scrutiny. See \Sref{subsubsec:notion_universality} concerning this point.}

Thus, to establish this result, we employ the following strategy. First of all, in \Sref{subsec:there}, we bypass the presence of interactive agents and we show that there is essentially a basic mathematical structure underlying the very notion of type structure (in presence of conditioning events), namely, a coalgebra (as in \Mref{def:coalgebra}). As a result, we establish the existence of a terminal coalgebra in \Mref{prop:coalgebraic_main_theorem}. Armed with this result, in \Sref{subsec:dale}, we tackle the presence of interactive agents and we show that the main bulk of the endeavour has already been established, since a minimal change allows us to obtain \Mref{prop:coalgebraic_main_theorem_type}, which is a translation of \Mref{th:main_theorem} in coalgebraic terms as set in \Mref{prop:coalgebraic_main_theorem}, with the caveat that we now take care of the presence of interactive agents. Finally, in \Sref{subsec:back_again}, we show that type structures are indeed coalgebras as defined in \Sref{subsec:dale}, thus, exploiting  \Mref{prop:coalgebraic_main_theorem_type} to establish \Mref{th:main_theorem} essentially as its immediate corollary. It is important to underline that, to proceed along the lines just described, we closely follow the path and the proofs in \cite{Viglizzo_2005a} (reproduced here with our notation for self-containment purposes), in themselves---as already mentioned---a translation in coalgebraic terms of those in \cite{Heifetz_Samet_1998}. Hence, from a technical standpoint, the crucial innovation of this paper lies in identifying the specific properties of the mathematical constructs we have to work with (i.e., product conditional measurable spaces sharing a family of conditioning events as in \Mref{def:space_sharing}), a point which has an immediate impact on the peculiarities of the categorical construct (i.e., the functor) we have to work with and on the proofs of all these results that explicitly refer to the presence of conditioning events, that are the `new' ones with respect to \cite{Viglizzo_2005a}.

Now, taking a very general view of the matter, it is important to stress that we see this work as having \emph{two} main applications.

The first concerns the result itself and how it can be used in game theory. Thus, in general, beyond being the reference framework for \citet[Section 2.5]{Bergemann_Morris_2005}, it should be observed that the construct obtained in \cite{Heifetz_Samet_1998} has been used to obtain the epistemic characterization of Interim Correlated Rationalizability as in \citet[Section 4.1]{Dekel_et_al_2007} and to study Rationalizability in games with incomplete information in general as in \citet[Section 1.3]{Battigalli_et_al_2011}. Hence, \Mref{th:main_theorem} (and the construction obtained here more in general) can provide the ground for similar works---in the spirit of \cite{Penta_2015} and \cite{Mueller_2016}---where there is the need to drop topological assumptions by also taking into account the presence of conditioning events possibly without any restriction, either topological or cardinal, on the nature of them. More specifically, in \Sref{sec:applications} we  provide examples of game-theoretical endeavours where our result, with its lack of topological or cardinality assumptions regarding the conditioning events  
can prove to be particularly useful. Thus, for example, in \Sref{subsec:topology-free_conditioning_events} we show how the lack of topological assumptions can play a role in \emph{psychological game theory},\footnote{See \cite{Battigalli_Dufwenberg_2022} for a survey of the literature.} which is a field where dynamic strategic interactions are a major topic of analysis\footnote{See \cite{Battigalli_Dufwenberg_2009} with respect to this point.} and infinite hierarchies of beliefs built on them are of special importance in light of the fact that the utility functions of the players \emph{depend} on---possibly higher order---beliefs of their co-players.

The second application, that we consider as important and potentially far-reaching as the previous one, concerns the introduction of the tools we employ to an `economics' audience. Indeed, whereas category theory and the theory of coalgebras can be used \emph{directly} to tackle specific technical and conceptual problems belonging to economic theory, as we do here and it has already been done in works referenced in \Sref{subsec:related_literature}, those fields can both prove to be extremely useful in a more indirect way. To see this, first of all, it is important to  recognize, in the words of \citet[Introduction, p.1]{Leinster_2014}, that
\begin{quote}
\emph{``Category theory takes a bird's eye view of mathematics. From high in the sky, details become invisible, but we can spot patterns that were impossible to detect from ground level.''}
\end{quote}
Thus, it is possible to use category theory in an indirect---heuristic---fashion. Here, we refer to the fact that being acquainted to  the categorical language and its results could in principle allow to see patterns \emph{``from high in the sky''}; this, in turn, could allow to conjecture certain solutions to specific problems proper of economic theory exactly via the fact that those solutions would be proper of more general settings, as showed via results in category theory. In other words, observing that something is true for many different categories all sharing certain `nice' properties  could prove to be useful to deal with specific problems, once the right category to address these problems has been identified and it turns out that this category has those very same `nice' properties. And here one interesting point is in order: this heuristic process should not ask to write results using the categorical language, which could potentially remain in the background.

\subsection{Related Literature}
\label{subsec:related_literature}

\enlargethispage{\baselineskip}

This work is related to various streams of literature. Of course, it is related to the literature focusing on the existence of large type structure, most notably \cite{Heifetz_Samet_1998}, in working in a topology-free setting, \cite{Battigalli_Siniscalchi_1999}, in addressing the presence of conditioning events,  \cite{Meier_2012}, in dealing with the problem of belief-completeness, and \cite{Viglizzo_2005a}, for the tools employed. In using machinery from category theory, it is related to an emerging literature in economic theory that employs these tools, such as \cite{Heinsalu_2014}, \cite{de_Oliveira_2018}, \cite{Galeazzi_Marti_2023}, \cite{Pivato_2024b}, \cite{Pivato_2024}, and \cite{Pivato_unpublished}. Also, regarding the usage of coalgebras,\footnote{\cite{Heifetz_1996} is a paper that should be---indirectly---related to this stream of literature via its relation with \cite{Aczel_1988}, where it is possible to find in Chapter 7 an explicit construction such as the one  in \Sref{subsec:there}.} it is related to \cite{Moss_Viglizzo_2004} and---in particular, once more---\cite{Viglizzo_2005a},\footnote{See also \cite{Viglizzo_2005}, \cite{Moss_Viglizzo_2006}, and \cite{Moss_2011}.} along with the aforementioned \cite{Heinsalu_2014}, \cite{Galeazzi_Marti_2023}, and \cite{Pivato_2024}, and unpublished work by Davide Ferri.\footnote{As an M.Sc. thesis at Bocconi University, starting from topological assumptions (in particular, starting from a Polish common domain of uncertainty as in \cite{Battigalli_Siniscalchi_1999}---personal communication).}

\subsection{Synopsis}

This paper is organized as follows. In \Sref{sec:mathematical_preliminaries}, we introduce the mathematical notions from measure theory and the theory of coalgebras we need for our endeavour. In \Sref{sec:measure-theoretic_type_structure}, we introduce type structures along with related notions. In \Sref{sec:universality}, we prove the existence of the topology-free universal type structure with conditioning events, while in \Sref{sec:discussion} we address some points related to our result and we discuss the relation between the present work and various works belonging to the same stream of literature. Finally, in \Sref{sec:applications}, we show various contexts where this object can be used. Regarding the appendices, we devote \Sref{app:category_theory_background} to a self-contained introduction to the notions from category theory needed for our purposes, we collect results from measure theory we employ to establish the universality in \Sref{app:measure-theoretic_results}, whereas we relegate to \Sref{app:proofs_universality} all the proofs of our results that cannot be found in the main body of this work.

\section{Mathematical Preliminaries}
\label{sec:mathematical_preliminaries}

\subsection{Measure Theory}
\label{subsec:measure_theory}

Let $\MS$ be a measurable space, that is, a set $M$ endowed with a $\sigma$-algebra $\sAlg_M$: in the following, whenever we refer  to a set $M$ as a measurable space without explicitly denoting its $\sigma$-algebra, it is assumed that $M$ is endowed with a $\sigma$-algebra $\sAlg_M$. Given a measurable space $\MS$, the $\sigma$-algebra $\sAlg_M$ is \emph{separative} if for every $x, x' \in M$ with $x \neq x'$ there exists an $E_x \in \sAlg_M$ such that $x \in E_x$ and $x' \notin E_x$. We let every product of measurable spaces be endowed with the product $\sigma$-algebra,\footnote{See \citet[Chapter 3.1, p.87]{Srivastava_1998}.} i.e., given an arbitrary product space $M := \prod_{\lambda \in \Lambda} M_\lambda$ with $(M_\lambda , \sAlg_\lambda)$ measurable and $\pi_\lambda$ denoting the projection function as canonically defined over the index set $\Lambda$, for every $\lambda \in \Lambda$, we endow $M$  with the  \emph{product $\sigma$-algebra} 
\begin{equation*}
\bigotimes_{\lambda \in \Lambda} \sAlg_\lambda := 
\sigma \big( \Set { \pi^{-1}_\lambda (E) | \lambda \in \Lambda , \ E \in \sAlg_\lambda } \big) ,
\end{equation*}
where we---alternatively---denote this $\sigma$-algebra with $\bigotimes_{\lambda \in \Lambda} \sAlg_\lambda$ (as we do above) or $\sAlg_{\prod_{\lambda \in \Lambda} M_\lambda}$. Given a measurable space $(M', \sAlg_{M'})$, a set $M$, and a function $f \in {M'}^M$, we let $\sigma (f)$ denote the smallest $\sigma$-algebra on $M$ that makes $f$ measurable. A \emph{measurable isomorphism} between two measurable spaces $(M, \sAlg_M)$ and $(M', \sAlg_{M'})$ is a bijective measurable function $f \in {M'}^M$ such that $f (E) \in \sAlg_{M'}$ for every $E \in \sAlg_M$ and $f^{-1} (E') \in \sAlg_M$ for every $E' \in \sAlg_{M'}$, where---to lighten the notation---we typically omit the reference to the $\sigma$-algebras that make a function measurable. Also, given an arbitrary set $M$, we let $\id_M$ denote the identity function on $M$, which is---trivially---a measurable isomorphism whenever $M$ is a measurable space.

We let $\Delta (M)$ denote the set of all $\sigma$-additive probability measures  over $M$ (henceforth, probability measures). Given a measurable space $(M, \sAlg_M)$ and a possibly uncountable\footnote{See \Sref{subsubsec:technical_assumptions} for a discussion of this point.} subset $\CEventsM \subseteq \sAlg_M \setminus \{ \emptyset\}$ of \emph{conditioning events}, we call the space $\MSC$ a \emph{conditional measurable space}. 

\begin{definition}[Conditional Probability System]
\label{def:CPS}
A \emph{conditional probability system}\footnote{\label{foot:CPS}This  corresponds to one of the primitive elements of a  \emph{conditional probability space} of \citet[Sections 1.2 \& 1.4]{Renyi_1955}. However, the name---that eventually stuck in the game-theoretic literature---actually comes from a related definition from \citet[Section 5, pp.336--337]{Myerson_1986}. See \citet[Section 3.2]{Hammond_1994} for a discussion of the relation between these two notions.} (henceforth, CPS) on a conditional measurable space $\MSC$ is a function
\begin{equation*}
\cps (\cdot | \cdot ) : \sAlg_M \times \CEventsM \to [0,1]
\end{equation*}
that satisfies the following axioms:
\begin{enumerate}[label=C\arabic*., leftmargin=*, itemsep=0.5ex]
\item For every $C \in \CEventsM$, $\cps (C | C) =1$;
\item For every $C \in \CEventsM$, $\cps (\cdot | C) \in \Delta (M)$;
\item For every $E \in \sAlg_M$ and $D, C \in \CEventsM$, if $E \subseteq D \subseteq C$, then $\cps (E|C) = \cps (E | D) \cdot \cps (D|C)$. 
\end{enumerate}
\end{definition}

We let $[\Delta(M)]^{\CEventsM}$ denote the set of all functions from $\CEventsM$ to $\Delta (M)$, while we let $\CPSM (M) \subseteq [\Delta(M)]^{\CEventsM}$ denote the set of CPSs on $\MSC$, with $\cps := \big( \cps (\cdot| C) \big)_{C \in \CEventsM} \in \CPSM (M)$. We let
\begin{equation*}
\gamma^{p}_C (E) := \Set { \cps \in \CPSM (M) | \cps  (E | C) \geq p },
\end{equation*}
for every  $E \in \Sigma_M$, $C \in \CEventsM$, and  $p \in [0,1]$.

\begin{definition}[{$\sigma$-Algebra on the Space of CPSs}]
\label{def:sAlg_CPS}
Given a conditional measurable space $\MSC$, the space $\CPSM (M)$ is endowed with the $\sigma$-algebra\footnote{\label{foot:sAlg_CPS}When $M$ is Polish and $\Delta (M)$ is endowed with the topology of weak convergence, this $\sigma$-algebra (modulo presence of conditioning events) is exactly the Borel $\sigma$-algebra induced by the topology of weak convergence (see \citet[Theorem 17.24, p.112]{Kechris_1995}). Additionally, when in presence of  $\CEventsM$, with this $\sigma$-algebra we have that  $\CPSM (M)$ is closed in $[\Delta (M)]^{\CEventsM}$, as established in \citet[Lemma 1, p.193]{Battigalli_Siniscalchi_1999} (see also \citet[Footnote 5, p.193]{Battigalli_Siniscalchi_1999}).}
\begin{equation}
\label{eq:sAlg_CPS}
\sAlg_{\CPSM (M)} := \sigma \big( %
\Set { \gamma^{p}_C (E) | E \in \Sigma_M, \ C \in \CEventsM, \ p \in [0,1] } \big).
\end{equation}
\end{definition}

Given a conditional measurable space $\MSC$, in this paper, we focus on a specific family of spaces tightly linked to $\MSC$, introduced with the idea of finding a `common' family of exogenously imposed conditioning events between product spaces, that we define next.\footnote{Where, in particular, the nature of the spaces in \Mref{def:type_morphism} is the main reason behind the need to introduce this construct. See also \Sref{sec:applications} for various examples regarding the nature of the conditioning events, where, in particular, it is important to stress, as can be observed from \Sref{subsec:nature_of_the_conditioning_events}, that these conditioning events do not need to represent verifiable information (e.g., information sets reached in a dynamic game), but they should be considered simply a tool to capture counterfactual reasoning.}

\begin{definition}[Product Conditional Measurable Spaces Induced by Conditioning Events]
\label{def:space_sharing}
Given a conditional measurable space $\MSC$ and a measurable space $(X, \sAlg_X)$, the space $\big(M \times X, \sAlg_M \otimes \sAlg_X, \CEvents_{M \times X} \big)$ is a \emph{product conditional measurable space induced by} $\MSC$ if
\begin{equation}
\label{eq:product_conditioning}
\mcal{C}_{M \times X} := \Set { D_X \in \sAlg_M \otimes \sAlg_X | \exists C \in \CEventsM : D_X = C \times X }.
\end{equation}
\end{definition}

When it does not lead to any ambiguity, we drop from \Mref{def:space_sharing} the reference to the family of conditioning events and we simply use the expression ``product conditional measurable space''. Also, two product conditional measurable spaces $\big(M \times X, \sAlg_M \otimes \sAlg_X, \CEvents_{M \times X} \big)$ and $\big(M \times Y, \sAlg_M \otimes \sAlg_Y, \CEvents_{M \times Y} \big)$ induced by $\CEventsM$ are said to \emph{share} $\CEventsM$. Now, observing the structure of $\CEvents_{M \times X}$ in \Mref{eq:product_conditioning}, we have that the conditioning events of spaces sharing $\CEventsM$ are always induced from the conditioning events in $\CEventsM$, which allows us to introduce the following notational conventions extensively employed throughout this work.

\begin{notation}[Common Conditioning Events]
\label{not:common_conditioning_events}
Given a conditional measurable space  $(M, \sAlg_M, \allowbreak \CEvents_{M})$, for every product conditional measurable space $(M \times X, \sAlg_M \otimes \sAlg_X, \CEvents_{M \times X})$ induced by $\CEventsM$, we write:
\begin{itemize}[leftmargin=*]
\item $(M \times X, \sAlg_M \otimes \sAlg_X, \CEventsM)$ instead of $(M \times X, \sAlg_M \otimes \sAlg_X, \CEvents_{M \times X})$, 

\item $\CPSM (M \times X)$ instead of $\Delta^{\CEvents_{M \times X}} (M \times X)$, and

\item $\big( \cps (\cdot | C )  \big)_{C \in \CEventsM} \in \CPSM (M \times X)$ instead of  $\big( \cps (\cdot | C \times X)  \big)_{C \in \CEventsM}$.
\end{itemize}
In particular, whenever we use the notation  $\CPSM (M \times X)$, it is understood that $M \times X$ is induced by $\CEvents_M$, i.e., $\CEvents_{M \times X}$ is defined as in \Mref{eq:product_conditioning}. 
\end{notation}

Given a conditional measurable space  $\MSC$, two product conditional measurable spaces  $(M \times X, \sAlg_M \otimes \sAlg_X, \CEvents_{M \times X})$ and $(M \times Y, \sAlg_M \otimes \sAlg_Y, \CEvents_{M \times Y})$ sharing $\CEventsM$, and a measurable function $f \in (M \times Y)^{M \times X}$,
the \emph{image measure with conditioning events} of $f$ is the function
\begin{equation*}
\pushf  :=  ( \pushf_{D_Y} )_{D_Y \in \CEvents_{M \times Y}} : \Delta^{\CEvents_{M \times X}} (M \times X) \to \Delta^{\CEvents_{M \times Y}} (M \times Y) ,
\end{equation*}
defined as
\begin{equation*}
\pushf_{D_Y} (\cps) (E) := \cps \big( f^{-1} (E) | D_X \big) ,
\end{equation*}
for every $D_Y \in \CEvents_{M \times Y}$ and $D_X \in \CEvents_{M \times X}$ such that there exists a $C \in \CEventsM$ with $\pi_M D_Y = \pi_M D_X = C$, CPS $\cps := \big( \cps (\cdot| D_X) \big)_{D_X \in \CEvents_{M \times X}} \in \Delta^{\CEvents_{M \times X}} (M \times X)$, and $E \in \sAlg_M \otimes \sAlg_Y$.\footnote{Obviously, in absence of conditioning events, this definition boils down to the standard one of image measure as in \citet[Chapter 13.12, p.483]{Aliprantis_Border_2006} or \citet[Chapter 3.6, p.190]{Bogachev_2007}.} We rephrase for future reference what we just introduced in the definition that follows, which exploits the notational conventions set forth in \Mref{not:common_conditioning_events}.

\begin{definition}[Image Measure with Conditioning Events]
\label{def:pushforward_conditioning}
Given a conditional measurable space  $\MSC$, two product conditional measurable spaces  $(M \times X, \sAlg_M \otimes \sAlg_X, \CEventsM)$ and $(M \times Y, \sAlg_M \otimes \sAlg_Y, \CEventsM)$ sharing $\CEventsM$, and a measurable function $f \in (M \times Y)^{M \times X}$,
the \emph{image measure with conditioning events} of $f$ is the function
\begin{equation*}
\pushf  :=  ( \pushf_{C} )_{C \in \CEventsM} : \CPSM (M \times X) \to \CPSM (M \times Y) ,
\end{equation*}
defined as
\begin{equation*}
\pushf_{C} (\cps) (E) := \cps \big( f^{-1} (E) | C \big) ,
\end{equation*}
for every $C \in \CEventsM$, $\cps := \big( \cps (\cdot| C ) \big)_{C \in \CEventsM} \in \CPSM (M \times X)$, and $E \in \sAlg_M \otimes \sAlg_Y$.
\end{definition}

In the following, it is understood that a given product conditional measurable space $(M \times X, \sAlg_M \otimes \sAlg_X, \CEventsM)$ induced by $\CEventsM$ is endowed with 
\begin{equation}
\label{eq:sAlg_product_cms}
\sAlg_{\CPSM (M \times X)}  := \sigma \big( %
\Set { \gamma^{p}_C (E) | E \in \sAlg_{M} \otimes \sAlg_{X}, \ C \in \CEventsM, \ p \in [0,1] } \big)
\end{equation}
naturally defined as an extension of \Mref{def:sAlg_CPS} to $(M \times X)$.

\subsection{Theory of Coalgebras}
\label{subsec:coalgebras}

In what follows, it is assumed a minimal knowledge of certain notions from category theory:\footnote{See \cite{Leinster_2014} for an introduction to the topic, \cite{MacLane_1998} for a comprehensive introduction, or \cite{Borceux_1994} for a more advanced treatment.} in particular, that of  category (along with the notions of objects and morphisms of a category, and of isomorphism and isomorphic objects), (full) subcategory, (endo)functor (along with the definition of four basic functors, i.e., the identity functor, the constant functor, the power set endofunctor, and the projection functor), and terminal object in a category, where---for self-containment purposes---the related definitions can all be found in \Sref{app:category_theory_background}. Also, we employ the following---common in the field---notational conventions.

\paragraph{Categorical Notational Conventions.} Given an arbitrary functor $\Functor : \Cat \to \Cat'$ with $\Ob (\Cat)$ denoting the objects of the category $\Cat$ and $\Cat (A, A')$ denoting the morphisms between two objects $A , A' \in \Ob (\Cat)$,\footnote{See \Mref{def:category} for the definition of category and the related notation as employed here.} we write $\Functor (X)$ to capture how the functor acts on an arbitrary $X \in \Ob (\Cat)$, whereas we write $\Functor f$ (without brackets) to capture how it acts on an arbitrary morphism $f \in \Cat (A, A')$. Also, given two arbitrary functors $\Functor$ and $\Functor'$, we omit brackets (unless needed to avoid ambiguous expressions) to capture how the first applies to the second, i.e., we write $\Functor \Functor'$. Regarding morphisms, given two arbitrary morphisms $f$ and $g$, we indifferently write ``$f g$'' or ``$f \circ g$'' (of course, assuming that the latter expressions are well-defined). Finally, in commutative diagrams, we represent unique morphisms via \emph{dashed} arrows.

\bigskip

As a matter of fact, this is everything we actually need for the next definition, which is going to prove to be crucial for our endeavour, i.e., that of a coalgebra.\footnote{See \cite{Jacobs_Rutten_1997} for an introductory article or \cite{Jacobs_2017} for a textbook on the topic.}

\begin{definition}[Coalgebra]
\label{def:coalgebra}
Given a category $\Cat$ and an endofunctor $\Functor$ on $\Cat$, a $\Functor$-\emph{coalgebra} is a tuple $\Coalg = \la \Car, \trans \ra$ where:
\begin{itemize}[leftmargin=*]
\item $\Car \in \Ob (\Cat)$ is its \emph{carrier};

\item the morphism $\trans : \Car \to \Functor (\Car)$ is its \emph{transition}.\footnote{\label{foot:co_convention}On the contrary, a $\Functor$-\emph{algebra} is a tuple $\mfrak{A} := \la \Car, \alpha \ra$ with $\alpha : \Functor (\Car) \to \Car$ (see also \citet[Chapter VI.2]{MacLane_1998}). The fact that the orientation of the morphism $\alpha$ is reversed with respect to the morphism $\trans$ is the reason behind the  name ``coalgebra''. Indeed, in category theory (and related fields), the prefix ``co'' is used  when the direction of the morphisms in a given category $\Cat$ is reversed: this amounts at working with the corresponding category $\Cat^{\op}$, whose objects coincide with the objects of $\Cat$ and such that $f^{\op} \in \Cat^{\op} (A', A)$ if $f \in \Cat (A, A')$. In particular, this has an immediate implication, which goes along the name of ``Principle of Duality'', that can be informally stated as follows:  every categorical definition, theorem, or proof has a dual, obtained by reversing all the morphisms (see \citet[Remark 1.1.10, p.16]{Leinster_2014}).} 
\end{itemize}
\end{definition}

Thus, a coalgebra is a structure that extracts information from a carrier $\Car$ via the transition $\trans$.\footnote{See \citet[Chapter 1]{Jacobs_2017} for an overview of how coalgebras arise in different contexts.}  In light of this, it is important to capture the way in which, given two coalgebras, the structure of the first is preserved when `moving' to the second. The next definition achieves this goal.

\begin{definition}[Coalgebra Morphism \& Coalgebra Isomorphism]
\label{def:coalgebra_morphism}
Given a category $\Cat$, an endofunctor $\Functor$ on $\Cat$, and two $\Functor$-coalgebras $\Coalg := \la \Car, \trans \ra$ and $\Coalg' := \la \Car', \trans' \ra$, a \emph{$\Functor$-coalgebra morphism} is a morphism $\cmor :  \Car \to \Car'$
such that the following diagram
\begin{equation*}
\begin{tikzcd}
\Car %
	\arrow[r, "\cmor"] %
	\arrow[d, "\trans"'] %
	& \Car' %
		\arrow[d, "\trans'"] \\
\Functor (\Car) %
	\arrow[r, "\Functor \cmor "'] %
	& \Functor (\Car')
\end{tikzcd}
\end{equation*}
commutes. If $\cmor$ is an isomorphism, then it is a $\Functor$-coalgebra isomorphism.
\end{definition}

\begin{notation}[Category of Coalgebras of an Endofunctor]
Given a category $\Cat$ and an endofunctor $\Functor$ on $\Cat$, we let $\CoAlg (\Functor)$ denote the category of $\Functor$-coalgebras, whose objects are $\Functor$-coalgebras and whose morphisms are $\Functor$-coalgebra morphisms.
\end{notation}

In this paper, we focus on the existence of a terminal object\footnote{See \Mref{def:terminal_object} for the definition of terminal object.} in the category of $\Functor$-coalgebra for a given endofunctor $\Functor$ on a category $\Cat$, i.e., the so-called terminal $\Functor$-coalgebra.

\begin{definition}[Terminal Coalgebra]
\label{def:terminal_coalgebra}
Given a category $\Cat$ and an endofunctor $\Functor$ on $\Cat$, a terminal $\Functor$-coalgebra $\overline{\Coalg} := \la \overline{\Car}, \overline{\trans} \ra$ is a terminal object in $\CoAlg (\Functor)$, i.e., it is a $\Functor$-coalgebra such that for every $\Functor$-coalgebra $\Coalg := \la \Car, \trans \ra$ there exists a unique $\Functor$-coalgebra morphism $\overline{\cmor} :  \Car \to \overline{\Car}$.
\end{definition}

When a terminal coalgebra exists in $\CoAlg (\Functor)$, then this object is \emph{essentially} unique, in light of the next well-known result from category theory, where we use the symbol ``$\cong$'' to capture the existence of an isomorphism between two objects in a category.\footnote{See \Mref{def:isomorphism} for the definition of isomorphism and \Mref{def:isomorphic_objects} for the definition of isomorphic objects.}

\begin{lemma}
\label{lem:uniqueness_ter_obj}
Given a category $\Cat$, terminal objects $\1, \1' \in \Ob (\Cat)$ are isomorphic, i.e., $\1 \cong \1'$.
\end{lemma}

Whereas---as pointed out above---the focus of this work is on establishing the existence of a terminal coalgebra given an endofunctor on a category, it is \emph{not} always the case that there exists a terminal coalgebra. Indeed, for example, given the category $\SET$ with $\PowerSet$ denoting the power set and  $\PowerSetF$ denoting the power set endofunctor,\footnote{See \Mref{ex:categories} for the definition of the category $\SET$ and \Mref{ex:endofunctors} for the definition of the power set endofunctor.} there does not exist a terminal $\PowerSetF$-coalgebra, i.e., there exists no terminal $\PowerSetF$-coalgebra $\overline{\Coalg} := \la \overline{\Car}, \overline{\trans} \ra$ such that $\overline{\Car} \cong \PowerSet (\overline{\Car})$, which is an immediate consequence of Cantor's theorem. However, when the terminal coalgebra on a given endofunctor on a category does exist, it is possible to exploit an important result from category theory and the theory of coalgebras known as \emph{Lambek's lemma},\footnote{In \cite{Lambek_1968}, there is no mentioning of the underlying coalgebraic nature of the result. See also \citet[Chapter 2.3, pp.69--70]{Jacobs_2017} for a textbook presentation of the result.} that we recall next.

\begin{lemma}[{\citet[Lemma 2.2, p.153]{Lambek_1968}}]
\label{lem:lambek}
If $\overline{\Coalg} := (\overline{\Car}, \overline{\trans})$ is a terminal $\Functor$-coalgebra, then the transition $\overline{\trans} : \overline{\Car} \to \Functor (\overline{\Car})$ is an isomorphism, i.e., $\overline{\Car} \cong \Functor (\overline{\Car})$.
\end{lemma}

Thus, regarding the interpretation of \Mref{lem:lambek}, by focusing on the representation of the result as establishing $\overline{\Car} \cong \Functor (\overline{\Car})$, it is important to observe that this amounts at finding a \emph{fixed point} for a given endofunctor $\Functor$.

\section{Type Structures}
\label{sec:measure-theoretic_type_structure}

In the following, we let $I$ denote a finite set of \emph{agents} (or individuals) and $0$ stand for what is called ``Nature'', with $0 \notin I$, from which we define $I_0 := I \cup \{ 0 \}$: we typically use the symbol ``$i$'' for a representative element of $I$ and ``$j$'' for a representative element of $I_0$. Concerning these sets when used for indexing purposes, we adopt the following standard conventions:
\begin{itemize}[leftmargin=*]
\item if $I_0$ is the index set, we let $X := \prod_{j \in I_0} X_j$ and $X_{-j} := \prod_{y \in I_0 \setminus \{j \}} X_y$;

\item if $I$ is the index set, we let $X := \prod_{i \in I} X_i$ and $X_{-i} := \prod_{y \in I \setminus \{i\}} X_y$.
\end{itemize}

Given a family of functions $(f_j)_{j \in I_0}$ of the form $f_j : X_j \to Y_j$ (with the same convention applied when the index set is $I$), the \emph{induced function} $f : X \to Y$ is defined as
\begin{equation*} 
f \big( (x_j )_{j \in I_0} \big) := \big( f_j (x_j) \big)_{j \in I_0}.
\end{equation*}
Finally, we employ the following non-standard convention, that considerably lightens our notation: given an agent $i \in I$ and an $\abs{I_0}$-tuple of functions $(f_0, f_i, f_{-i})$, we let $f_{\pm i} := (f_0, f_{-i})$.

\subsection{Basic Definition}
\label{subsec:standard_type_structure}

For our endeavour, first of all, we fix a conditional measurable space $\TSC$ that captures the common domain of uncertainty of interest. We now provide a formal definition of what a type structure is on $\TSC$.

\begin{definition}[Type Structure]
\label{def:type_structure}
A \emph{type structure} on a conditional measurable space $(\State, \sAlg_\State, \CEventsT)$ is a tuple
\begin{equation*}
\mscr{T} := \la (T_i, \beta_i )_{i \in I} \ra
\end{equation*}
where, for every $i \in I$,
\begin{itemize}[leftmargin=*]
\item $T_i$ is a measurable space, called the \emph{type space}\footnote{In the literature, the word ``space'' is often used to refer both to a tuple of objects such as $\mscr{T}$ \emph{and} the type spaces $T_i$, for every $i \in I$. Here, we distinguish these constructs by using the word ``structure'' for the tuple and the word ``space'' for the set of types.} of agent $i$;
\item $\beta_i := (\beta_{i, C})_{C \in \CEventsT} : T_i \to \CPST (\State \times T_{-i})$ is a measurable function, called the \emph{belief function} of agent $i$.
\end{itemize}
\end{definition}

Thus, given that $t := (t_i)_{i \in I}$, an element $(\theta, t)$ is called a \emph{state of the world}, with $\State \times T$ called the set of states of the world or \emph{state space}, while a $t_i \in T_i$ is called an \emph{epistemic type} (henceforth, type) of agent $i \in I$, for every $i \in I$. In particular, for every $i \in I$ and type $t_i \in T_i$, the belief function $\beta_i (t_i)$ captures the belief of type $t_i$ regarding $\State \times T_{-i}$, for every conditioning event $C \in \CEventsT$, i.e., $\beta_{i, C} (t_i) \in [\Delta (\State \times T_{-i})]^{\{C\}} =  \Delta (\State \times T_{-i})$, for every $C \in \CEventsT$.

\subsection{Relation Between Type Structures}
\label{subsubsec:type_morphisms}

Having formalized the notion of type structure, we now provide a formal definition of a morphism between two type structures on a conditional measurable space $(\State, \sAlg_\State, \CEventsT)$ that preserves the properties of the spaces belonging to the two structures.

\begin{definition}[Type Morphism \& Type Isomorphism]
\label{def:type_morphism}
Given a conditional measurable space $\TSC$, two type structures  $\mscr{T} := \la (T_i, \beta_i )_{i \in I} \ra$ and $\mscr{T}' := \la (T'_i, \beta'_i )_{i \in I} \ra$ on the conditional measurable space $(\State, \sAlg_\State, \CEventsT)$, the induced  measurable function $\tmor := (\tmor_j)_{j \in I_0}$ with $\tmor_0 : \State \to \State$ and $\tmor_i : T_{i} \to T'_{i}$, for every $i \in I$,  is a \emph{type morphism} if
\begin{enumerate}[leftmargin=*, label=\arabic*)]
\item $\tmor_0 := \id_\State$;

\item for every $i \in I$, $\beta'_i \circ \tmor_i = \widehat{\tmor_{\pm i}} \circ \beta_i$, i.e., the diagram
\begin{equation*}
\begin{tikzcd}[column sep = huge]
T_i %
	\arrow[r, "\tmor_i"] %
	\arrow[d, "\beta_i"'] %
	& T'_i %
		\arrow[d, "\beta'_i"] \\
\CPST (\Theta \times T_{-i}) %
	\arrow[r, "\widehat{\tmor_{\pm i}}"'] %
	& \CPST (\Theta \times T'_{-i})
\end{tikzcd}
\end{equation*}
commutes.
\end{enumerate}
If $(\tmor_j)_{j \in I_0}$ is a measurable isomorphism, then the type morphism is a \emph{type isomorphism}.
\end{definition}

In particular, Condition (2) in \Mref{def:type_morphism} can be alternatively formalized by saying that 
\begin{equation*}
\beta'_{i, C} (\tmor_i (t_i)) (E) = \beta_{i, C} (t_i) \big( \tmor^{-1}_{\pm i} (E) \big) 
\end{equation*} 
for every $i \in I$, $t_i \in T_i$, $C \in \CEventsT$, and $E \subseteq \State \times T'_{-i}$.

\subsection{Infinite Hierarchies of Beliefs}
\label{subsec:IHBs}

Given a conditional measurable space $\TSC$ and a type structure $\mscr{T} : = \la (T_i, \beta_i)_{i \in I} \ra$ appended to it, for every $i \in I$, we can unpack the information contained in a type $t_i \in T_i$. This can be achieved by means of infinite hierarchies of beliefs. 

The idea behind the construction of infinite hierarchies of beliefs is to construct beliefs of increasing order, and, for every belief order, a corresponding family of conditioning events.

\begin{definition}[The Hierarchical Space]
\label{def:hierarchical_space}
Let $H^{n}_0 := \State$ for every $n \in \bN := \{0, 1, 2, \dots \}$ and for every $i \in I$ proceed with the following inductive construction with $n \in \bN$:
\begin{center}
\begin{tabular}{cc}
$ H^{0}_{i} := \{ \bullet \}$, 		&%
$\CEvents^{0}_i :=  \CEventsT$, \\
$\vdots$			& 	$\vdots$ \\
$ H^{n+1}_{i} := H^{n}_i \times \Delta^{\mcal{C}^{n}_i} (H^{n}_{\pm i} ) $, &%
$\CEvents^{n+1}_i := \Set {  D \subseteq H^{n}_{\pm i} | %
\exists C \in \CEvents^{0}_i : D = C \times H^{n}_{- i} } $,   \\
$\vdots$		&	 	$\vdots$ \\
\end{tabular}
\end{center}
The \emph{hierarchical space of agent $i$} is 
\begin{equation*}
H_{i} :=  H^{0}_i \times \prod_{\ell \in \bN} \Delta^{\CEvents^{\ell}_i} (H^{\ell}_{\pm i}) ,
\end{equation*}
whereas $H := \prod_{i \in I} H_i$ is the \emph{hierarchical space}.\footnote{It is understood that in this definition, in the part that pertains the conditioning events, we exploit \Mref{not:common_conditioning_events}.}
\end{definition}

We now introduce a family of functions, one for every agent $i \in I$, to extract information from the hierarchical space.

\begin{definition}[Hierarchy Functions]
\label{def:hierarchy_function}
Given a type structure $\mscr{T} := \la (T_i , \beta_i)_{i \in I} \ra$ and an arbitrary agent $i \in I$, the function 
\begin{equation*}
\hier^{\beta, n}_{i} := (\hier^{\beta, n}_{i, C})_{C \in \CEventsT} : T_i \to H^{n}_{i}
\end{equation*}
is the \emph{$n^{\text{th}}$-order hierarchy function} of $i \in I$, inductively defined as follows, for every $n \in \bN$, $C \in \CEventsT$, and $t_i \in T_i$:
\begin{itemize}[leftmargin=*]
\item  ($n = 0$) Let $\hier^{\beta, 0}_{i, C}$ be uniquely defined as $\hier^{\beta, 0}_{i, C} : T_i \to \{ \bullet\} $;

\item ($n \geq 0$) Assume that $\hier^{\beta, n}_C := \Big( \id_{\State, C}, (\hier^{\beta, n}_{i, C} )_{i \in I} \Big)$ has been defined, with $\id_{\State,C} := \id_\State$ for every $C \in \CEventsT$, and let $\hier^{\beta, n+1}_{i, C}$ be defined as 
\begin{align*}
\hier^{\beta, n+1}_{i, C} (t_i) %
& := \bigg( \hier^{\beta, n}_{i, C} (t_i), \, \beta_{i, C} (t_i) \circ  \Big( \id_{\State, C}, (\hier^{\beta, n}_{-i, C})^{-1} \Big) \bigg) , \\
& \textcolor{white}{:}= \bigg( %
\hier^{\beta, 0}_{i, C} (t_i), \, %
\beta_{i, C} (t_i) \circ \Big( \id_{\State, C}, (\hier^{\beta, 0}_{-i, C})^{-1} \Big) , %
\, \dots, \, %
\beta_{i, C} (t_i) \circ \Big( \id_{\State, C}, (\hier^{\beta, n}_{-i, C})^{-1} \Big)  %
\bigg) , 
\end{align*}
for every $C \in \CEventsT$.
\end{itemize}
Thus, the \emph{hierarchy function of agent $i$} is the function $\hier^{\beta}_i := (\hier^{\beta}_{i, C})_{C \in \CEventsT} : T_i \to H_i$ such that $\hier^{\beta, n}_{i, C} = \pi_{n} \circ \hier^{\beta}_{i, C}$, for every $n \in \bN$. Finally, the \emph{hierarchy function} is the unique induced function
\begin{equation*}
\hier^{\beta}  := (\hier^{\beta}_{i, C} )_{ i \in I, C \in \CEventsT} : T \to H ,
\end{equation*}
where an element $\hier^{\beta}_C (t)$ is the \emph{hierarchy description of $t \in T$ at $C$}.\footnote{See \cite{Friedenberg_Meier_2011} for an analysis of the relation between this notion and that of type morphism.}
\end{definition}

\subsection{Large Type Structures}
\label{subsec:large_type_structures}

In light of what we introduced in the previous sections, we can now collect the definitions of large type structures that we employ in this paper.

The notion of terminality we use goes back to  \citet[Section 4, p.19]{Armbruster_Boge_1979}\footnote{\label{foot:AB79}Given the general lack of availability of this paper, we report here what we consider the relevant part with respect to the point above. Thus, interestingly, given that in \cite{Armbruster_Boge_1979} an \emph{oracle system} is essentially a type structure, in \citet[Section 4, p.19]{Armbruster_Boge_1979}, we find that \emph{``the canonical oracle system is a terminal object in the category of oracle systems and $n$-tuples of commuting maps''}, where the authors use the expression ``canonical oracle system'' essentially to refer to the universal type structure built from infinite hierarchies of coherent beliefs (as a projective limit) and the expression ``$n$-tuples of commuting maps'' to refer to an $n$-tuple of type isomorphisms, with $n \in \bN$ being the cardinality of the set of agents. We are grateful to Michael Greinecker for having made available to us the original article.} and \citet[Section 2, p.196]{Boge_Eisele_1979}, that employ this terminology remaining true to the original categorical\footnote{The same terminology can be found in \cite{Vassilakis_1991}, \cite{Vassilakis_1992}, and \cite{Pinter_2010}, that refer explicitly to a categorical reformulation of the problem at hand.} formulation of the problem at hand.\footnote{It should be pointed out how two other notions of terminality can be found in the literature, namely, in \citet[p.93]{Siniscalchi_2008} (built on the intuition in \citet[Remark 2, p.201]{Battigalli_Siniscalchi_1999}) and \citet[Definition 2.6, p.114]{Friedenberg_2010}. We are grateful to Gabriel Ziegler for having raised our attention to this point.}

\begin{definition}[Terminal Type Structure]
\label{def:terminal_type_structure}
A type structure $\otbar{\mscr{T}} := \la (\otbar{T}_i , \otbar{\beta}_i )_{i \in I} \ra$ is \emph{terminal} if for every type structure $\mscr{T}$ on $(\State, \sAlg_\State, \CEventsT)$ there exists a unique type morphism from $\mscr{T}$ to $\otbar{\mscr{T}}$.
\end{definition}

The next notion has been introduced in the literature as \emph{completeness}\footnote{With \cite{Siniscalchi_2008} adopting the same term. By employing this terminology, we follow \cite{Meier_2012}. Regarding this notion, see \Sref{subsubsec:notion_belief-completeness}.} in \cite{Brandenburger_2003}.

\begin{definition}[Belief-Complete Type Structure]
\label{def:belief-complete_type_structure}
A type structure $\otbar{\mscr{T}} := \la (\otbar{T}_i , \otbar{\beta}_i )_{i \in I} \ra$ is \emph{belief-complete} if the function $\otbar{\beta}_i$ is surjective, for every $i \in I$.
\end{definition}

The next definition, introduced in \citet[Definition 2.4, p.6]{Mertens_Zamir_1985},  formally captures the idea that, given a type structure $\mscr{T} := \la (T_i , \beta_i )_{i \in I} \ra$, the hierarchy function $\hier^{\beta}_i$ is injective, for every $i \in I$, i.e., for every $t_i , t'_i \in T_i$, if $t_i \neq t'_i$, then $\hier^{\beta}_i (t_i) \neq \hier^{\beta}_i (t'_i)$,\footnote{See in general the discussion in \citet[Section 2, p.6]{Mertens_Zamir_1985}.} where, in particular, it should be observed how \citet[Proposition 2, p.2123]{Liu_2009} establishes that the injectivity of the (induced) hierarchy function of a given type structure actually characterizes  its non-redundancy.

\begin{definition}[Non-redundancy]
\label{def:non-redundancy}
Given a type structure $\otbar{\mscr{T}} := \la (\otbar{T}_i , \otbar{\beta}_i )_{i \in I} \ra$ and the corresponding (induced) hierarchy function $\hier^{\otbar{\beta}} := \big( \hier^{\otbar{\beta}}_i \big)_{i \in I}$, $\otbar{\mscr{T}}$ is \emph{non-redundant} if the  $\sigma$-algebra $\sigma (\hier^{\otbar{\beta}_i}_i)$ on $\otbar{T}_i$ is separating, for every $i \in I$.\footnote{This notion has been introduced in \citet[Definition 2.4, p.6]{Mertens_Zamir_1985}. See in general the discussion in \citet[Section 2, p.6]{Mertens_Zamir_1985} to motivate it and \cite{Liu_2009}.}
\end{definition}

Thus, extending a terminology introduced in \cite{Siniscalchi_2008} and hinted in \citet[Section 3.4]{Di_Tillio_2008}, we now define the so-called universal type structure.\footnote{See \Sref{subsubsec:notion_universality} for a discussion of this notion by considering that an alternative notion of universality can be found in the literature, as in \citet[Section 11.i]{Brandenburger_Keisler_2006} and \citet[Section 5.a]{Friedenberg_2010}.} 

\begin{definition}[Universal Type Structure]
\label{def:universal_type_structure}
A type structure $\otbar{\mscr{T}} := \la (\otbar{T}_i , \otbar{\beta}_i )_{i \in I} \ra$ is  \emph{universal} if it is terminal, belief-complete, and non-redundant.
\end{definition}

\section{Universality}
\label{sec:universality}

The main goal of this paper is to show that, for every conditional measurable space $(\State, \sAlg_\State, \CEventsT)$, there exists a type structure appended to it that is universal as in \Mref{def:universal_type_structure}.

In this section we provide a proof of the following theorem, where, for the case of a measurable space in absence of conditioning events,  terminality, uniqueness, and non-redundancy have been proved by \cite{Heifetz_Samet_1998}, whereas the belief-completeness has been proved by \cite{Meier_2012}, \cite{Moss_Viglizzo_2004}, and \cite{Viglizzo_2005a}.

\begin{theorem}
\label{th:main_theorem}
For every conditional measurable space $\TSC$, there exists a type structure $\mscr{T}^* := \la (T^{*}_i, \beta^{*}_i )_{i \in I} \ra$ on $(\State, \sAlg_\State, \CEventsT)$ such that:
\begin{enumerate}[leftmargin=*, label=\arabic*)]
\item  {\emph{Terminality:}} $\mscr{T}^*$ is terminal;

\item {\emph{Uniqueness:}} $\mscr{T}^*$ is unique up to measurable type isomorphism;

\item {\emph{Belief-Completeness:}} $\mscr{T}^*$ is belief-complete;

\item {\emph{Non-Redundancy:}} $\mscr{T}^*$ is non-redundant.
\end{enumerate}
The type structure $\mscr{T}^*$ is the \emph{Universal Type Structure} on $(\State, \sAlg_\State, \CEventsT)$, which is unique up to measurable type isomorphism.
\end{theorem}

To establish this result, we employ the following strategy. First of all, in \Sref{subsec:there}, we bypass the presence of interactive agents and we show that there is essentially a basic mathematical structure underlying the very notion of type structure (in presence of conditioning events), which is nothing more than a coalgebra for an appropriate endofunctor. As a result, we establish the existence of a terminal coalgebra for this endofunctor  in \Mref{prop:coalgebraic_main_theorem}. Armed with this result, in \Sref{subsec:dale}, we tackle the presence of interactive agents and we show that the main bulk of the endeavour has---essentially---already been established. Indeed, a minimal change in the definition of the appropriate functor is enough to establish \Mref{prop:coalgebraic_main_theorem_type}, which is a translation of \Mref{th:main_theorem} in coalgebraic terms as set in \Mref{prop:coalgebraic_main_theorem} (i.e., taking into account the presence of interactive agents). Finally, in \Sref{subsec:back_again}, we simply show that type structures are indeed coalgebras for the appropriate endofunctor defined in \Sref{subsec:dale}, thus exploiting the obtained result to establish \Mref{th:main_theorem} essentially as its immediate corollary.

Before embarking in proceeding along the lines of the path sketched above, it is important to point out that the sections that follow have been written in a modular fashion. As a result, the reader who does not want to delve into the details of the coalgebraic canonical construction along with its structural properties as described in details in \Sref{subsec:there} can skip that section and move to \Sref{subsec:dale}, where there is essentially a description of type structures as a certain kind of coalgebras, which is actually the very point established in \Sref{subsec:back_again} which delivers the theorem above.

\subsection{There (in the Coalgebraic Framework)\dots}
\label{subsec:there}

The very first step of our endeavour is to introduce the category that is going to be the main building block behind our undertaking. 

\begin{definition}[{Category $\Meas$}] 
\label{def:Meas}
The category $\Meas$ is the category of measurable spaces, where
\begin{itemize}[leftmargin=*]
\item the \emph{objects} of $\Meas$ are measurable spaces;

\item the \emph{morphisms} of $\Meas$ are measurable functions.
\end{itemize}
\end{definition}

Now, given that we fix a conditional measurable space $\MSC$, where---of course---we have that $M \in \Ob (\Meas)$, the first two functors needed for our endeavour are the constant functor $\Cons_M$ and the identity functor $\Id$.\footnote{See \Mref{ex:endofunctors} in \Sref{app:category_theory_background} for their definitions, where it should be observed how, from a typographical standpoint, we distinguish the identity \emph{endofunctor} from the identity \emph{morphism} via the usage in the former of an \emph{uppercase} initial letter (i.e., ``I'').} As a matter of fact, they are the building blocks of the derived functor $\ProdFM$, which is the functor defined as the binary product of the constant functor $\Cons_M$ and the identity functor $\Id$, i.e., $\ProdFM := \Cons_M \times \Id$, where
\begin{itemize}[leftmargin=*]
\item for every $X  \in \Ob (\Meas)$, $\ProdFM (X) := M \times X$,
\item and  
\begin{equation*}
\label{eq:ProdF}
\ProdFM f : M \times X \to M \times Y
\end{equation*}
such that $\ProdFM f := (\id_M, f)$, for every $f \in \Meas (X, Y)$.
\end{itemize}

However, the category $\Meas$ is somewhat too `large' to capture the peculiarities of product conditional measurable spaces and functions between them. The definition that follows introduces a (full) subcategory\footnote{See \Mref{def:subcategory} for the definition of (full) subcategory.} of $\Meas$ which---on the contrary---is perfectly suited for this objective.

\begin{definition}[{Subcategory $\MeasM$}] 
\label{def:MeasM}
Given a conditional measurable space $\MSC$, the full subcategory $\MeasM$ of $\Meas$ consists of
\begin{itemize}[leftmargin=*]
\item product conditional measurable spaces sharing $\CEventsM$ and

\item the measurable functions between product conditional measurable spaces sharing $\CEventsM$.
\end{itemize}
\end{definition}

Now, given a (conditional) measurable space $\MSC$, building on \Mref{def:pushforward_conditioning}, we introduce the endofunctor $\CPSFM$ on $\MeasM$ defined as
\begin{itemize}[leftmargin=*]
\item for every $M \times X  \in \Ob (\MeasM)$, $\CPSFM (M \times X) := \CPSM (M \times X)$,
\item and  
\begin{equation*}
\label{eq:CPSF}
\CPSFM f := \pushf : \CPSM (M \times X) \to \CPSM (M \times Y)
\end{equation*}
as in \Mref{def:pushforward_conditioning}, for every $f \in \MeasM (M \times X, M \times Y)$.
\end{itemize}
That this is---indeed---an endofunctor\footnote{That is, a morphism on the same category that satisfies Axioms (AC)--(AI) in \Mref{def:functor}.} is immediate in light of the measurability of the image measure. 

\begin{definition}[Type Functor]
The \emph{type functor} $\TypeF$ is the morphism defined as  
\begin{equation*}
\TypeF  := \CPSFM \ProdFM = \CPSFM ( \Cons_M \times \Id) ,
\end{equation*}
such that 
\begin{itemize}[leftmargin=*]
\item for every $X  \in \Ob (\Meas)$, $\TypeF (X) := \CPSM (M \times X)$,
\item and  
\begin{equation*}
\TypeF f : \CPSM (M \times X) \to \CPSM (M \times Y)
\end{equation*}
as $\TypeF f = \widehat{\ProdFM f}$, for every $f \in \Meas (X , Y)$.
\end{itemize}
\end{definition}

Given the setting above, we can now state a coalgebraic version of \Mref{th:main_theorem}, where the remainder of this section is devoted to establish this result.

\begin{proposition}
\label{prop:coalgebraic_main_theorem}
Fix a conditional measurable space $\MSC$. %
\begin{enumerate}[leftmargin=*, label=\arabic*)]
\item {\emph{Existence:}} there exists a terminal $\TypeF$-coalgebra  $\TCoalg := \la  \Terminal , \belt \ra$;

\item {\emph{Uniqueness:}} the terminal $\TypeF$-coalgebra  $\TCoalg := \la  \Terminal , \belt \ra$  is unique up to $\TypeF$-coalgebra isomorphism;

\item {\emph{Isomorphic Transitions:}} the transition $\belt$ of the terminal $\TypeF$-coalgebra  $\TCoalg := \la \Terminal , \belt \ra$ is an isomorphism.
\end{enumerate}
\end{proposition}

In other words, in line with the interpretation provided in \Sref{subsec:coalgebras} of terminal coalgebras as fixed point of a given functor, we want to establish the existence of an object $\Terminal \in \Ob (\CoAlg (\TypeF))$ such that $\Terminal \cong \TypeF (\Terminal) \equiv \CPSM ( M \times \Terminal)$, i.e., such that the measurable spaces $\Terminal$ and $\CPSM ( M \times \Terminal)$ are isomorphic.\footnote{See \citet[Section 2, p.30]{Moss_2011}.}

\subsubsection{Canonical Construction of the Projective Limit}
\label{subsubsec:canonical_construction_of_the_projective_limit}

The proof of \Mref{prop:coalgebraic_main_theorem} moves from the following coalgebraic canonical construction, which builds on \citet[Section A]{Adamek_Koubek_1979}.\footnote{In \cite{Adamek_Koubek_1979}, the result is established for algebras, but it applies by duality---as in \Sref{foot:co_convention}--to coalgebras as well, as emphasized in \cite{Barr_1993} and \citet[Example, p.60]{Adamek_Koubek_1995}. The intuition for this construction has to be found---in light of the tight relation between posets and categories---in the construction of a least fixed point for directed complete partial orders or lattices, originally identified in \citet[Section 5, p.552]{Scott_1976} in the context of $\lambda$-calculus (see also \citet[Chapter 4.6, p.223]{Jacobs_2017}). \cite{Smyth_Plotkin_1982} builds a categorical framework to deal with the aforementioned constructions. It should be also pointed out that  \cite{Worrell_2005} is a paper that establishes the result we are after by performing a different construction (see \citet[Section 5, p.406]{Viglizzo_2005a} regarding this point).} The crucial element to perform this canonical construction is to work with a category with a terminal object.\footnote{For example, this canonical construction cannot be used with bipointed sets (i.e., sets of the form $\{ \bot \} \sqcup X \sqcup \{ \top \}$ with $X$ an arbitrary set), as employed in \cite{Freyd_2008} in his coalgebraic description of $[0 ,1]$, since the corresponding category does \emph{not} have a terminal object.} It turns out that the category $\Meas$ has a terminal object, namely, the measurable space $(\1 , \sAlg_{\1})$, which is a singleton set $\{ \bullet \}$ endowed with the discrete (and trivial) $\sigma$-algebra, where---of course---working with $(\1 , \sAlg_{\1})$ as a terminal object is without loss of generality, since all terminal objects in a given category are isomorphic from \Mref{lem:uniqueness_ter_obj}. 

Given what is written above, the first step of the construction consists in identifying the functors upon which our functor of interest $\TypeF$ is built, to the then collect them: the idea  is to decompose a given functor in `more basic' functors that should act as the \emph{ingredients} upon which the functor is built.\footnote{As in \cite{Jacobs_2001} (see also \citet[Chapter 6.5, p.379]{Jacobs_2017}), which is the path taken in \cite{Moss_Viglizzo_2004} and \cite{Viglizzo_2005a}. See also \Sref{subsubsec:Viglizzo_2005}.} Thus, given that we obtain  $\{ \TypeF, \ProdFM, \Cons_M, \Id \}$, for every $\Ing \in \{ \TypeF, \ProdFM, \Cons_M, \Id \}$, we build the corresponding $\Ing$-based $\opchain$-chains,\footnote{Alternatively called ``$\opchain$-sequence'', where the name of this construct comes from the fact that this is a sequence over the first infinite ordinal number $\omega$. See \Mref{def:opchain_endofunctor} for the definition of $\opchain$-chain on an arbitrary endofunctor.} where a \emph{ $\Ing$-based $\opchain$-chain on the endofunctor $\TypeF$} is a diagram of the form 
\begin{equation}
\label{cd:opchain_ingredient}
\begin{tikzcd}[column sep = large]
\Ing (\1) %
	& \Ing \TypeF^{1} (\1) %
		\arrow[l, "\Ing !"'] %
		& \Ing \TypeF^{2} (\1)  %
			\arrow[l, "\Ing \TypeF^{1} !"'] %
			& \dots. %
				\arrow[l, "\Ing \TypeF^{2} !"']
\end{tikzcd}
\end{equation}
Thus, in particular, by spelling out the details:
\begin{itemize}[leftmargin=*] 
\item when $\Ing := \TypeF$, we have
\begin{equation}
\label{cd:opchain_functor}
\begin{tikzcd}
\TypeF^{1} (\1) %
	& \TypeF^{2} (\1) %
		\arrow[l, "\TypeF^{1} !"'] %
		& \TypeF^{3} (\1) %
			\arrow[l, "\TypeF^{2} !"'] %
			& \dots, %
				\arrow[l, "\TypeF^{3} !"']
\end{tikzcd}
\end{equation}
which corresponds to
\begin{equation*}
\begin{tikzcd}
\CPSM (M \times \1) %
	& \CPSM \big(M \times (\CPSM (M \times \1) \big) %
		\arrow[l, "\TypeF^{1} !"'] %
		& \CPSM \Big(M \times \CPSM \big(M \times (\CPSM (M \times \1) \big) \Big)  %
			\arrow[l, "\TypeF^{2} !"'] %
			& \dots; %
				\arrow[l, "\TypeF^{3} !"']
\end{tikzcd}
\end{equation*}

\item when $\Ing := \ProdFM$, we have
\begin{equation*}
\begin{tikzcd}[column sep = large]
\ProdFM (\1) %
	& \ProdFM \TypeF^{1} (\1) %
		\arrow[l, "\ProdFM !"'] %
			& \ProdFM \TypeF^{2} (\1)  %
				\arrow[l, "\ProdFM \TypeF^{1} !"'] %
					& \dots, %
						\arrow[l, "\ProdFM \TypeF^{2} !"']
\end{tikzcd}
\end{equation*}
which corresponds to
\begin{equation*}
\begin{tikzcd}[column sep = huge]
M \times \1 %
	& M \times \TypeF^{1} (\1) %
		\arrow[l, "({\id_M ,} !)"'] %
		& M \times \TypeF^{2} (\1)  %
			\arrow[l, "({\id_M ,}  \TypeF^{1} ! )"'] %
			& \dots ; %
				\arrow[l, "({\id_M ,}  \TypeF^{2} !)"']
\end{tikzcd}
\end{equation*}

\item when $\Ing := \Cons_M$, we have
\begin{equation*}
\begin{tikzcd}[column sep = large]
\Cons_M (\1) %
	& \Cons_M \TypeF^{1} (\1) %
		\arrow[l, "\Cons_M !"'] 
		& \Cons_M \TypeF^{2} (\1) %
			\arrow[l, "\Cons_M \TypeF^{1} !"'] %
			& \dots, %
				\arrow[l, "\Cons_M \TypeF^{2} !"']
\end{tikzcd}
\end{equation*}
which corresponds to
\begin{equation*}
\begin{tikzcd}[column sep = large]
M %
	& M  %
		\arrow[l, "\id_M"'] %
		& M  %
			\arrow[l, "\id_M"'] %
			& \dots ; %
				\arrow[l, "\id_M"']
\end{tikzcd}
\end{equation*}

\item when $\Ing := \Id$, we have
\begin{equation*}
\begin{tikzcd}[column sep = large]
\Id (\1) %
	& \Id \TypeF^{1} (\1) %
		\arrow[l, "\Id  !"'] %
		& \Id \TypeF^{2} (\1)  %
			\arrow[l, "\Id  \TypeF^{1} !"'] %
			& \dots , %
				\arrow[l, "\Id  \TypeF^{2} !"']
\end{tikzcd}
\end{equation*}
which corresponds to
\begin{equation*}
\begin{tikzcd}
\1 & \TypeF^{1} (\1) \arrow[l, dashed, "!"'] & \TypeF^{2} (\1)  \arrow[l, "\TypeF^{1} !"'] & \dots . \arrow[l, "\TypeF^{2} !"']
\end{tikzcd}
\end{equation*}
\end{itemize}

Now, given a $\Ing$-based $\opchain$-chain on the endofunctor $\TypeF$, 
the  \emph{$\Ing$-based projective limit}\footnote{See \Mref{def:projective_limit} for the definition of limit in category theory: the notion of projective limit employed here can be considered a special instance of the latter---more general---definition  (with the understanding that, in category theory, it is possible to find the expressions ``projective limit'' or ``inverse limit'' as synonyms of ``limit'').} of the $\Ing$-based  $\opchain$-chain on the endofunctor $\TypeF$ is the set 
\begin{equation}
\label{eq:Ing_projective_limit}
\Projective_\Ing := \Set { (x_\ell)_{\ell \in \bN} \in \prod_{\ell \in \bN} \Ing \TypeF^\ell (\1) | %
\forall n \in \bN \ \Big( \Ing \TypeF^{n} ! (x_{n+1}) = x_n \Big) } 
\end{equation}
endowed with the $\sigma$-algebra 
\begin{equation*}
\label{eq:Ing_algebra_terminal}
\sAlg_{\Projective_\Ing} := %
\sigma \big( \Set { \pi^{-1}_n (E) | n \in \bN , \ E \in \sAlg_{\Ing \TypeF^n (\1)} } \big) ,
\end{equation*}
where  $\pi^{\Ing}_n : \prod_{\ell \in \bN} \Ing \TypeF^\ell (\1) \to \Ing \TypeF^{n} (\1)$ denotes the projection on $n \in \bN$ as canonically defined.

\begin{remark}
\label{rem:well-definedness_ingredients}
For every $n \in \bN$, $\pi^{\ProdFM}_n (\Projective_{\ProdFM}) = M \times \TypeF^n (1)$ and $\pi^{\Cons_M}_n (\Projective_{\Cons_M}) = M^n$.
\end{remark}

Thus, we can now explicitly introduce the notion of $\opchain$\emph{-chain on the endofunctor} $\TypeF$ and that of projective limit of the $\opchain$-chain on the endofunctor $\TypeF$:  even if they are simply the $\Ing$-based $\opchain$-chain on the endofunctor $\TypeF$ and the  \emph{$\Ing$-based projective limit} of the $\Ing$-based  $\opchain$-chain on the endofunctor $\TypeF$ with $\Ing := \Id$,  they deserve dedicated definitions in light of their importance for this construction.

\begin{definition}[$\opchain$-Chain on the Endofunctor $\TypeF$]
\label{def:opchain}
An $\opchain$\emph{-chain on the endofunctor} $\TypeF$ is a diagram of the form
\begin{equation}
\label{cd:opchain}
\begin{tikzcd}
\1 & \TypeF^{1} (\1) \arrow[l, dashed, "!"'] & \TypeF^{2} (\1)  \arrow[l, "\TypeF^{1} !"'] & \dots, \arrow[l, "\TypeF^{2} !"']
\end{tikzcd}
\end{equation}
with:
\begin{itemize}[leftmargin=*]
\item concerning objects,
\begin{align*}
\TypeF^{0} (\1) & := \1, \\ 
\TypeF^{1} (\1) & :=  %
\CPSM (M \times \1)  , \\ 
\vdots \\
\TypeF^{n+1} (\1) & :=   \TypeF (\TypeF^{n} (\1)) ,
\end{align*}
where, for every $m \in \bN$, the space $\big( \TypeF^{m} (\1) , \sAlg_{\TypeF^{m} (\1)} \big)$ is a measurable space;

\item concerning morphisms and their composition, 
\begin{align*}
\TypeF^{0} ! & :=  !, \\ 
\TypeF^{1} ! & : \CPSM (M \times \CPSM (M \times \1)) \to \CPSM (M \times \1), \\ 
\vdots \\
\TypeF^{n+1} ! & :=  \TypeF \TypeF^{n} ! .
\end{align*}
\end{itemize}
\end{definition}

One point should be observed regarding \Mref{def:opchain}, namely, that it is essentially in the spirit of the definition of \emph{hierarchies} in \citet[Section 5, p.335]{Heifetz_Samet_1998} and of \Mref{def:hierarchical_space}, which is a recursive definition with a base step that is a singleton set (which in our context is the terminal object $\1$).

\begin{remark}
Since $\TypeF^1 (\1) = \TypeF (\1)$ is nonempty, the morphism $! : \TypeF (\1) \to \1$ is measurable and surjective, from which it follows the measurability and surjectivity of $\TypeF^{n} !$, for every $n \geq 1$.
\end{remark}

In the following, we let  $\pi_n : \prod_{\ell \in \bN} \TypeF^\ell (\1) \to \TypeF^n (\1)$ denote the projection operator on $\prod_{\ell \in \bN} \TypeF^\ell (\1)$ as canonically defined, for every $n \in \bN$. 

\begin{definition}[Projective Limit of the  $\opchain$-chain on $\TypeF$]
\label{def:projective_limit}
The  \emph{projective limit} of the  $\opchain$-chain on the endofunctor $\TypeF$ is the set 
\begin{equation}
\label{eq:projective_limit}
\Projective := \Set { (x_\ell)_{\ell \in \bN} \in \prod_{\ell \in \bN} \TypeF^\ell (\1) | %
\forall n \in \bN \ \Big(  \TypeF^{n} ! (x_{n+1}) = x_n \Big) }.
\end{equation}
endowed with the $\sigma$-algebra 
\begin{equation*}
\label{eq:algebra_terminal}
\sAlg_\Projective := %
\sigma \big( \Set { \pi^{-1}_n (E) | n \in \bN , \ E \in \sAlg_{\TypeF^n (\1)} } \big).
\end{equation*}
\end{definition}

\subsubsection{Defining the Cone}
\label{subsubsec:defining_the_cone}

Having obtained the $\Ing$-based projective limit in \Mref{eq:Ing_projective_limit}, we now introduce a family of morphisms that allow us to `unpack' the information  that is contained in a given $\TypeF$-coalgebra $\Coalg := \la \Car , \trans \ra$ exploiting the $\Ing$-based projective limits.

Thus, given a $\TypeF$-coalgebra $\Coalg := \la \Car , \trans \ra$ and an arbitrary $\Ing \in \{ \TypeF, \ProdFM, \Cons_M, \Id \}$,  a \emph{$\Ing$-based cone}\footnote{See \Mref{def:cone_general} for the definition of cone.} for the $\TypeF$-coalgebra $\Coalg$ (or $\Ing$-based $\Coalg$-cone) is a recursively defined collection of morphisms
\begin{itemize}[leftmargin=*]
\item $\Ing \cone^{\trans}_{0, C} := \ ! : \Ing (\Car) \to \Ing(\1)$ and

\item $\Ing \cone^{\trans}_{n, C} : \Ing(\Car) \to \Ing \TypeF^{n} (\1)$ such that $\Ing \cone^{\trans}_{n+1, C} = \Ing \TypeF \cone^{\trans}_{n, C} \circ \trans$, for every $n \in \bN$,
\end{itemize}
with $C \in \CEventsM$ arbitrary and $\Ing \cone^{\trans}_{n} := \Rounds { \Ing \cone^{\trans}_{n, C} }_{C \in \CEventsM}$ for every $n \in \bN$,  such that\footnote{The following is what is deemed in the literature a \emph{universal property} (see \citet[Introduction]{Leinster_2014} and \citet[Chapter 1.2, p.7]{Jacobs_2017}, or \citet[Chapter I.5, p.33]{Aluffi_2009}).} there exists a unique morphism  
\begin{equation*}
\cone^{\trans}_{\Ing, C} : \Ing (\Car) \to \Projective_\Ing 
\end{equation*}
such that $\Ing \cone^{\trans}_{n, C} = \pi^{\Ing}_n \circ \cone^{\trans}_{\Ing, C}$, for every $C \in \CEventsM$, with $\cone^{\trans}_{\Ing} := (\cone^{\trans}_{\Ing, C})_{C \in \CEventsM}$. 

\begin{remark}
\label{rem:Ing_cone}
For every $n \in \bN$, $\Ing \cone^{\trans}_{n} = (\Ing \TypeF^{n} ! ) \circ \Ing \cone^{\trans}_{n+1}$.\footnote{See \citet[Section 2.2, p.398]{Viglizzo_2005a} concerning the details of the case $\Ing := \Id$.} 
\end{remark}

The reason behind the fact that  this construction is called ``cone'' essentially lies in how the commutative diagram
\begin{equation}
\label{eq:Ing_cone_diagram_representation}
\begin{tikzcd}[row sep = huge]
  &              &              & \Ing (\Car) \arrow[llldd, "\Ing \cone^{\trans}_0"'] \arrow[lldd, "\Ing \cone^{\trans}_1" description] \arrow[ldd, "\Ing \cone^{\trans}_2"] \arrow[rdd, "\Ing \cone^{\trans}_n" description] \arrow[rrrdd, "\cone^{\trans}_\Ing"] &    &       &                                                                                                                          \\
  &              &              &                                                                                                                      &    &       &                                                                                                                          \\
\Ing (\1) & \Ing \TypeF^{1} (\1) \arrow[l, "\Ing !"] & \Ing \TypeF^{2} (\1) \arrow[l, "\Ing \TypeF^{1} !"] & \dots 	& \Ing \TypeF^{n} (\1) & \dots & \Projective_\Ing \arrow[llllll, "\pi^{\Ing}_{0}", bend left] \arrow[lllll, "\pi^{\Ing}_{1}" description, bend left] \arrow[llll, "\pi^{\Ing}_{2}"', bend left] \arrow[ll, "\pi^{\Ing}_n"', bend left]
\end{tikzcd}
\end{equation}
looks like, with the commutative diagram
\begin{equation}
\label{eq:Ing_cone_diagram}
\begin{tikzcd}[row sep = large, column sep = huge]
\Ing (\Car) %
	\arrow[r, "\Ing\trans"] %
	\arrow[d, "\Ing \cone^{\trans}_n"'] %
	\arrow[dr, "\Ing \cone^{\trans}_{n+1}", pos=0.7] %
	& \Ing \TypeF (\Car) %
		\arrow[d, "{\Ing \TypeF \cone^{\trans}_{n} }"] %
\\
\Ing \TypeF^{n} (\1) %
	& \Ing \TypeF^{n+1} (\1) %
		\arrow[l, "\Ing \TypeF^{n} !"] %
\end{tikzcd}
\end{equation}
capturing the definition along with the property captured in \Mref{rem:Ing_cone}, with $n \in \bN$ arbitrary.

The next definition is simply that of the $\Id$-based $\Coalg$-cone: it deserves its own definition since it is the translation in this context of the \emph{description map} of \citet[Section 5, p.336]{Heifetz_Samet_1998} and of \Mref{def:hierarchy_function}.

\begin{definition}[Cone for $\TypeF$-coalgebra $\Coalg$]
\label{def:cone}
Given a $\TypeF$-coalgebra $\Coalg := \la \Car , \trans \ra$,  a \emph{cone} for the $\TypeF$-coalgebra $\Coalg$ (or $\Coalg$-cone) is a recursively defined collection of morphisms
\begin{itemize}[leftmargin=*]
\item $\cone^{\trans}_{0, C} := \ ! : \Car \to \1$ and

\item $\cone^{\trans}_{n, C} : \Car \to \TypeF^{n} (\1)$ such that $\cone^{\trans}_{n+1, C} = \TypeF \cone^{\trans}_{n, C} \circ \trans$, for every $n \in \bN$,
\end{itemize}
with $C \in \CEventsM$ arbitrary and $\cone^{\trans}_{n} := \Rounds { \cone^{\trans}_{n, C} }_{C \in \CEventsM}$ for every $n \in \bN$,  such that there exists a unique morphism  
\begin{equation*}
\cone^{\trans}_C : \Car \to \Projective 
\end{equation*}
such that $\cone^{\trans}_{n, C} = \pi_n \circ \cone^{\trans}_{C}$, for every $C \in \CEventsM$, with  $\cone^{\trans} := \Rounds { \cone^{\trans}_{C} }_{C \in \CEventsM}$.
\end{definition}

In the spirit of \Mref{eq:Ing_cone_diagram}, the commutative diagram 
\begin{equation}
\label{eq:cone_diagram}
\begin{tikzcd}[row sep = large, column sep = huge]
\Car %
	\arrow[r, "\trans"] %
	\arrow[d, "\cone^{\trans}_n"'] %
	\arrow[dr, "\cone^{\trans}_{n+1}", pos=0.7] %
	& \TypeF (\Car) %
		\arrow[d, "{\TypeF \cone^{\trans}_{n} }"] %
	& \hspace{-3.2cm} = \CPSM ( M \times \Car )
\\
\TypeF^{n} (\1) %
	& \TypeF^{n+1} (\1) %
		\arrow[l, "\TypeF^{n} !"] %
		& \hspace{-1.9cm} = \CPSM \big( M \times \TypeF^{n} (\1) \big) %
\end{tikzcd}
\end{equation}
provides a perspicuous representation of the properties captured in  \Mref{def:cone} along with \Mref{rem:Ing_cone} by setting $\Ing := \Id$ (with the understanding that the same process can lead to a commutative diagram in the spirit of \Mref{eq:Ing_cone_diagram_representation}), where $n \in \bN$ is arbitrary.

Having defined the notion of cone, it is crucial for our purposes to establish that, given two $\TypeF$-coalgebras and a $\TypeF$-coalgebra morphism between the two, cones `preserve' the properties of these $\TypeF$-coalgebras in a precise sense, as established in \citet[Proposition 5.1, p.336]{Heifetz_Samet_1998} for what are called there ``(hierarchies) descriptions''.

\begin{lemma}
\label{lem:coalgebra_morphism_preservation}
Given two $\TypeF$-coalgebras  $\Coalg := \la \Car, \trans \ra$ and $\Coalg' := \la \Car', \trans' \ra$ and a $\TypeF$-coalgebra morphism $\cmor  : \Car \to \Car'$, $\cone^{\trans} = \cone^{\trans'} \circ \cmor$ is a $\TypeF$-coalgebra morphism.
\end{lemma}

\subsubsection{Intermezzo: Peculiarities of the Topology-Free Coalgebraic Construction}
\label{subsubsec:intermezzo}

We now revise what we obtained in the previous section. Essentially, this can be compactly captured by two commutative diagrams.\footnote{We would like to thank an anonymous referee for pointing out the usefulness of these commutative diagrams.}  In particular, recalling \Mref{eq:cone_diagram} and the fact that $\TypeF (\Car) = \TypeF^{1} (\Car)$, with the commutative diagram 
\begin{equation}
\label{eq:commutative_diagram}
\begin{tikzcd}[row sep = large, column sep = large]
\Car \arrow[r, "\trans"] \arrow[d, "!"] \arrow[dd, "\cone^{\trans}"', bend right=40] 	& \TypeF^{1} (\Car) \arrow[r, "\TypeF^{1} \trans"] \arrow[d, "\TypeF^{1} !"] &	\TypeF^{2} (\Car)  \arrow[r, "\TypeF^{2} \trans"] \arrow[d, "\TypeF^{2} !"] &	\TypeF^{3} (\Car)  \arrow[r, "\TypeF^{3} \trans"] \arrow[d, "\TypeF^{3} !"]	& \dots \\
\1 	& \TypeF^{1} (\1) \arrow[l, dashed, "!"] 	& \TypeF^{2} (\1) \arrow[l, "\TypeF^{1} !"] 	& \TypeF^{3} (\1) \arrow[l, "\TypeF^{2} !"] 	& \dots \arrow[l, "\TypeF^{3} !"] \\
\Projective \arrow[u, "\pi_{0}"'] \arrow[ru, "\pi_{1}"'] \arrow[rru, "\pi_{2}"', bend right=10] \arrow[rrru, "\pi_{3}"', bend right=20] & 	& 	& 	&                 
\end{tikzcd}
\end{equation}
we capture the cone $\cone^{\trans}$ and the $\opchain$-chain on $\TypeF$, whereas with the commutative diagram 
\begin{equation}
\label{eq:Ing_commutative_diagram}
\begin{tikzcd}[row sep = large, column sep = large]
\Ing (\Car) \arrow[r, "\Ing \trans"] \arrow[d, "\Ing !"] \arrow[dd, "\Ing \cone^{\trans}"', bend right=40] 	& \Ing \TypeF^{1} (\Car) \arrow[r, "\Ing \TypeF^{1} \trans"] \arrow[d, "\Ing \TypeF^{1} !"] &	\Ing \TypeF^{2} (\Car)  \arrow[r, "\Ing \TypeF^{2} \trans"] \arrow[d, "\Ing \TypeF^{2} !"] &	\Ing \TypeF^{3} (\Car)  \arrow[r, "\Ing \TypeF^{3} \trans"] \arrow[d, "\Ing \TypeF^{3} !"]	& \dots \\
\Ing (\1) & \Ing \TypeF^{1} (\1) \arrow[l, "\Ing !"] 	& \Ing \TypeF^{2} (\1) \arrow[l, "\Ing \TypeF^{1} !"] 	& \Ing \TypeF^{3} (\1) \arrow[l, "\Ing \TypeF^{2} !"] 	& \dots \arrow[l, "\Ing \TypeF^{3} !"] \\
\Projective_\Ing \arrow[u, "\Ing \pi_{0}"'] \arrow[ru, "\Ing \pi_{1}"'] \arrow[rru, "\Ing \pi_{2}"', bend right=10] \arrow[rrru, "\Ing \pi_{3}"', bend right=20] & 	& 	& 	&                 
\end{tikzcd}
\end{equation}
we capture what happens when we apply a $\Ing \in \{ \TypeF, \ProdFM, \Cons_M, \Id \}$ to the previous commutative diagram.

Now, if the functor $\TypeF$ would preserve $\opchain$-limits,\footnote{See \Mref{def:preservation_limits} for the definition of functor that preserves $\opchain$-limits. Also, the expression ``$\opchain$-continuous'' can alternatively be found to refer to the same property.}  we would naturally obtain a corresponding terminal coalgebra, with corresponding terminal carrier and terminal transition,\footnote{For example, see \citet[Proposition 4.6.1, p.224]{Jacobs_2017}.} which---as a matter of fact---is the `standard' way of proceeding to obtain a terminal coalgebra when---indeed---a functor preserves $\opchain$-limits. The intuition behind the result is the following and starts by assuming the existence of a limit $\Limit$ for an $\opchain$-chain on an endofunctor $\Functor$ (on an arbitrary category $\Cat$) along with the corresponding cone (e.g., in line with the limit $\Projective$ in \Mref{eq:commutative_diagram} for our case).\footnote{See also \citet[Section 10, p.43]{Rutten_2000} and \citet[Section 2.1, p.614]{Moss_Viglizzo_2006}.} Now, if the endofunctor $\Functor$ would preserve  $\opchain$-limits, it would follow that $\Functor (\Limit)$ would also be a limit, e.g., in our case we would have that $\TypeF (\Projective)$ would be a limit itself of the corresponding $\opchain$-chain.\footnote{As it is established in \citet[Lemma 2, p.765]{Smyth_Plotkin_1982} (for the case of algebras, as in \Sref{foot:co_convention}) or \citet[Proposition 4.6.1, pp.224-225]{Jacobs_2017} (for a textbook presentation).} 

However, the main problem behind the present topology-free construction arises exactly at this stage. Indeed, the functor $\TypeF$ does \emph{not} preserve $\opchain$-limits. In particular, this is the result of the  fact that the functor $\CPSFM$ (upon which $\TypeF$ is built) does not preserve limits in the form of $\opchain$-chains, which is an immediate consequence of \citet[Section 2.2, p.399]{Viglizzo_2005a} in light of the fact that product conditional measurable spaces are a `special' case of measurable spaces, i.e., $\Delta (M) = \CPSM (M)$ for an arbitrary $M$ when $\CEventsM$ is a singleton.\footnote{See  \Sref{subsubsec:consistency} with respect to this point, in particular concerning how it translates in the context of type structures.} 

Thus, having identified the problem, the solution lies in two stages
\begin{enumerate}[leftmargin=*,label=\arabic*)]
\item First of all, we obtain the carrier of the terminal $\TypeF$-coalgebra in an `indirect' fashion, by collecting all the states that belong to a carrier of a $\TypeF$-coalgebra, for \emph{every} possible carrier of a $\TypeF$-coalgebra. In other words, we `carve out' from $\Projective$ those elements that are the actually relevant ones for our purposes in light of the fact that they actually `belong' to some $\TypeF$-coalgebra.

\item Moving from the terminal carrier obtained in the previous step, we opportunely define the terminal transition by exploiting the `ingredients' of the functor $\TypeF$ in the form of all the constructs we have introduced for them in \Sref{subsubsec:defining_the_cone}, i.e., $\Ing$-based $\Coalg$-cones and $\Ing$-based projective limits.
\end{enumerate} 
Hence, the next two sections are devoted to these two steps.

\subsubsection{Obtaining the Terminal Carrier}
\label{subsubsec:terminal_carrier}

Building on the apparatus developed in \Sref{subsubsec:canonical_construction_of_the_projective_limit} and \Sref{subsubsec:defining_the_cone} and on the intuition provided by  \Sref{subsubsec:intermezzo}, we now define for every $\Ing \in \{ \TypeF, \ProdFM, \Cons_M, \Id \}$ a subset of the projective limit $\Projective_\Ing$, namely,
\begin{equation}
\label{eq:Ing_terminal_carrier}
\Terminal_\Ing := \Set { \terminal \in \Projective_\Ing | %
\exists \Coalg := \la \Car, \trans \ra \in \Ob(\CoAlg(\TypeF)) \ %
\exists \car \in \Car : \ \terminal = \cone^{\trans}_{\Ing} (\car) },
\end{equation} 
which is endowed with the $\sigma$-algebra $\sAlg_{\Terminal_\Ing}$ inherited from $\Projective_\Ing$. It is worth mentioning how, already at this stage, the paraphernalia introduced in \Sref{subsubsec:canonical_construction_of_the_projective_limit} and \Sref{subsubsec:defining_the_cone} for the $\Ing \in \{ \TypeF, \ProdFM, \Cons_M, \Id \}$ play a role in \Mref{eq:Ing_terminal_carrier}.

Thus, we can now define what is going to turn out as the carrier of the terminal $\TypeF$-coalgebra we are interested in, namely,
\begin{equation}
\label{eq:terminal_carrier}
\Terminal := \Set { \terminal \in \Projective | %
\exists \Coalg := \la \Car, \trans \ra \in \Ob(\CoAlg(\TypeF)) \ %
\exists \car \in \Car : \ \terminal = \cone^{\trans} (\car) },
\end{equation} 
which is endowed with the $\sigma$-algebra $\sAlg_{\Terminal}$ inherited from $\Projective$ and it is simply \Mref{eq:Ing_terminal_carrier} with $\Ing := \Id$, i.e., $\Terminal = \Terminal_\Id$.

\subsubsection{Obtaining the Terminal Transition}
\label{subsubsec:terminal_transition}

The importance of $\Terminal_{\Ing}$, for every $\Ing \in \{ \TypeF, \ProdFM, \Cons_M, \Id \}$,  lies in the fact that we employ the following strategy to obtain the terminal transition:
\begin{enumerate}[leftmargin=*, label=\arabic*)]
\item we show the existence of a measurable morphism $\zetap: \Terminal \to \Terminal_{\TypeF}$;

\item we show the existence of a measurable morphism $\zetat : \Terminal_{\TypeF} \to \TypeF (\Terminal)$;

\item we let $\belt : \Terminal \to \TypeF (\Terminal)$ be defined as $\belt := \zetat \circ \zetap$.
\end{enumerate}
However, and rather crucially, Step (2) of the project relies on showing the existence of measurable morphisms $\belt_\Ing$ and $\rhot_\Ing$, for every $\Ing \in \{ \TypeF, \ProdFM, \Cons_M, \Id \}$. Essentially, the strategy revolves around the idea of capturing the very nature of the functor $\TypeF$ via morphisms  $\belt_\Ing$ (which capture the `nature' of the functors upon which it is built)  \emph{and} morphisms $\rhot_\Ing$ (which encapsulate the order in which the functor `unpacks' the functors that are its building blocks). In a more compact way, we want to obtain measurable spaces and measurable morphisms such that the following diagram
\begin{equation}
\label{cd:transition_general}
\begin{tikzcd}
\Car %
	\arrow[rr, "{\trans}"] %
	\arrow[d, "{\cone^{\trans}}"'] &  & %
			\TypeF ( \Car) = \CPSM (M \times \Car) %
				\arrow[d, %
				start anchor={[shift={(-1.3cm,0cm)}]}, %
				end anchor={[shift={(-1.3cm,0cm)}]},  %
				"{\TypeF \cone^{\trans}}"] %
				\arrow[ld, %
				start anchor={[shift={(-0.6cm,0cm)}]}, %
				end anchor={[shift={(0cm,0.15cm)}]}, %
				"{\cone^{\trans}_\TypeF}"'] %
				\arrow[ldd, bend left=105, %
					start anchor={[shift={(1.4cm,0.1cm)}]}, %
					end anchor={[shift={(1.5cm,0pt)}]}, %
					"{\TypeF \cone^{\trans}_n}"] %
\\
\Terminal %
	\arrow[r, "{\zetap}"] %
	\arrow[rd, end anchor={[shift={(0.7cm,-0.1cm)}]}, "{\pi_{n+1}}"'] %
	& \Terminal_\TypeF %
		\arrow[r, pos=0.7, "{\zetat}"] %
		\arrow[d, end anchor={[shift={(0.6cm,0pt)}]}, "{\pi^{\TypeF}_{n}}"] %
		& \TypeF ( \Terminal) = \CPSM (M \times \Terminal)  %
			\arrow[ld, %
			start anchor={[shift={(-0.3cm,0cm)}]}, %
			end anchor={[shift={(0.5cm,-0.1cm)}]}, "{\TypeF \pi_{n}}"] \\
	& \hspace{1.5cm} \TypeF^{n+1} (\1)  &                       
\end{tikzcd}
\end{equation}
\vspace{-1.2cm}

\noindent commutes, with the understanding that  $\zetat$ is the result of the following diagram 

\begin{equation}
\label{cd:transition_ingredients}
\begin{tikzcd}[row sep = large, column sep = huge]
\Terminal_\TypeF = \Terminal_{\CPSFM \ProdFM} %
	\arrow[r, "\belt_{\CPSFM \ProdFM}"] %
	\arrow[rrd, "\zetat"'] 
	& \CPSM (\Terminal_{\ProdFM}) %
		\arrow[r, "\CPSFM \belt_{\ProdFM}"] %
		& \CPSM (\Terminal_{\Cons_M} \times \Terminal_\Id ) %
		\arrow[d, "\CPSFM \big( \belt_{\Cons_M} {,} \belt_{\Id} \big)"] \\
\Terminal %
	\arrow[rr, "\belt"'] %
	\arrow[u, "\zetap"] %
	&	& \CPSM (M \times \Terminal )  
\end{tikzcd}
\end{equation}
commuting, i.e., we want to define $\zetat$ in such a way that  \Mref{cd:transition_ingredients} commutes, where every $\belt_\Ing$ for $\Ing \in \{ \TypeF, \ProdFM, \Cons_M, \Id \}$ is defined in \Sref{app:proofs_universality} and it is important to observe that the expressions ``$\CPSM (\Terminal_{\ProdFM})$'' and ``$\CPSM (\Terminal_{\Cons_M} \times \Terminal_{\Id})$'' in \Mref{cd:transition_ingredients} are both well-defined in light of \Mref{rem:well-definedness_ingredients}.  In particular, we want to define $\zetat$ as 
\begin{align}
\label{eq:definition_zetat}
\zetat & := \rhot_{\CPSFM \ProdFM} \notag \\%
	& \textcolor{white}{:}= 
\Big( \CPSFM \rhot_{\ProdFM} \Big) %
			\belt_{\CPSFM \ProdFM} \notag \\
	& \textcolor{white}{:}= \bigg( %
	\CPSFM  \Big( \big( \belt_{\Cons_M} , \belt_\Id \big) %
		\belt_{\ProdFM} \Big) \bigg) %
			\belt_{\CPSFM \ProdFM}
\end{align}
where the idea behind \Mref{eq:definition_zetat} is that, for example, in the passage from $\CPSM (\Terminal_{\Cons_M} \times \Terminal_\Id )$ to $\CPSM (M \times \Terminal)$, the functor $\CPSFM$ has already been employed in the previous step and as a result should not be written down, where the details can be found---once more---in \Sref{app:proofs_universality} (in particular, in the proof of \Mref{lem:crucial} below).\footnote{See \citet[Example 2, p.405]{Viglizzo_2005a} for a similar reading of a commutative diagram in the same context.}  

Thus, first of all, we establish Step (1) of the program above by proving the existence of a measurable morphism $\zetap : \Terminal \to \Terminal_{\TypeF}$.

\begin{lemma}
\label{lem:terminal_transition}
There exists a measurable morphism $\zetap : \Terminal \to \Terminal_{\TypeF}$ such that $\zetap \circ \cone^{\trans} = \cone^{\trans}_{\TypeF} \circ \trans$, for every $\TypeF$-coalgebra $\Coalg := \la \Car, \trans \ra$.
\end{lemma}

More succinctly, \Mref{lem:terminal_transition} can be captured by stating that the following diagram
\begin{equation*}
\begin{tikzcd}
\Car %
	\arrow[r, "\trans"] %
	\arrow[d, "\cone^{\trans}"']
	& \TypeF (\Car) %
		\arrow[d, "{\cone^{\trans}_\TypeF}"]\\
\Terminal %
	\arrow[r, "\zetap"']
	& \Terminal_{\TypeF}
\end{tikzcd}
\end{equation*}
commutes, where it is understood that, given that the actual codomain of $\cone^{\trans}$ is $\Projective$, this is an innocuous abuse of notation which subsumes the existence of an inclusion map from $\Terminal$ to $\Projective$.

Then, we establish Step (2) of the strategy above, with the understanding that all the---background---work whose intuition is sketched above regarding the existence of measurable morphisms $\belt_\Ing$, for every $\Ing \in \{ \TypeF, \ProdFM, \Cons_M, \Id \}$, can be found in \Sref{app:proofs_universality}.

\begin{lemma}
\label{lem:crucial}
There exists a measurable morphism 
\begin{equation*}
\zetat := ( \zetat_{C} )_{\CEventsM} : \Terminal_{\TypeF} \to \TypeF (\Terminal) = \CPSM (M \times \Terminal)
\end{equation*} 
such that:
\begin{enumerate}[leftmargin=*, label=\arabic*)]
\item $\zetat \circ \cone^{\trans}_{\TypeF} = \TypeF \cone^{\trans}$, for every $\TypeF$-coalgebra $\Coalg := \la \Car, \trans \ra$; 
    
\item $\TypeF \pi_{n} \circ \zetat = \pi^{\TypeF}_{n}$, for every $n \in \bN$.
\end{enumerate}
\end{lemma}

Once more, succinctly, \Mref{lem:crucial} amounts at establishing the commutativity of the following diagram 
\begin{equation}
\label{cd:crucial}
\begin{tikzcd}
	& \CPSM (M \times \Car) %
		\arrow[ld, 
		start anchor={[shift={(-0.45cm,0.1cm)}]}, %
		end anchor={[shift={(0cm,0pt)}]}, %
		"\cone^{\trans}_{\TypeF}"'] %
		\arrow[rd, 
		start anchor={[shift={(0.3cm,0.1cm)}]}, %
		end anchor={[shift={(-0.3cm,0pt)}]}, %
		"\TypeF \cone^{\trans}"] & %
\\
\Terminal_{\TypeF} %
	\arrow[rr, "\zetat"'] %
	\arrow[rd, 
	start anchor={[shift={(0cm,0cm)}]}, %
	end anchor={[shift={(-0.1cm,0pt)}]}, %
	"\pi^{\TypeF}_n"'] & %
	& \CPSM (M \times \Terminal) %
		\arrow[ld, "\TypeF \pi_n", start anchor={[shift={(-0.3cm,0cm)}]}] \\
	& \TypeF^{n+1} (\1) &  %
\end{tikzcd}
\end{equation}
where it should be recalled that $\TypeF^{n+1} (\1) =  \CPSM ( M \times \TypeF^n (\1))$.

Finally, leveraging on \Mref{lem:terminal_transition} and \Mref{lem:crucial}, we now employ what we obtained with the previous lemmata by defining the \emph{terminal transition} $\belt$ as
\begin{equation}
\label{eq:transition_terminal}
\belt := \zetat \circ \zetap .
\end{equation}

\subsubsection{Establishing Terminality}
\label{subsubsec:terminality}

We now collect what we established in \Sref{subsubsec:terminal_carrier} and \Sref{subsubsec:terminal_transition} in a dedicated definition.

\begin{definition}[The $\TypeF$-Coalgebra $\TCoalg$]
The $\TypeF$-Coalgebra 
\begin{equation*}
\TCoalg := \la \Terminal , \belt \ra
\end{equation*}
is the $\TypeF$-Coalgebra whose carrier $\Terminal$ is defined as in \Mref{eq:terminal_carrier} and whose transition  $\belt$ is defined as in \Mref{eq:transition_terminal}.
\end{definition}

What now remains to do is to establish the terminality of $\TCoalg$. Thus, first of all we start from the following lemma, whose proof  is immediate in light of the fact that the following diagram
\begin{equation*}
\begin{tikzcd}
\Car %
	\arrow[r, "\trans"] %
	\arrow[d, "\cone^{\trans}"'] %
	& \TypeF (\Car) %
		\arrow[d, "\cone^{\trans}_{\TypeF}"'] %
		\arrow[rd, "\TypeF \cone^{\trans}"] &   \\
\Terminal %
	\arrow[r, "\zetap"'] %
	\arrow[rr, bend right=45, "\belt"]
	& \Terminal_{\TypeF} %
		\arrow[r, "\zetat"'] %
		& \TypeF (\Terminal) %
\end{tikzcd}
\end{equation*}
commutes from \Mref{lem:terminal_transition} and \Mref{lem:crucial}, which is in the spirit of \citet[Proposition 5.3, p.337]{Heifetz_Samet_1998}.

\begin{lemma}[Cone as Coalgebra Morphism]
\label{lem:coalgebra_morphism}
For every $\TypeF$-coalgebra $\Coalg := \la \Car, \trans \ra$, $\cone^{\trans}$ is a $\TypeF$-coalgebra morphism.
\end{lemma}

\begin{notation}[Cone for $\TypeF$-Coalgebra $\TCoalg$]
We let $\cone$ denote the cone of $\TCoalg := \la \Terminal , \belt \ra$.
\end{notation}

We now establish the fact that the cone $h$ is nothing more than the identity function on $\Terminal$, which is analogous to \citet[Lemma 5.4, p.338]{Heifetz_Samet_1998}.

\begin{lemma}
\label{lem:terminal_identity}
Given the $\TypeF$-coalgebra $\TCoalg$, $\cone = \id_{\Terminal}$.
\end{lemma}

We can now state the terminality of $\TCoalg$ in the following lemma, which is analogous to \citet[Theorem 5.5, p.338]{Heifetz_Samet_1998}.

\begin{lemma}
\label{lem:terminal_existence}
The $\TypeF$-coalgebra $\TCoalg := \la \Terminal , \belt \ra$ is a terminal $\TypeF$-coalgebra.
\end{lemma}

Finally, we provide the proof of \Mref{prop:coalgebraic_main_theorem}.

\begin{proof}[Proof of \Mref{prop:coalgebraic_main_theorem}]
We establish the points in the order in which they are presented in the statement of the result.
\begin{enumerate}[leftmargin=*, label=\arabic*)] 
\item {\emph{Existence:}} This is an immediate consequence of \Mref{lem:terminal_existence}. 

\item {\emph{Uniqueness:}} The uniqueness (in this case up to $\TypeF$-coalgebra isomorphism) of a terminal object comes from \Mref{lem:uniqueness_ter_obj}.

\item {\emph{Isomorphic Transitions:}} The fact that $\belt$ is an isomorphism is an immediate consequence of Lambek's lemma, i.e., \Mref{lem:lambek}.
\end{enumerate}
Thus, what is written above establishes the result.
\end{proof}

\subsection{\dots In the Dale (between the Two Worlds)\dots}
\label{subsec:dale}

In the previous section, we established the existence of a terminal $\TypeF$-coalgebra in the category $\Meas$. However, we do not have to lose sight of the fact that what we are after is actually a proof of \Mref{th:main_theorem}, which is a theorem regarding the existence of a type structure, i.e., an object that is intrinsically related to the presence of \emph{multiple} interactive agents. Thus, quite simply, $\Meas$ is not rich enough for our purposes. Indeed, we need to work in a theoretical framework that captures the presence of multiple interactive agents. Thus, the definition that follows accomplishes exactly this.

\begin{definition}[{Category $\MeasI$}] 
\label{def:MeasI}
The category $\Meas^{I}$ is the category of measurable spaces, where
\begin{itemize}[leftmargin=*]
\item the \emph{objects} of $\MeasI$ are an $\abs{I}$-fold product of measurable spaces;

\item the \emph{morphisms} of $\MeasI$ are an $\abs{I}$-fold product of measurable functions.
\end{itemize}
\end{definition}

Hence, in order to get closer to the original problem addressed for type structures built on a common domain of uncertainty $\TSC$, from now on we focus on a given conditional measurable space $\TSC$. Also, with an innocuous abuse of notation, we employ the same symbols we used in the previous section (modulo usage of $\State$ instead of $M$), with the caveat that---with a minimal change in comparison to the previous section---now we let $\ProdFT := \Cons_\State \times \Proj_{-i}$, i.e., $\ProdFT$ is defined by using  the constant functor $\Cons_\State$  on $\State$ and the projection functor $\Proj$ indexed by $I$.\footnote{See \Mref{ex:product_categories} for its definition.} Thus, given a conditional measurable space $\TSC$, we let the morphism
\begin{equation*}
\TypeFp  : \MeasI \to \MeasI
\end{equation*}
such that $\TypeFp := \CPSFT \ProdFT$, i.e.,
\begin{equation*}
\TypeFp  := \CPSFT \ProdFT = \CPSFT ( \Cons_\State \times \Proj_{-i}) ,
\end{equation*}
be defined as,
\begin{itemize}[leftmargin=*]
\item for every $(\Type_i)_{i \in I}  \in \Ob (\Meas^{I})$,
\begin{equation*}
\TypeFp \big( (\Type_i)_{i \in I} \big) := %
\Big( \CPST (\State \times \Type_{-i} ) \Big)_{i \in I} ,
\end{equation*}
\item and  
\begin{equation}
\label{eq:endofunctor_well-defined}
\TypeFp \cmor  := %
\Big( \widehat{(\id_\State , \cmor_{-i})} : \CPST ( \State \times \Type_{-i}) \to \CPST ( \State \times \Type_{-i}') \Big)_{i \in I} 
\end{equation}
for every $\cmor := (\cmor_i)_{i \in I}  \in \Meas^{I} \big( (\Type_i)_{i \in I}, (\Type'_i)_{i \in I} \big)$,
\end{itemize}
where it is immediate to establish that this morphism is indeed a functor.

\begin{notation}
We let $\HierF_i := \Proj_i \TypeFp$, i.e., $\HierF_i (\Type_i) := \Proj_i \TypeFp \big( (\Type_i)_{i \in I} \big)$ and $\HierF_i \cmor_i := \Proj_i \TypeFp \big(  (\cmor_i)_{i \in I} \big)$.
\end{notation}

It has to be observed that \Mref{eq:endofunctor_well-defined} is well-defined in light of the definition of $\TypeFp$, which implies that, given a certain profile $\Type := (\Type_i)_{i \in I}$, the application of the functor $\ProdFT$ leads to $\State \times \Type$, which---in turn---is related to the fact that, for every $i \in I$ with $\cmor_i : \Type_i \to \Type'_i$, $\HierF_i \cmor_i = \widehat{(\id_\State , \cmor_{-i})}$ is well-defined, because the image measure $\widehat{(\id_\State , \cmor_{-i})}$ is defined on the induced function 
\begin{equation*}
(\id_\State, \cmor_{-i}) : \State \times \Type_{-i} \to \State \times \Type'_{-i}
\end{equation*}
which is a product conditional measurable space as needed for \Mref{def:pushforward_conditioning}.

\begin{definition}[$\TypeFp$-Coalgebra]
\label{def:type_coalgebra}
A $\TypeFp$-coalgebra is a tuple
\begin{equation*}
\Coalg := \la (\Type_i , \bel_i )_{i \in I} \ra
\end{equation*}
where $\Type_i$ is the $i$-carrier of the $\TypeFp$-coalgebra with transition $\bel_i : \Type_i \to \CPST ( \State \times \Type_{-i})$, for every $i \in I$.
\end{definition}

We can now reformulate \Mref{prop:coalgebraic_main_theorem} for the present setting where the focus is on the category $\MeasI$.

\begin{proposition}
\label{prop:coalgebraic_main_theorem_type}
Fix a conditional measurable space $\TSC$. %
\begin{enumerate}[leftmargin=*, label=\arabic*)]
\item {\emph{Existence:}} there exists a terminal $\TypeFp$-coalgebra  $\TCoalg := \la ( \Terminal_i , \belt_i )_{i \in I} \ra$;

\item {\emph{Uniqueness:}} the terminal $\TypeFp$-coalgebra  $\TCoalg := \la ( \Terminal_i , \belt_i )_{i \in I} \ra$  is unique up to $\TypeFp$-coalgebra isomorphism;

\item {\emph{Isomorphic Transitions:}} the transition $\belt_i$ of the terminal $\TypeFp$-coalgebra  $\TCoalg := \la ( \Terminal_i , \belt_i )_{i \in I} \ra$ is an isomorphism, for every $i \in I$.
\end{enumerate}
\end{proposition}

Whereas we do not give the details of the proof, which essentially proceeds along the lines of that of \Mref{prop:coalgebraic_main_theorem} with the caveat that a considerably heavier notation is needed to deal with the presence of multiple agents, we sketch the path that should be taken by proceeding along the lines of  what has been done in \Sref{subsec:there}. Thus, in the following, everything starts from the---by now classical---canonical construction, with the understanding that $\TypeFp (\1)$ (and the like) is used to denote the application of the endofunctor $\TypeFp$ to an $(\abs{I} -1)$--tuple of $\1$ (endowed with the discrete $\sigma$-algebra), whereas $\TypeFp !$ (and the like) is  used to denote the application of the endofunctor $\TypeFp$ to an $(\abs{I} -1)$--tuple of morphisms $!$.

It is useful to `unpack' the information contained in \Mref{def:opchain} by showing how the $\opchain$-chains look for  a  $\Type_i$, with $i \in I$ arbitrary. Hence, we have 
\begin{equation*}
\begin{tikzcd}
\1 %
	& \HierF^{1}_i (\1) %
		\arrow[l, dashed, "!"'] %
		& \HierF^{2}_i (\1)   %
			\arrow[l, "\HierF^{1}_i !"'] %
			& \dots, %
				\arrow[l, "\HierF^{2}_i !"']
\end{tikzcd}
\end{equation*}
which, by spelling out the details, amounts at
\begin{equation*}
\begin{tikzcd}
\1 %
	& \CPST \Big( \State \times  \prod_{j \in I \setminus \{i\}} \1  \Big) %
		\arrow[l, dashed, "!"'] %
		& \CPST \Big( \State \times \prod_{j \in I \setminus \{i\}} \Big( \CPST (\State \times \prod_{j \in I \setminus \{i\}} \1) \Big) \Big)  %
			\arrow[l, "\HierF^{1}_i !"'] %
			& \dots, %
				\arrow[l, "\HierF^{2}_i !"']
\end{tikzcd}
\end{equation*}
in light of the fact that $\prod_{j \in I \setminus \{i\}} \1 = \1^{\abs{I}-1}$. Once more, it is immediate to observe by inspection that this is in the spirit of the construction of \citet[Section 5, p.335]{Heifetz_Samet_1998} in presence of conditioning events.

Getting into the details of the profile $\Projective := (\Projective_i)_{i \in I}$ and decomposing the information contained in \Mref{eq:projective_limit}, we have
\begin{equation*}
\Projective_i  := \Set { (x_\ell)_{\ell \in \bN} \in \prod_{\ell \in \bN} \HierF^{\ell}_i (\1) | %
\forall n \in \bN \ \Big(  \HierF^{n}_i ! (x_{n+1}) = x_n \Big) } .
\end{equation*}

Now, given a $\TypeFp$-coalgebra $\Coalg := \la (\Type_i, \bel_i )_{i \in I} \ra$, we define the notion of \emph{cone} over that $\TypeFp$-coalgebra, whose role is to map the information `implicitly' contained in the carrier $(\Type_i)_{i \in I}$ to  $(\Projective_i)_{i \in I}$ as defined above. Thus, given a $\TypeFp$-coalgebra $\Coalg := \la (\Type_i , \bel_i )_{i \in I} \ra$,  a \emph{cone} for the $\TypeFp$-coalgebra $\Coalg$ is a recursively defined collection of morphisms, for every $i \in I$ and $n \in \bN$,
\begin{itemize}[leftmargin=*]
    \item $\cone^{\bel}_{i, 0} := ! : \Type_i \to \1$,
    
    \item $\cone^{\bel}_{i, n} : \Type_i  \to \HierF^{n}_i (\1)$ such that $\cone^{\bel}_{i, n+1} = \HierF_{i}  \cone^{\bel}_{i, n} \circ \bel_i$,    
\end{itemize}
such that there exists a unique morphism  
\begin{equation*}
\cone^{\bel}_i : \Type_i \to \Projective_i 
\end{equation*}
such that $\cone^{\bel}_{i, n} (\type_j) = \pi_n \circ \cone^{\bel}_i (\type_i)$, for every $i \in I$, $\type_i \in \Type_i$, and $n \in \bN$. Finally, building on \Mref{def:cone}, we can now define, for every $i \in I$, the carrier of the terminal $\TypeFp$-coalgebra we are interested in, namely,
\begin{equation*}
\Terminal_i := \Set { \terminal_i \in \Projective_i | %
\exists \Coalg := \la (\Type_i, \bel_i)_{i \in I} \ra \in \Ob(\CoAlg(\TypeFp)) \ %
\exists \type_i \in \Type_i : \ \terminal_i = \cone^{\bel}_i (\type_i) },
\end{equation*} 
which is endowed with the $\sigma$-algebra $\sAlg_{\Terminal_i}$ inherited from $\Projective_i$, with $\Terminal := (\Terminal_i)_{i \in I}$.

Having obtained the structure of $\Terminal := (\Terminal_i)_{i \in I}$, constructing the terminal transition $\belt := (\belt_i)_{i \in I}$ and establishing the terminality of the corresponding $\TypeFp$-coalgebra proceeds along the same path taken in \Sref{subsubsec:terminal_transition} and \Sref{subsubsec:terminality}. In particular, given that we let $\cone := (\cone_i)_{i \in I}$ denote the cone of the terminal $\TypeFp$-coalgebra $\TCoalg := \la (\Terminal_i, \belt_i)_{i \in I} \ra$, we can establish the following crucial---and obviously related---properties of $\cone_i$, for every $i \in I$.

\begin{remark}
\label{rem:separativity_injectivity}
For every $i \in I$, the $\sigma$-algebra on $\Terminal_i$ is separative, with $\cone_i = \id_{\Terminal_i}$ obviously injective.
\end{remark}

\subsection{\dots and Back Again (to Type Structures)}
\label{subsec:back_again}

We can now go back to our original problem, namely, the proof of \Mref{th:main_theorem}. In light of \Sref{subsec:dale}, it should be observed how  the notation we use in this section for the elements of a type structures and those of a coalgebra are actually the same. Thus, the proof of the next result is immediate by inspection. It has to be observed that this also---implicitly---establishes that cones (as in \Sref{subsec:dale}) are hierarchy functions as in \Mref{def:hierarchy_function}.

\begin{lemma}[Type Structures as Coalgebras]
\label{lem:type_coalgebra}
Given a conditional measurable space $(\State, \sAlg_\State, \CEventsT)$, a type structure $\mscr{T} := \la (T_i, \beta_i )_{i \in I} \ra$ is a $\TypeFp$-coalgebra $\Coalg := \la (T_i, \beta_i )_{i \in I}  \ra$.
\end{lemma}

Crucially, we have to establish that type morphisms and coalgebra morphisms are actually the same, which we achieve next.

\begin{lemma}[Type Morphisms as Coalgebra Morphisms]
\label{lem:type_coalgebra_morphism}
Given a conditional measurable space $(\State, \sAlg_\State, \CEventsT)$ and two arbitrary type structures $\mscr{T} := \la (T_i, \beta_i )_{i \in I} \ra$, and $\mscr{T}' := \la (T'_i, \beta'_i )_{i \in I} \ra$ on $(\State, \sAlg_\State, \CEventsT)$ sharing $\CEventsT$, a type morphism $\tmor := (\tmor_j)_{j \in I_0}$  is a $\TypeFp$-coalgebra morphism.
\end{lemma}

In light of what we established in \Sref{subsec:there} and \Sref{subsec:dale} and the lemmata above, we can finally provide the proof of \Mref{th:main_theorem}.

\begin{proof}[Proof of \Mref{th:main_theorem}]
From \Mref{prop:coalgebraic_main_theorem_type}, we have these immediate implications concerning \Mref{th:main_theorem}:
\begin{enumerate}[leftmargin=*, label=\arabic*)] 
\item {\emph{Terminality:}} From Point (1) of \Mref{prop:coalgebraic_main_theorem_type}, \Mref{lem:type_coalgebra}, and \Mref{lem:type_coalgebra_morphism}, there exists a terminal type structure
\begin{equation*}
\mscr{T}^* := \la (T^{*}_i, \beta^{*}_i )_{i \in I} \ra
\end{equation*}
on $(\State, \sAlg_\State, \CEventsT)$;

\item {\emph{Uniqueness:}} From Point (2) of \Mref{prop:coalgebraic_main_theorem_type}, this terminal type structure $\mscr{T}^*$ is \emph{the} unique up to measurable type isomorphism terminal type structure on $(\State, \sAlg_\State, \CEventsT)$;

\item {\emph{Belief-Completeness:}} From Point (3) of \Mref{prop:coalgebraic_main_theorem_type}, the terminal type structure $\mscr{T}^*$ is belief-complete, i.e., the belief function $\beta^{*}_i$ is surjective, for every $i \in I$;

\item {\emph{Non-Redundancy:}} From \Mref{lem:type_coalgebra} and \Mref{rem:separativity_injectivity}, $\mscr{T}^*$ is non-redundant by construction.
\end{enumerate}
Being terminal, belief-complete, and non-redundant, $\mscr{T}^*$ is the Universal Type Structure on $(\State, \sAlg_\State, \CEventsT)$ as in \Mref{def:universal_type_structure}, which, in light of its uniqueness up to measurable type isomorphism, establishes the result.
\end{proof}

\section{Discussion of the Construction}
\label{sec:discussion}

\subsection{On Technical Aspects of the Construction}
\label{subsec:on_technical_aspects_of_the_construction}

\subsubsection{On Type Structures}
\label{subsubsec:type_structures}

Regarding our definition of type structure as in \Mref{def:type_structure}, it should be observed that we could have defined the belief function of agent $i$ as $\beta_i :T_i \to \CPST (\State \times T)$, for every $i \in I$, as in \citet[Definition 3.1, pp.329--330]{Heifetz_Samet_1998}. With such a definition, it is typically imposed the additional requirement that  $\marg_{T_i} \beta_{i, C} (t_i) = \delta_{t_i}$, for every $t_i \in T_j$ and $C \in \CEventsT$,\footnote{In particular, this is actually Condition (3) in \citet[Definition 3.1, pp.329--330]{Heifetz_Samet_1998}.} with $\marg$ denoting the marginal operator as canonically defined and $\delta_{t_i}$ denoting the Dirac measure on $t_i$. It has to be observed that this property is not necessary for the construction: \citet[Section 5, p.41]{Heifetz_Mongin_2001} and \citet[Definition 8, p.12]{Meier_2012} distinguish type structures which possess this property from those which do not,\footnote{See also the discussion of this point in a coalgebraic context in \citet[Chapter 7.1, pp.79--81]{Viglizzo_2005}.} where in their terminology a type structure that satisfies this requirement is called a \emph{``Harsanyi type structure''}.\footnote{With the caveat that these authors employ the word ``space'' instead of ``structure''.}

\subsubsection{On Type Morphisms}
\label{subsubsec:type_morphisms}

Concerning our definition of type morphism, as it is widespread in the literature, given an arbitrary conditional measurable space $\TSC$ and a type structure $\mscr{T}$, we defined it in \Mref{def:type_morphism} by introducing a function $\vartheta_0 := \id_\State$. Interestingly, this is not \emph{explicitly} present in the notion of $\TypeFp$-coalgebra morphism. Now, a similar issue is present in \cite{Moss_Viglizzo_2004}, as emphasized in \citet[Section 2, pp.284--285]{Moss_Viglizzo_2004}, where the authors point out that their notation remains silent regarding the underlying (common) domain of uncertainty. Of course, our notation does not---and cannot---remain silent, since emphasizing the common domain of uncertainty is crucial to properly deal with product conditional measurable spaces sharing a given set of conditioning events. Thus, as a matter of fact, our proof takes care of the existence of the function $\vartheta_0$ in light of the definition of $\ProdFT$, which is based on a binary product involving $\Cons_\State$, that---in turn---induces a morphism $\id_\State$.

\subsubsection{On the Family of Conditioning Events}
\label{subsubsec:technical_assumptions}

The present construction does not put any restriction on the  conditioning events belonging to a given conditional measurable space $\MSC$. Thus, it is important to make a comparison with the original construction obtained in \citet[Section 2]{Battigalli_Siniscalchi_1999} by paying special attention to the assumptions made there and the rationale behind them. In particular, in \citet[Section 2.1, p.191]{Battigalli_Siniscalchi_1999}, it is assumed that, given a conditional measurable space $\MSC$, $M$ is a Polish space, $\sAlg_M$ is its Borel $\sigma$-algebra, and the events in $\CEventsM$ are clopen (i.e., closed and open) and at most countable.

Regarding the clopeness of the conditioning events in $\CEventsM$, this assumption is used in \citet[Proof of Lemma 1, p.224]{Battigalli_Siniscalchi_1999} to establish that $\Delta^{\CEventsM} (M)$ is a closed subset of $[\Delta (M)]^{\CEventsM}$. Concerning the cardinality assumption, assuming an at most countable family of conditioning events is crucial in  \citet[Proof of Proposition 1, pp.196--197]{Battigalli_Siniscalchi_1999}, where it is used to prove the existence of---using our notation---a homeomorphism $f : H_c \to \Delta^{\CEventsM} (M \times H)$, where $H$ is the set of infinite hierarchies of beliefs not necessarily coherent, whereas $H_c$ is the set of infinite hierarchies of beliefs that are coherent. In particular, this assumption is needed to establish, via arguments from \citet[Section 11.A, p.68]{Kechris_1995} concerning Borel sets, that every $n^{\text{th}}$-order hierarchy of CPSs belonging to a CPS over $\Delta^{\CEventsM} (M \times H)$ satisfies Axiom (C3) from \Mref{def:CPS}.

It is immediate to observe that, while the clopeness assumption simply has no bite in our topology-free framework, the cardinality assumption is not needed to establish our result. Hence, our construction does not put any restriction on the nature of the conditioning events.\footnote{We are extremely grateful to an anonymous referee for having questioned the need of the countability assumption we had in a previous version of the paper. In doing so, we realized we could go back to work with `unrestricted' conditioning events in the spirit of \cite{Guarino_2017} and \cite{Fukuda_2024a} (and, with respect to this point, it is worthwhile mentioning that neither cardinality nor topological assumptions are made in the original \cite{Renyi_1955}).}

\subsection{Relation to Universality Notions}
\label{subsec:relation_to_universality_notions}

\subsubsection{On the Notion of Kolmogorov's Consistency}
\label{subsubsec:consistency}

The point made in \Sref{subsubsec:intermezzo}, that is built on \citet[Section 2.2, p.399]{Viglizzo_2005a}, is essentially related to the impossibility of obtaining an analog of the Kolmogorov's Extension Theorem\footnote{See \citet[Theorem 15.26, p.522]{Aliprantis_Border_2006}.} for an infinite product of measurable spaces, as shown by \cite{Sparre-Andersen_Jessen_1948}, \citet[Section 7, p.42]{Dieudonne_1948}, and \citet[Chapter IX.49(3), p.214]{Halmos_1950}, which is---in turn---related to the result obtained in \cite{Heifetz_Samet_1999} concerning the fact that there are coherent infinite hierarchies of beliefs that are not types, i.e., when we do not start from topological assumptions, it is possible to obtain a coherent infinite hierarchy of beliefs such that there exists no extension over the infinite product (which would be essentially the type, as employed in economics).\footnote{See \cite{Fukuda_2024b} for a work that obtains types as coherent infinite hierarchies of beliefs in a topology-free setting via the usage of ordinal numbers and transfinite recursion.} In other words, by letting $C$ denote the space of coherent infinite hierarchies of beliefs over a common domain of uncertainty $\State$ with $T^*$ the corresponding space of \emph{all} types, we have $T^* \subseteq C$, with strict inclusion $T^* \subset C$ when we work without topological assumptions.\footnote{In particular, see \citet[Section 5]{Heifetz_Samet_1999}.} Relatedly, it is important to point out that  it has been shown in  \cite{Schubert_2009} that starting with appropriate topological assumptions (and relying on a suitable version of the Kolmogorov's Extension Theorem), it is possible to build an endofunctor that preserves limits in the form of $\opchain$-chains and allows to obtain the corresponding terminal coalgebra via the canonical hierarchical construction of $\opchain$-chains presented in \Sref{subsubsec:canonical_construction_of_the_projective_limit}, with the caveat that the intuition provided in \Sref{subsubsec:intermezzo} can now be used.\footnote{In particular, \cite{Schubert_2009} works in the subcategory $\bm{\mathsf{SB}}$ of $\Meas$ comprised of standard Borel spaces (i.e., those measurable spaces  $(X, \mcal{B}_X)$ such that there exists a Polish topology $\tau_X$ on $X$ with $\mcal{B}_X$ being the family of Borel sets generated by $\tau_X$) with the subprobability endofunctor $S$, where a subprobability measure $\mu$ on a standard Borel space $(X, \mcal{B}_X)$ is a $\sigma$-additive functional on $\mcal{B}_X$ such that $\mu (X) \in [0,1]$ (see also \citet[Section 2.4, p.33]{Moss_2011}).}

\subsubsection{%
\texorpdfstring{On Belief-Completeness \& Relation to \cite{Meier_2012}}%
{On Belief-Completeness \& Relation to Meier (2012)}}
\label{subsubsec:notion_belief-completeness}

For the `standard' case without conditioning events, the isomorphic nature of the belief functions in the construction in \cite{Heifetz_Samet_1998} along with---as a corollary---its belief-completeness has been established for the first time in \citet[Theorem 4, p.29]{Meier_2012},\footnote{The first version of this (in the words of \citet[Footnote 14, p.1674]{Aumann_Heifetz_2002}, where there is also a brief---nonetheless informative---description of part of the results therein) \emph{``beautiful, path-breaking paper''} goes back to 2001, thus establishing the result before \cite{Moss_Viglizzo_2004}.} where the author builds an infinitary probability logic to capture reasoning in type structures.\footnote{See also \cite{Heifetz_1997}, \cite{Aumann_1999b}, and \cite{Heifetz_Mongin_2001}.} Thus, to relate this work to the present one, it is worthwhile to recall the path taken in \cite{Meier_2012} to obtain the result. Hence, \cite{Meier_2012} starts by addressing belief structures, where, given a measurable space $(\State, \sAlg_\State)$,  a \emph{belief structure}\footnote{See \citet[Definition 2.2, p.4]{Mertens_Zamir_1985} or \citet[Chapter 10]{Maschler_et_al_2013}.} (as in \citet[Definition 7, p.11]{Meier_2012}, in its full generality) is a tuple $\mscr{B} := \la \Theta, \Omega , \bm{\theta}, (\tau_i )_{i \in I}  \ra$ where $\bm{\theta} \in \State^{\Omega}$ is a measurable function and $\tau_i \in [\Delta (\Omega)]^\Omega$ is measurable, for every $i \in I$. Thus, first of all, completeness (in the logical sense as in \citet[Chapter 4]{Blackburn_et_al_2001}) is established for the logical system introduced to deal with belief structures; then, it is established the terminality of the constructed structure canonically built from appropriate formulas; finally, it is showed that the canonical structure has a product structure (in \citet[Proposition 4(3), p.28]{Meier_2012}) \emph{and} it is belief-complete (in \citet[Theorem 4, p.29]{Meier_2012}).

In light of the previous paragraph, three points are in order. The first one is that, in this work, we obtain the belief-completeness of the universal type structure $\mscr{T}^*$ for free, as a corollary of Lambek's lemma. As a matter of fact, we see this as a positive side of performing the endeavour in a coalgebraic framework, which can be exploited by practitioners working on the existence of large interactive structures in topology-free settings to obtain all relevant results at once. The second one is that, whereas we built the terminal $\TypeF$-coalgebra via so-called final sequences, it is possible to obtain the same result via tools from coalgebraic modal logic, as done in \cite{Moss_Viglizzo_2004},\footnote{See also \cite{Moss_Viglizzo_2006}. Regarding coalgebraic logic, see \cite{Jacobs_2001}, \cite{Kurz_2001}, and \cite{Roessiger_2001}  or \citet[Chapter 6.5]{Jacobs_2017} for a textbook presentation.} where it would be interesting to see the relation between these logics and those as the one obtained in \cite{Meier_2012}.  Finally, obtaining the belief-completeness of our structure is one of the differences between the present work and \citet[Section C.1, Online Appendix]{Fukuda_2024a}, where only terminality is established, the other being that the focus in that work is on belief structures.\footnote{See also \cite{DiTillio_et_al_2014}.}

\subsubsection{On the Notion of Universality}
\label{subsubsec:notion_universality}

First of all, we want to emphasize one point about the terminology we use, namely our---somewhat non-standard---definition of universality with respect to how the term is used in the literature on large interactive structures. As mentioned in  \Sref{subsec:large_type_structures}, we extend a terminology that goes back to \citet[p.93]{Siniscalchi_2008} according to which a type structure is universal if terminal and belief-complete. However, starting from \citet[Theorem 2.9, pp.7--8]{Mertens_Zamir_1985},\footnote{For example, see also \citet[Definition 3.3, p.331]{Heifetz_Samet_1998}.} the notion of universal type structure has been associated with the idea that any other type structure can be uniquely embedded into the universal one. In other words, the term ``universal'' has been used as a synonym for ``terminal''.  Interestingly, and somewhat reconciling the usage of both terms for the same object, it should be pointed out how this very definition of universality fits the categorical usage of the word ``universal'' in the opening quotation of this work by \cite{Dieudonne_1989}: every other type structure can indeed be built from the \emph{universal} one.

In second place, it is important to address the relation between the notion of terminality and its connection to the notion of universality---as described above---in its categorical interpretation and the other notions of belief-completeness and non-redundancy. Thus, in particular, it is natural to ask ourselves if terminality actually \emph{implies} all these properties.\footnote{We are extremely grateful to an anonymous referee for having raised this point.} The answer is: \emph{it depends}! Indeed, the problem lies in how we look at this issue. If we approach it from a purely mathematical standpoint, terminality does not necessarily imply all the other properties due to what happens in the topology-free case, where the construction of the terminal type structure as performed in \cite{Heifetz_Samet_1998} does \emph{not} deliver automatically belief-completeness (obtained---as pointed out in \Sref{subsec:motivation_results} and \Sref{subsubsec:notion_belief-completeness}---for the first time in \cite{Meier_2012}). However, it is also important to recognize that, exactly due to `categorical' reasons, terminality from a conceptual standpoint does `imply' all the other properties and, once certain tools are employed (for example, in the topology-free case without conditioning events, infinitary probability logic as in \cite{Meier_2012} or coalgebraic methods as in \cite{Viglizzo_2005a}), indeed it turns out that the terminal type structure \emph{is} the universal one. In particular, with respect to this point, it could be possible to introduce a new notion of universality, call it \emph{``Categorical Universality''}, which would correspond to the idea of a type structure being terminal (and unique up to type isomorphism), non-redundant, and such that all the belief functions are \emph{isomorphisms} in the relevant category. Indeed, this would capture exactly the properties we obtain via the coalgebraic canonical construction, with belief-completeness being an immediate consequence.

Now, it seems that the issues pointed out above strongly support the usage of coalgebraic methods for the construction of large interactive structures in light of the fact that, via these tools, terminality \emph{implies} all the properties exactly as in the case of the topological constructions, where the usual canonical construction\footnote{With the caveat that, in the topological case, there are essentially two canonical constructions as in the terminology introduced in \cite{Battigalli_et_al_Forthcoming}: i.e., the \emph{bottom-up constructions}, with the coherency requirement imposed at the outset in the spirit of \cite{Mertens_Zamir_1985}, and the \emph{top-down constructions}, where first  infinite hierarchies of beliefs are constructed and then the coherency requirement is imposed on them, like in \cite{Brandenburger_Dekel_1993}.} of a type structure that is terminal automatically delivers its universality. In particular, in the spirit of what pointed out in \Sref{subsubsec:consistency}, it is worth emphasizing that the coalgebraic canonical construction presented in \Sref{subsubsec:canonical_construction_of_the_projective_limit} can be---in principle---delivered in both the topological and the topology-free case,\footnote{See \citet[Section 2, p.257]{Heinsalu_2014} for a similar point.} thus, turning out to be a bridge between constructs obtained via procedures at the present stage seemingly different.\footnote{And it is worth mentioning with respect to the notion of Categorical Universality set forth above that homeomorphisms are---indeed---isomorphisms in the category $\Top$ of topological spaces and continuous functions between them.}

\subsection{Relation to the Literature}
\label{subsec:relation_to_the_literature}

\subsubsection{%
\texorpdfstring%
{Relation to \cite{Heifetz_Samet_1998}}%
{Relation to Heifetz \& Samet (1998)}%
}
\label{subsubsec:HS98}

Concerning the construction in \citet[Section 5]{Heifetz_Samet_1998}, there is a tight relation between the construction performed here and the one delivered therein. Indeed, the importance of the construction in \citet[Section 5]{Heifetz_Samet_1998} cannot be emphasized enough, since, starting from the usage of a singleton set, it is essentially coalgebraic in nature. Indeed, it is possible to find a tight link between all the results obtained \citet[Section 5]{Heifetz_Samet_1998} and those in \cite{Viglizzo_2005a}.  However, as it has already been pointed out in \Sref{subsec:motivation_results}, it is important to stress that the construction in \citet[Section 5]{Heifetz_Samet_1998} does \emph{not} deliver all the universality properties at once, which is the reason why we employ coalgebraic tools in the present work.\footnote{See also \Sref{subsubsec:notion_universality} with respect to this point.}

\subsubsection{%
\texorpdfstring%
{Relation to \cite{Viglizzo_2005a}}%
{Relation to Viglizzo (2005b)}%
}
\label{subsubsec:Viglizzo_2005}

Regarding \cite{Viglizzo_2005a}, it is important to emphasize two---somewhat related---points. The first one is that the result (and corresponding proof) in \cite{Viglizzo_2005a} is actually more general than \Mref{prop:coalgebraic_main_theorem}. Indeed, \citet[Theorem 1, p.404]{Viglizzo_2005a} states that there exists a terminal $\Functor$-coalgebra, for \emph{every} endofunctor $\Functor$ on $\Meas$ which is build from the identity, the constant, and an appropriately defined $\bm{\Delta}$ functor (as $\bm{\Delta} (X) := \Delta (X)$, for every $X \in \Ob (\Meas)$, and $\bm{\Delta} f$ as the image measure of a morphism $f$ belonging to $\Meas$) with a closure over binary products and coproducts, i.e., disjoint unions. Thus, for example, this result tells us that there exists a terminal $\Functor$-coalgebra for the---admittedly cumbersome---functor $\Functor := \Id \times \Cons_M \times \bm{\Delta} ( \Id \sqcup \Cons_M)$. Now, our \Mref{prop:coalgebraic_main_theorem} does not deliver anything like that, but it is rather tailor-made to deal with type structures with conditioning events, which considerably changes the nature of the problem, the point being that, for example, a minimal departure from the theoretical problem we faced such as the expression $\CPSFM ( \Id)$ would actually be ill-defined. Nonetheless, in the spirit of the minimal adaptation made in \Sref{subsec:nature_of_the_conditioning_events} of the path sketched in \Sref{subsec:dale}, we can obtain a result concerning the flexibility of the form of the functor we want to focus on as in \cite{Viglizzo_2005a}, leveraging on the definition of $\belt$ as $\belt := \zetat \circ \zetap$ and always taking into account the peculiarities of the functor $\CPSFM$ introduced here. In particular, this is a byproduct of the fact that the path we follow and the proofs of the results we deliver in \Sref{subsec:there} are (modulo notation) those in \cite{Viglizzo_2005a},\footnote{That we rewrite here with our notation for self-containment purposes.} with the notable exception of \Mref{lem:cps}, which---even if along the lines of the corresponding \citet[Lemma 8, pp.402--403]{Viglizzo_2005a}---deals with the presence of conditioning events and, as a result, shows how the crucial new technical element of the present work lies in identifying the peculiarities that arise in presence of conditioning events, and of \Mref{lem:crucial}, which `fixes'---contrary, as pointed out above, to \cite{Viglizzo_2005a}---the `form' of the functor we are interested in. 

Now, the generality of \cite{Viglizzo_2005a} is not a peculiarity of this very paper, but rather of the nature of the problems addressed in the stream of literature to which the paper belongs, which can be captured by the question if it is possible to build a terminal coalgebra for a given collection of endofunctors built from `more basic' endofunctors on a given category. Thus, this is related to the second point, namely, that this paper and the literature to which it belongs tell us something about results from interactive epistemology at large. As an example, we know from \citet[Proposition 1, p.32]{Brandenburger_2003} that there exists no belief-complete interactive---possibility---structure of the form $\mscr{P} := \la (T_i, \wp_i)_{i \in I} \ra$, where $\wp_i : T_i \to \PowerSet (\State \times T_{-i})$.  However, strictly speaking, this is \emph{neither} a consequence of the functors upon which the functor $\PowerSetF$ acts (in the spirit of \Sref{subsec:there}), e.g., $\PowerSetF$-coalgebras for a functor of the form $\PowerSetF (\Cons_\State \times \Id)$, \emph{nor} it is a result specific of interactive structures built on the functor $\PowerSetF$ (in the spirit of \Sref{subsec:dale}), i.e., $\PowerSetF$-coalgebras for a functor of the form $\PowerSetF (\Cons_\State \times \Proj_{-i})$, but rather it is a more basic issue concerning the very functor $\PowerSetF$, as pointed out in  \Sref{subsec:coalgebras}, which shows us that the impossibility runs deeper than the product structure we typically deal with in interactive epistemology, i.e., the problem is \emph{functorial}.

\subsubsection{%
\texorpdfstring{Relation to \cite{Heinsalu_2014}}%
{Relation to Heinsalu (2014)}}
\label{subsubsec:Heinsalu}

\cite{Heinsalu_2014} is an important paper in the literature on the construction of large type structures: beyond addressing the conceptually challenging problem of such a construction for the case of unawareness in a static setting, this is the first paper belonging to the game-theoretical literature where coalgebraic methods are employed. Thus, it is essential to identify how this paper is different from the present one. Now, bypassing the fact that \cite{Heinsalu_2014} deals with a static setting in presence of unawareness, whereas our focus is on a setting with conditioning events without unawareness, the major difference between these two papers lies in the background work. \cite{Heinsalu_2014} is crucially based on the background work and the results obtained in \cite{Viglizzo_2005a} and the relevant functor for the enterprise is ready-made. On the contrary, here, we have to identify the properties of the spaces we are dealing with along with the appropriate functors to address the problem. Indeed, a major point of the present work lies in identifying the peculiarities of product conditional measurable spaces (in general) and product conditional measurable spaces sharing a family of conditioning events (in particular), which leads to the definition of a new---tailor-made for the problem at hand---functor, i.e., the functor $\CPSFM$ upon which the functor  $\TypeF := \CPSFM \ProdFM$ is based. As such, the present work, by additionally spelling out the details of the---relevant---proofs from \cite{Viglizzo_2005a} can prove to be a useful introduction to categorical and coalgebraic methods.

\section{Applications}
\label{sec:applications}

On general and conceptual grounds, as hinted in \Sref{subsec:motivation_results}, our construction can be used as a `foundation' for the construction obtained in \citet[Section 2]{Battigalli_Siniscalchi_1999}. In order to see this point, all the relevant information has already been provided in \Sref{foot:sAlg_CPS}, whose details we now spell out. Thus, starting from a $\sigma$-algebra over $\Delta^{\CEventsM} (M)$ as in \Mref{def:sAlg_CPS}, we do provide a `conceptual' foundation  in terms of what a hypothetical individual could consider as a relevant event. And that is the crucial point: indeed, when we start from topological assumptions, as in  \cite{Battigalli_Siniscalchi_1999}, with $\Delta^{\CEventsM} (M)$ endowed with the topology of weak convergence, it turns out from  \citet[Theorem 17.24, p.112]{Kechris_1995}, opportunely adapted to the presence of conditioning events, that the two $\sigma$-algebras are one and the same. Thus, the events a hypothetical individual can conceive are `the same' and---in a sense---we do not lose anything by working with topological assumptions in terms of \emph{expressibility}.\footnote{Regarding this notion, see for example \citet[Section 1.3]{Battigalli_et_al_2011}.} It is actually in this spirit that the present construction is used in \cite{Meier_Perea_2023}\footnote{With the caveat that the authors employ the term ``universal'' to refer to what we call here ``terminal'' (see \citet[Section 4.1, p.20]{Meier_Perea_2023}).} for the epistemic characterization of their solution concept called ``Forward and Backward Rationalizability''. And, once more, as already pointed out in \Sref{subsec:motivation_results}, it is in this spirit that the present construction can be used for endeavours along the lines of \cite{Bergemann_Morris_2005}, \cite{Dekel_et_al_2007}, or \cite{Battigalli_et_al_2011} when working with conditioning events (similar, for example, to \cite{Penta_2015} and \cite{Mueller_2016}).

However, crucially, as a matter of fact, we do lose something by working with topological assumptions when in presence of conditioning events, since the present work actually generalizes the one in \citet[Section 2]{Battigalli_Siniscalchi_1999}, where there is the need to assume an at most countable family of clopen (i.e., closed and open) conditioning events contrary to the present setting where the conditioning events are unrestricted.\footnote{See \Sref{subsubsec:technical_assumptions} for the details behind the need for these assumptions in \cite{Battigalli_Siniscalchi_1999} and our possibility of dropping them altogether.} Thus, in the next sections, we provide examples of how our construction can prove to be useful in various game-theoretical contexts by exploiting the two relaxations of the assumptions in \citet[Section 2]{Battigalli_Siniscalchi_1999} we introduce, namely:
\begin{itemize}[leftmargin=*]
\item lack of topological assumptions regarding the conditioning events;

\item lack of cardinality assumptions regarding the conditioning events.
\end{itemize}
Additionally, we also show how the notion of conditioning events we employ proves to be amenable to various interpretations, a point which turns out to be useful---for example---for the study of environments in presence of lack of topological assumptions regarding a parameter space and `unverifiable' conditioning events.

Now, whereas we provide examples of endeavours which simply cannot be performed \emph{without}  the present type structure in light of the need to free the conditioning events from possible restrictions, one point is in order: the universal type structure built in \Sref{sec:universality} \emph{in itself} does not provide any insight on the analyses that can be performed with it. Indeed, recalling the quote in \Sref{subsec:motivation_results} from \citet[Section 8, pp.1672--1673]{Aumann_Heifetz_2002}, this is a framework that acts as a tool for potential analyses. However, as such, it can indeed allow to perform new analyses that can lead to new insights.\footnote{We are grateful to an anonymous referee for having pointed out the need to further clarify this point.}

\subsection{Topology-Free Conditioning Events}
\label{subsec:topology-free_conditioning_events}

In this section, we show how our lack of topological assumptions on the conditioning events can prove to be useful in game-theoretical contexts. 

For this purpose, we now introduce a minimum of game-theoretical notation. Thus, in particular, a dynamic game with perfect recall and possibly simultaneous moves in its extensive form representation\footnote{See \citet[Definition 200.1, Chapter 11.1.2]{Osborne_Rubinstein_1994} for a similar definition (under a different name), with the caveat that the latter does not explicitly allow for simultaneous moves (see  \citet[Definition 200.1, Chapter 6.3.2]{Osborne_Rubinstein_1994} for the corresponding extension).} is a tuple
\begin{equation}
\label{eq:dynamic_game}
\Gamma := \la I , (\State_i, A_i)_{i \in I} , X, Z , (H_i, S_i, u_i)_{i \in I} \ra
\end{equation} 
where, given that the notation employed here has to be taken as standalone with respect to the rest of the paper,  $I$ now denotes the set of players, and, for every $i \in I$, we let $\State_i$ denote the set of \emph{non-epistemic types} (e.g., payoff types) of player $i$ and $A_i$ her set of \emph{actions}. The set $X$ is the set of \emph{histories}, where a history $x$ is either the empty sequence $\la \varnothing \ra$ (i.e., the initial history), or it is a sequence $(a^1, \dots, a^K)$,  where $a^k := (a^{k}_{i})_{i \in I}$ with $a^{k}_i \in A_i$ for every $i \in I$ and for every $1 \leq k \leq K$. We let $A(x) : = \prod_{i \in I} A_i (x)$ denote the set of actions available to the players at history $x$, with player $i$ \emph{active} at history $x$ if $\abs{A_i (x)} \geq 2$. The set $Z \subseteq X$ is the set of \emph{terminal} histories. Given that $H_i$ denotes the set of \emph{information sets} of player $i$ with $H := \bigcup_{i \in I} H_i$ and that we extend  to information sets the notation introduced above of actions available at a history, a \emph{strategy} of player $i$ is a function $s_i : H_i \to \bigcup_{\overline{h} \in H_i} A_i (\overline{h})$ such that $s_i (h) \in A_i (h)$, for every $h \in H_i$. We let $S_i$ denote the set of strategies of player $i$, with  $S_{-i} := \prod_{j \in I \setminus \{i\}} S_j$ and $S := \prod_{j \in I} S_j$. Also, we let $S (h)$ denote the set of strategy profiles that allow information set $h \in H$, with $S_i (h) := \pi_{S_i} S (h)$ and $S_{-i} (h) := \pi_{S_{-i}} S (h)$.  Finally, given that $\Xi_i := \State_i \times S_i$, with $\Xi_{-i}$, $\Xi$, and  $\Xi_{-i} (h)$ accordingly defined as above, we let $u_i \in \Re^{\Xi} $ be a Bernoulli (unique up to positive affine transformation) utility function. 

Now, our construction covers the game-theoretical setting described above by freeing it from the necessity of working only with clopen conditioning events. Indeed, given that we let $(\Xi , \sAlg_\Xi)$ be the relevant measurable space of interest, in the present context, we identify a player $i$'s set of conditioning events (topologically unrestricted) as the collection of sets $\Xi_{-i} (h)$, for every $h \in H_i$, i.e.,
\begin{equation*}
\CEvents_i := \Set { \Xi_{-i} (h) | h \in H_i },
\end{equation*}
with $\CEvents_{\Xi} := \prod_{i \in I} \CEvents_i$. Thus, building the universal type structure on the conditional measurable space $(\Xi , \sAlg_{\Xi}, \CEvents_{\Xi})$, we can capture a situation where a player $i$ has beliefs over her own strategies and non-epistemic types for every possible information set in the game, including those where she is not active. Now, if we additionally want to work with a type structure where every player $i$ has beliefs only over \emph{her} domain of uncertainty $\Xi_{-i}$ and these beliefs arise only at those information set where she is active, i.e., those information sets belonging to $H_i$, we simply have to `carve out' from the previously obtained universal type structure the corresponding belief-closed type structure where the only conditioning events are those in $\CEvents_i$ and the only relevant domain of uncertainty is given by $\Xi_{-i}$, for every $i$, where a \emph{belief-closed} type structure\footnote{The following notion is a translation in the framework of type structures of \citet[Definition 2.15, p.12]{Mertens_Zamir_1985}. \cite{Battigalli_et_al_Forthcoming} refer to these constructs as \emph{self-evident events}.} of a given type structure $\mscr{T} := \la (T_i , \beta_i)_{i \in I} \ra$ is a type structure $\widetilde{\mscr{T}} := \la (\widetilde{T}_i , \widetilde{\beta}_i)_{i \in I} \ra$ such that $\supp \widetilde{\beta}_{i, C} (\widetilde{t}_i) \subseteq \Xi_{-i} \times \widetilde{T}_{-i}$,  for every $i \in I$, $C \in \CEvents_i$, and $\widetilde{t}_i \in \widetilde{T}_i$ with $\widetilde{T}_i \subseteq T_i$. As a result, we would obtain, as mentioned above, a belief-closed type structure of the original one, which---nonetheless---would be itself universal in its own rights given the imposed restrictions.

Having a framework that allows to deal with the potential presence of topologically unrestricted conditioning events is particularly important because, as \citet[Section 5.6, p.758]{Battigalli_Tebaldi_2019} point out, working with clopen conditioning events in the spirit of \citet[Section 2]{Battigalli_Siniscalchi_1999} can be restrictive in certain applications. For example, given a player $i \in I$ and assuming that $\State_i$ is uncountable  and that player $i$ obtains a signal regarding $\theta_{-i} \in \State_{-i}$, even if the signal is discrete, it is not necessarily the case that $\Xi_{-i} (h)$ is clopen for a given information set $h \in H_i$.\footnote{See the similar discussion in \citet[Section 6]{De_Vito_2023}.}

In particular, the present construction can prove to be a crucial tool in \emph{psychological game theory}. Indeed, as pointed out in  \Sref{subsec:motivation_results}, this is a field where dynamic strategic interactions are of special interest (as argued in \citet[Section 2]{Battigalli_Dufwenberg_2009}) and  where infinite hierarchies of beliefs play a fundamental role in light of the fact that the utility functions of the players depend on---possibly higher order---beliefs of their co-players. Thus, for example, in \citet[Section 5.6, p.758]{Battigalli_Tebaldi_2019}, it is sketched a situation where, given two players $i$ and $j$ involved in a face-to-face interaction, there could be \emph{observable} features of $j$ that may signal $j$'s first-order beliefs to player $i$. Thus, letting $\mu^{1}_j$ denote a first-order belief of $j$, in this scenario, the tuple $(\theta_j , s_j, \mu^{1}_j)$ compatible with a given information set in the dynamic game corresponding to the interaction under scrutiny may not be clopen. As such, no formal analysis of this scenario could be performed without relying on the construction developed here, which allows to drop the topological assumptions regarding the conditioning events altogether.

\subsection{Uncountability of the Conditioning Events}
\label{subsec:uncountability_of_the_conditioning_events}

Having shown how the lack of topological assumptions on the conditioning events can be useful, we now discuss possible endeavours where it is the lack of cardinality assumptions on the family of conditioning events that is going to be crucial.

In particular, we focus on how the present construction can be used in the context of \emph{epistemic game theory}. For this purpose, it has to be recalled that the notion of \emph{Rationality and Common Strong Belief in Rationality} (henceforth, RCSBR) introduced in \citet[Section 4.2, p.372]{Battigalli_Siniscalchi_2002} is the epistemic solution concept that characterizes \emph{Strong Rationalizability}\footnote{Alternatively called ``Extensive-Form Rationalizability'' (as, for example, in \citet[Definition 5, p.373]{Battigalli_Siniscalchi_2002}). Here, we adopt a terminology used for example in \cite{Battigalli_Siniscalchi_2003}, which has the virtue of distinguishing this one from other forms of Rationalizability that can be implemented in the analysis of dynamic games represented in their extensive-form (i.e., Initial (or Weak) Rationalizability \emph{{\`a} la} \cite{Ben-Porath_1997}, Backward Rationalizability \emph{{\`a} la} \cite{Penta_2015}, or Forward and Backward Rationalizability \emph{{\`a} la} \cite{Meier_Perea_2023}).} of \citet[Definition 9, p.1042]{Pearce_1984} and \citet[Definition 2, p.46]{Battigalli_1997} in a belief-complete type structure. This is particularly appealing since Strong Rationalizability is considered a non-equilibrium based solution concept that captures \emph{forward induction reasoning} predictions.\footnote{See \cite{Govindan_Wilson_2009} for an analysis of forward induction in the context of  equilibrium-based solution concepts.}

Recently, two papers have profitably employed RCSBR and Strong Rationalizability in applications: in particular, in \cite{Friedenberg_2019}, RCSBR is used to study behavior in bargaining, while  Strong Rationalizability is employed in \citet[Proposition 2, p.495]{Guo_Shmaya_2021} to `refine' the results obtained via Bayes Nash Equilibrium in the study of the behavior of a platform that provides probabilistic forecasts to a customer in presence of costly miscalibration. Crucially with respect to our endeavour, both papers point out, respectively, in \citet[Remark A.1, pp. 1855--1856]{Friedenberg_2019} and \citet[Section 4.2, p.494]{Guo_Shmaya_2021}, a technical limitation of RCSBR and Strong Rationalizability, namely that both solution concepts are defined for an at most countable family of conditioning events, which are \emph{information sets} in \cite{Friedenberg_2019} and \emph{messages} in \cite{Guo_Shmaya_2021}. Now, the present work does not provide an answer to solve the issue just mentioned. However, it is not actually possible to provide an answer to this very issue \emph{without} the present work. 

Indeed, in light of its universality properties, the Universal Type Structure of \cite{Battigalli_Siniscalchi_1999} is the type structure that is employed to obtain the behavioral predictions corresponding to RCSBR when no \emph{a priori} assumptions are made on the infinite hierarchies of beliefs held by the players, where---in particular---it is its belief-completeness that is typically exploited for this purpose (as in \citet[Proposition 6, p.373]{Battigalli_Siniscalchi_2002}). However, that object cannot deal with uncountably many conditioning events. Thus, with respect to this point, our construction can be fruitfully used as the epistemic type structure that is going to be the main building block for a definition of RCSBR based on possibly uncountably many conditioning events, which, given an appropriate definition of Strong Rationalizability (for possibly uncountably many conditioning events), could actually epistemically characterize the aforementioned new notion of Strong Rationalizability exploiting standard arguments in epistemic game theory based on the belief-completeness of this type structure. This, in turn, would lead---among other things---to the possibility of properly addressing the points (described above) raised in \citet[Remark A.1, pp. 1855--1856]{Friedenberg_2019} and \citet[Section 4.2, p.494]{Guo_Shmaya_2021}.

\subsection{Nature of the Conditioning Events}
\label{subsec:nature_of_the_conditioning_events}

\enlargethispage{\baselineskip}

In both \Sref{subsec:topology-free_conditioning_events} and \Sref{subsec:uncountability_of_the_conditioning_events}, we consider conditioning events as observable events (e.g., information sets) belonging to a dynamic game. However, as natural as this is, interestingly, this is \emph{not} the only way in which we can interpret our \emph{abstract} conditioning events. Thus, with respect to this point, as a case study we focus on \cite{Guo_Yannelis_2022}, where an analysis of belief-free full implementation in the spirit of \cite{Bergemann_Morris_2009} and \cite{Bergemann_Morris_2011} is performed in presence of possible coalitions, with the understanding that a coalition can be an example of a conditioning event, and---incidentally---it is also an example of how our topology-free construction with unrestricted conditioning events can be a foundation for `robustness' endeavours in the spirit of \cite{Bergemann_Morris_2005} when in presence of conditioning events and---in particular---counterfactual reasoning.\footnote{We would would like to thank the editor for suggesting this application.}

To see this, the following is---modulo notation and terminology---the framework of \citet[Section 2, pp.555--556]{Guo_Yannelis_2022}, where a \emph{payoff environment} is a tuple
\begin{equation*}
\mscr{E} := \la I, X, A , (\State_i , u_i )_{i \in I} \ra
\end{equation*}
with $I$ being a set of \emph{agents}, $X$ a set of \emph{feasible deterministic outcomes}, $A := \Delta (X)$ a set of \emph{feasible outcomes}, whereas $\State_i$ is the space of payoff types of player $i$, with $\State := \prod_{j \in I} \State_j$, and $u_i : A \times \State \to \Re$ is her Bernoulli (unique up to positive affine transformation) utility function. Given a payoff environment $\mcal{E}$, an information structure appended on $\mcal{E}$ is a tuple $\mscr{Y} := \la (Y_i, \varsigma_i , \varpi_i )_{i \in I}$, where, for every $i \in I$, $Y_i$ is agent $i$'s \emph{information space}, $\varsigma_i : Y_i \to \State_i$ is her \emph{payoff type function}, and $\varpi_i : Y_i \to \Delta (Y_{-i})$ is her \emph{belief function}.\footnote{In \citet[Section 2, p.556]{Guo_Yannelis_2022}, the authors actually use the expression ``type structure'' to refer to this object and use $T_i$ to denote the type space of an arbitrary player $i$. Here, we opt to call this object ``information structure'' with $Y_i$ being agent $i$'s \emph{information space} to avoid an overloading of the previous expressions and notation.} Finally, to capture the presence of coalitions and corresponding beliefs of a player $i$, given that $\mcal{I} := 2^I \setminus \{\emptyset\}$ denotes the set of \emph{coalitions} and that
\begin{equation*}
\varpi_{i, J} (y_i) \big( y_{S\setminus \{i\}} \big) := %
\marg_{ Y_{S\setminus\{i\}}} \varpi_i (y_i) (y_{-i}),
\end{equation*} 
where $J \in \mcal{I}$ with $i \in J$ and $y := (y_i)_{i \in I} \in Y$ (with $\marg$ denoting the marginal operator as canonically defined), a \emph{belief revising rule} specifies a posterior belief $\varpi_i (y_i) (y_{-i}) \in \Delta (Y_{-i})$ with $\varpi_{i, J} (y_i) ( y_{J \setminus \{i\}} ) = 1$ whenever $\varpi_{i, J} (y_i) \big( y_{J \setminus \{i\}} \big) = 0$, where the posterior belief is defined via the Bayes rule (i.e., Axiom (C3) in \Mref{def:CPS}) whenever $\varpi_{i, J} (y_i) \big( y_{J \setminus \{i\}} \big) > 0$.

Now, it turns out that our construction in  \Sref{subsec:dale} allows to obtain the topology-free universal type structure in presence of unrestricted conditioning events  $\mscr{T}^* := \la (T^{*}_i , \beta^{*}_i)_{i \in I} \ra$  for the setting just described, where an arbitrary type structure $\mscr{T}$ is essentially equivalent to an information structure $\mscr{Y}$ given appropriate restrictions. Thus, in particular, given that $\CEventsI := \Coalitions$ with $\ProdFI := \Cons_I \times \Cons_\State \times \Proj_{-i}$, we need to define the new functor $\TypeC := \CPSFC \ProdFI$, i.e.,
\begin{equation*}
\TypeC  := \CPSFC \ProdFI = \CPSFC ( \Cons_I \times \Cons_\State \times \Proj_{-i}) ,
\end{equation*}
where the appropriate definitions regarding the way in which the functor acts on objects and morphisms follow the path described in \Sref{subsec:dale}, which, as a result, allows us to obtain a belief function
\begin{equation*}
\beta^{*}_i := (\beta^{*}_{i, J})_{J \in \Coalitions} : T_i \to \Delta^{\Coalitions} (I \times \Theta \times T_{-i}) .
\end{equation*}
Given the steps sketched above, it remains to carve out (in the spirit of \Sref{subsec:topology-free_conditioning_events}) the resulting structure, a belief-closed type structure with the desired properties which is universal in its own rights given these properties, where---in particular---we have to impose that $\beta^{*}_i$ is such that $\marg_{J} \beta^{*}_{i, J} (t_i) = \delta_{J}$ and $\marg_{\State_i} \beta^{*}_{i, J} (t_i) = \delta_{\{\theta_i\}}$ for a $\theta_i \in \State_i$, for every $t_i \in T_i$ and $J \in \mcal{I}$, where $\delta_J$ and $\delta_{\{\theta_i\}}$ denote the Dirac measure as canonically defined. Indeed, this delivers exactly what we want in light of the obvious relation between $\mscr{Y} := \la (Y_i, \varsigma_i , \varpi_i )_{i \in I}$ and an arbitrary belief-closed type structure $\mscr{T} := \la (T_i , \beta_i)_{i \in I} \ra$ with $\beta_i$ defined along the lines of $\beta^{*}_i$ above.

One natural question is why the type structure built here should be used in the context presented in this section. Now, the reason lies in a point which overlaps with the issue presented in \Sref{subsec:uncountability_of_the_conditioning_events} concerning the cardinality of the set of conditioning events: indeed, as soon as the set of agents $I$ in $\mscr{E}$ is uncountable, it becomes crucial to use a construction along the lines of  the present one.

\appendix

\section*{Appendices}

\section{Category Theory Background}
\label{app:category_theory_background}

In this section, for self-containment reasons, we provide an introduction to those elements of category theory that we need for our purposes. Thus, first of all, we define what a category actually is.

\begin{definition}[Category]
\label{def:category}
A \emph{category} $\Cat$ consists of the following data:
\begin{itemize}[leftmargin=*]
\item a collection of \emph{objects}, denoted by $\Ob (\Cat)$;

\item for every $A, A' \in \Ob (\Cat)$, a collection of \emph{morphisms} (or \emph{arrows}) from $A$ to $A'$, denoted by $\Cat (A, A')$;

\item for every $A, A', A'' \in \Ob (\Cat)$, a \emph{composition law} 
\begin{equation*}
\Cat (A, A') \times \Cat (A', A'')  \to \Cat (A, A''), 
\end{equation*}
where the composite of $f \in \Cat (A, A')$ and $f' \in \Cat (A', A'')$ is denoted $f' \circ f$;

\item for every $A \in \Ob (\Cat)$, a morphism $\id_A \in \Cat (A, A)$, called the \emph{identity on} $A$. 
\end{itemize}
These data satisfy the following axioms:
\begin{itemize}[leftmargin=*]
\item \emph{associativity:} for every $f \in \Cat (A, A')$, $f' \in \Cat (A', A'')$, $f'' \in \Cat (A'', A''')$,
\begin{equation*}
(f'' \circ f') \circ f = f'' \circ (f' \circ f);
\end{equation*}
\item \emph{identity:} for every $f \in \Cat (A, A')$, 
\begin{equation*}
f \circ \id_A = f = \id_{A'} \circ f.
\end{equation*}
\end{itemize}
\end{definition}

Having introduced the notion of category, it is typical to provide actual examples of categories, since it could be stated that one of the major points of embracing a `categorical' approach is the fact that it is possible to identify properties shared by seemingly exceedingly different constructs. 

\begin{example}[label=ex:categories, name=Examples of Categories]
We now present two examples of categories that show how flexible \Mref{def:category} actually is.
\begin{itemize}[leftmargin=*]
    \item The category $\SET$ is the category whose objects are sets and whose morphisms are functions between sets. Relevant for our purposes, given $A, A' \in \Ob (\SET)$, we have $A \times A' \in \Ob (\SET)$.
    
        \item The category $\Top$ is the category whose objects are topological spaces and whose morphisms are continuous functions between topological spaces. \qedhere
\end{itemize}
\end{example}

\begin{definition}[Subcategory]
\label{def:subcategory}
Given a category $\Cat$, a \emph{subcategory} $\SubCat$ of $\Cat$ consists of:
\begin{itemize}[leftmargin=*]
\item a subclass $\Ob (\SubCat)$ of $\Ob (\Cat)$ and

\item a subclass $\SubCat (A, A')$ of $\Cat (A, A')$
\end{itemize}
such that $\SubCat$ is closed under composition and identities. A subcategory is \emph{full} if $\SubCat (A, A') = \Cat (A, A')$, for every $A, A' \in \Ob (\SubCat)$.
\end{definition}

Given two categories, it is possible to have morphisms between the two. The next definition captures exactly what are the properties that these morphisms, called functors, need to satisfy in order to be `well-behaved'.

\begin{definition}[Functor]
\label{def:functor}
Given two categories $\Cat$  and $\Cat'$, a \emph{functor} $\Functor: \Cat \to \Cat'$ between $\Cat$ and $\Cat'$ consists of:
\begin{itemize}[leftmargin=*]
\item a morphism $\Ob (\Cat) \to \Ob (\Cat')$, written $A \mapsto \Functor (A)$;

\item for every $A, A' \in \Ob (\Cat)$, a morphism
\begin{equation*}
\Cat (A, A') \to \Cat' ( \Functor (A), \Functor (A')),
\end{equation*}
written $f \mapsto \Functor (f)$. 
\end{itemize}
These data satisfy the following axioms:
\begin{itemize}[itemsep=0.5ex]
\item[AC)] $\Functor (f' \circ f) = \Functor (f') \circ \Functor (f)$, with $f \in  \Cat (A, A')$ and $f' \in \Cat (A', A'')$,
\item[AI)] $\Functor (\id_A) = \id_{\Functor (A)}$, for every $A \in \Ob (\Cat)$.
\end{itemize}
\end{definition}

The notion of functor captures the crucial idea of \emph{functoriality}. Thus, given two categories $\Cat$ and $\Cat'$, if we define a morphism $\Functor$ between them and we prove that $\Functor$ actually is a functor, i.e., it satisfies Axioms (AC) and (AI), then it turns out that $\Functor$ does not simply act on the elements $\Ob (\Cat)$, but also on its morphisms in an appropriate way.\footnote{Regarding this notion, see the discussion (and the examples) in \citet[Section 4]{Jacobs_Rutten_1997}.}

\begin{example}[Product Categories \& Projection Functors]
\label{ex:product_categories}
Given a \emph{product} category $\prod_{\lambda \in \Lambda} \Cat_\lambda$, the \emph{projection functor} $\Proj_\lambda$ is the functor defined as
\begin{itemize}[leftmargin=*]
    \item $(A_\lambda)_{\lambda \in \Lambda} \mapsto A_\lambda$, for every $(A_\lambda)_{\lambda \in \Lambda} \in \Ob \big( \prod_{\lambda \in \Lambda} \bm{\mathsf{C}}_\lambda \big)$,
    
    \item $(f_\lambda)_{\lambda \in \Lambda} \mapsto f_\lambda$, for every $(f_\lambda)_{\lambda \in \Lambda} \in \Cat_\lambda \big( (A_\lambda)_{\lambda \in \Lambda} , (A'_\lambda)_{\lambda \in \Lambda} \big)$,
\end{itemize}
for every $\lambda \in \Lambda$. Now, let $\SET^2 := \SET \times \SET$ and assume that $A, A'$ are exactly the same sets considered in \Mref{ex:categories}. It is important to observe that $A \times A' \in \Ob (\SET)$ is not the same as $A \times A' \in \Ob (\SET) \times \Ob (\SET)$ since the two expressions capture objects belonging to different categories.
\end{example}

In this paper, we focus on a specific class of functors, namely, those functors that map a category to itself.

\begin{definition}[Endofunctor]
Given a category $\Cat$, an \emph{endofunctor on} $\Cat$ is a functor $\Functor : \Cat \to \Cat$.
\end{definition}

\begin{example}[label=ex:endofunctors, name={Three Endofunctors}]
We now provide three examples of endofunctors: the first two are phrased in rather abstract terms and are extensively used in our construction, while the third is an example of a functor in the category $\SET$.
\begin{itemize}[leftmargin=*]
\item Given an arbitrary category $\Cat$, the \emph{identity (endo)functor} $\Id : \Cat \to \Cat$ is defined as $\Id (A) := A$, for every $A \in \Ob (\Cat)$, and $\Id (f) := f$, for every $f \in \Cat (A, A')$, with $A, A' \in \Ob (\Cat)$;

\item Given an arbitrary category $\Cat$, the \emph{constant endofunctor given $X$}, denoted $\Cons_X : \Cat \to \Cat$, is defined as $\Cons_X (A) := X$, for every $A \in \Ob (\Cat)$, and $\Cons_X (f) := \id_X$, for every $f \in \Cat (A, A')$, with $A, A' \in \Ob (\Cat)$.

\item Building on the category $\SET$, we now provide an example of an endofunctor on $\SET$. Thus, let $\PowerSetF: \SET \to \SET$ be defined as:
\begin{itemize}[leftmargin=*]
    \item regarding objects, $\PowerSetF (A) := \PowerSet (A)$, for every $A \in \Ob (\SET)$, with $\PowerSet$ denoting the power set;
    
    \item regarding morphisms,
\begin{equation*}
\PowerSetF (f) : \PowerSet (A) \to \PowerSet (A') 
\end{equation*}
such that $ \PowerSetF (f) (B) := f(B)$, for every $f \in \SET (A, A')$ and $B \in \PowerSet (A)$.
\end{itemize}
The resulting morphism is the so-called \emph{power set (endo)functor}.\footnote{For completeness, this is called the \emph{covariant} power set functor: see \citet[Chapter 1.2, p.22]{Leinster_2014} for an explanation of this terminology.} \qedhere
\end{itemize}
\end{example}

Having established the setting above, it is possible to define when two objects in a given category are---in a sense to be precisely defined---the same. The two definitions that follow accomplish exactly this objective.

\begin{definition}[Isomorphism]
\label{def:isomorphism}
Given a category $\Cat$ and objects $A, A' \in \Ob (\Cat)$, $f \in \Cat (A, A')$ is an \emph{isomorphism} if there exists a morphism $f' \in \Cat (A', A)$ such that the following diagram
\begin{equation*}
\begin{tikzcd}
A %
	\arrow[r, "f"] %
	\arrow[rr, bend right=35, "f' \circ f = \id_A"']
	& A' %
		\arrow[r, "f'"] %
		\arrow[rr, bend left=35, "f \circ f' = \id_{A'}"]
		& A %
			\arrow[r, "f"] %
				& A'
\end{tikzcd}
\end{equation*}
commutes.
\end{definition}

\begin{definition}[Isomorphic Objects]
\label{def:isomorphic_objects}
Given a category $\Cat$ and objects $A, A' \in \Ob (\Cat)$, $A$ and $A'$ are \emph{isomorphic}, denoted $A \cong A'$, if there exists an isomorphism between $A$ and $A'$.
\end{definition}

It is the existence of a special kind of objects in a category, namely, the so-called terminal ones, that plays a major role in this paper.

\begin{definition}[Terminal Object]
\label{def:terminal_object}
Given a category $\Cat$, an object $\1 \in \Ob (\Cat)$ is  \emph{terminal} if for every $A \in \Ob (\Cat)$ there exists a unique morphism $! \in \Cat (A, \1)$ such that $!: A \to \1$, i.e., \begin{tikzcd} A \arrow[r, dashed, "!"] & \1 \end{tikzcd}, for every $A \in \Ob (\Cat)$.
\end{definition}

We close this section by introducing the missing notions from category theory that are needed for the purpose of our construction. Thus, in particular, in the following we take an index set $\Lambda$ and we treat it as a category\footnote{Technically, this is a \emph{small category}, i.e., a category whose both objects and morphisms form a set, i.e., not a class (see \citet[Chapter 3.2, p.75]{Leinster_2014} for the notion of small category and \citet[Chapter 1, pp.5--6]{Jech_2006} for the notion of class in set theory). Thus, the set $\bN$ can be taken as the index set.} 

\begin{definition}[Diagram]
Given an index set $\Lambda$ and an arbitrary category $\Cat$, a functor $\Diagram$ from $\Lambda$ to $\Cat$ is a \emph{diagram}. 
\end{definition}

\begin{definition}[$\opchain$-Chain for the Endofunctor $\Functor$]
\label{def:opchain_endofunctor}
Given a category $\Cat$ and an endofunctor $\Functor$ on $\Cat$, an $\opchain$\emph{-chain on the endofunctor} $\Functor$ is a diagram of the form
\begin{equation*}
\label{cd:opchain_endofunctor}
\begin{tikzcd}
\1 & \Functor^{1} (\1) \arrow[l, dashed, "!"'] & \Functor^{2} (\1)  \arrow[l, "\Functor^{1} !"'] & \dots . \arrow[l, "\Functor^{2} !"']
\end{tikzcd}
\end{equation*}
\end{definition}

\begin{definition}[Cone]
\label{def:cone_general}
Given an index set $\Lambda$, an arbitrary category $\Cat$, and a diagram $\Diagram$ from $\Lambda$ to $\Cat$,  a \emph{cone}\footnote{See \citet[Chapter 5.1, p.118]{Leinster_2014} for the definition of cone.} 
on $\Diagram$ is an object $A \in \Ob (\Cat)$ together with a family of maps
\begin{tikzcd}
\big( A \arrow[r, "f_\lambda"] & \Diagram_\lambda \big)_{\lambda \in \Lambda} ,
\end{tikzcd}
such that the following diagram commutes
\begin{equation*}
\begin{tikzcd}
\hspace{1.5cm} A %
	\arrow[d, "f_\lambda"', %
		start anchor={[shift={(0.55cm,0cm)}]}, %
		end anchor={[shift={(0.1cm,0cm)}]}] %
	\arrow[dr, "f_\lambda'", %
		start anchor={[shift={(0.55cm,0cm)}]}, %
		end anchor={[shift={(-0.5cm,0.3cm)}]}] \\%
\Diagram_\lambda %
	& \hspace{-0.7cm} \Diagram_{\lambda'} %
		\arrow[l, "\Diagram u", %
		start anchor={[shift={(-0.7cm,0cm)}]}, %
		end anchor={[shift={(-0.1cm,0cm)}]}]
\end{tikzcd}
\end{equation*}
for every $\lambda' \overset{u}{\longrightarrow} \lambda$. 
\end{definition}

\begin{definition}[Limit]
\label{def:limit}
A \emph{limit} $\Limit$ is a cone %
\begin{tikzcd}
\big( \Limit \arrow[r, "f_\lambda"] & \Diagram_\lambda \big)_{\lambda \in \Lambda} 
\end{tikzcd}
with the property that there exists a unique morphism 
\begin{tikzcd}
A \arrow[r, dashed, "\overline{f}"] & \Limit 
\end{tikzcd}
such that $\pi_\lambda \circ \overline{f} = f_\lambda$, for every $\lambda \in \Lambda$.\footnote{As usual, here,  $\pi_\lambda$ denotes the projection operator.}
\end{definition}

\begin{definition}[Preservation of Limits by Functors]
\label{def:preservation_limits}
A functor $\Functor$ \emph{preserves limits} if, given a limit  %
\begin{tikzcd}
\big( \Limit \arrow[r, "f_\lambda"] & \Diagram_\lambda \big)_{\lambda \in \Lambda} ,
\end{tikzcd}
we have that %
\begin{tikzcd}
\Big( \Functor \Limit \arrow[r, "\Functor f_\lambda"] & \Functor \Diagram_\lambda \Big)_{\lambda \in \Lambda} 
\end{tikzcd}
is a limit.
\end{definition}

\section{Measure-Theoretic Results}
\label{app:measure-theoretic_results}

We recall that, given a measurable space $\MS$, a $\pi$\emph{-system} is a nonempty family of subsets $\PiSystem \subseteq \PowerSet (M)$ that is closed under finite intersections. The next result is an immediate consequence of Dynkin's $\pi$--$\lambda$ Lemma\footnote{See \citet[Lemma 4.10]{Aliprantis_Border_2006}.} applied to all the conditioning events in $\CEventsM$.
 
\begin{applemma}
\label{lem:pi_system}
Given a product conditional measurable space $(M \times X, \sAlg_M \otimes \sAlg_X, \CEventsM)$, where $\sAlg_{M} \otimes \sAlg_{X} := \sigma (\PiSystem)$ with $\PiSystem$ an arbitrary $\pi$-system, and two CPSs $\cps, \cps' \in \CPSM (M \times X)$, if $\cps (D|C) = \cps' (D|C)$, for every $D \in \PiSystem$ and $C \in \CEventsM$, then $\cps (E|C) = \cps' (E|C)$, for every $E \in \sigma (\PiSystem)$ and $C \in \CEventsM$.
\end{applemma}

The next remark is an analog of \citet[Lemma 2, p.396]{Viglizzo_2005a}, which in turn is in the spirit of \citet[Equation 3.2, p.330]{Heifetz_Samet_1998}.

\begin{remark}
\label{rem:measurability}
Given a conditional measurable space  $\MSC$, two product conditional measurable spaces  $(M \times X, \sAlg_M \otimes \sAlg_X, \CEventsM)$ and $(M \times Y, \sAlg_M \otimes \sAlg_Y, \CEventsM)$ sharing $\CEventsM$, and a measurable function $f \in (M \times Y)^{(M \times X)}$,
\begin{equation*}
\gamma^{p}_C \big( f^{-1} (E) \big) = \widehat{f}^{-1} \big( \gamma^{p}_C (E) \big) ,
\end{equation*}
for every $E \in \sAlg_{M \times Y}$, $C \in \CEventsM$, and $p \in [0, 1]$.
\end{remark}

Finally, the next lemma is a version of \citet[Lemma 4.5, p.334]{Heifetz_Samet_1998} which takes into account the presence of conditioning events.

\begin{applemma}
\label{lem:HS98}
Given a product conditional measurable space $(M \times X, \sAlg_M \otimes \sAlg_X, \CEventsM)$ and an algebra $\Alg_{M \times X}$ such that $\sAlg_{M \times X} := \sigma (\Alg_{M \times X})$, if 
\begin{equation*}
\Alg_{\CPSM (M \times X)} := %
\sigma \big(  \Set { \xi^{p}_C (E) | %
E \in \Alg_M \otimes \Alg_X , \ p \in [0,1], \ C \in \CEventsM } \big)  
\end{equation*}
on $\CPSM ( M \times X)$, then $\Alg_{\CPSM (M \times X) } = \sAlg_{\CPSM (M \times X)}$.
\end{applemma}

\begin{proof}
Observe that the extension of \Mref{def:sAlg_CPS} to $\CPSM (M \times X)$, i.e., $\sAlg_{\CPSM (M \times X)}$, coincides with the extension of \Mref{def:sAlg_CPS} to $[\Delta (M \times X)]^{\CEventsM}$, which is a collection of copies of $\Delta (M \times X)$. Hence, the result follows from \citet[Lemma 4.5, p.334]{Heifetz_Samet_1998} applied to $[\Delta (M \times X)]^{\{C\}} \equiv \Delta (M \times X)$, for every $C \in \CEventsM$.
\end{proof}

\section{\texorpdfstring{Proofs of \Sref{sec:universality}}%
{Proofs of Section 4}}
\label{app:proofs_universality}

\begin{proof}[Proof of \Mref{lem:coalgebra_morphism_preservation}]
We establish the result by proving that 
\begin{equation}
\label{eq:inductive_proof}
\pi_n \circ \cone^{\trans} = \pi_n \circ \cone^{\trans'} \circ \cmor
\end{equation}
proceeding inductively over $n \in \bN$, first by considering the following diagram
\begin{equation*}
\begin{tikzcd}
\Car %
	\arrow[rr, "\cmor"]  %
	\arrow[rd, "\cone^{\trans}"] %
	\arrow[rdd, "\cone^{\trans}_{n}"', end anchor={[shift={(-0.3cm,0pt)}]}] &	%
	& \Car' %
		\arrow[ld, "\cone^{\trans'}"'] %
		\arrow[ldd, "\cone^{\trans'}_{n}", end anchor={[shift={(0.3cm,0pt)}]}] %
\\
	& \Projective %
		\arrow[d, "\pi_n"] & %
\\
	& \TypeF^{n} (\1)  &  %
\end{tikzcd}
\end{equation*}
and observing that \Mref{eq:inductive_proof} is equal to $\cone^{\trans}_{n} = \cone^{\trans'}_{n} \circ \cmor$ from \Mref{def:cone}.
\begin{itemize}[leftmargin=*]
\item ($n = 0$) We have that $\cone^{\trans'}_{0} \circ \cmor = !_{\Car'} \circ \cmor = !_{\Car} = \cone^{\trans}_{0}$.

\item ($n \geq 0$) Assume that the result has been established for $n$, i.e., $\cone^{\trans}_{n} = \cone^{\trans'}_{n} \circ \cmor$. Thus, we have that
\begin{align*}
\cone^{\trans'}_{n+1} \circ \cmor & = \TypeF \cone^{\trans'}_{n} \circ \trans' \circ \cmor \\
& = \TypeF \cone^{\trans'}_{n} \circ  \TypeF \cmor \circ \trans \\
& = \TypeF (\cone^{\trans'}_{n} \circ \cmor) \circ  \trans \\
& = \TypeF \cone^{\trans}_{n} \circ  \trans \\
& = \cone^{\trans}_{n+1},
\end{align*}
that is made perspicuous by the following diagram
\begin{equation*}
\begin{tikzcd}[row sep = huge, column sep = huge]
\Car %
	\arrow[r, "\cmor"] %
	\arrow[d, "\trans"'] %
	\arrow[rrd, "\cone^{\trans}_{n+1}" description, pos=0.3] %
	& \Car' %
		\arrow[d, "\trans'", pos=0.2] %
		\arrow[rd, "\cone^{\trans'}_{n+1}"] %
		\arrow[r, "\cone^{\trans'}_{n}"] %
			& \TypeF^{n} (\1) %
\\
\TypeF (\Car) %
	\arrow[r, "\TypeF \cmor"'] %
	\arrow[rr, "\TypeF \cone^{\trans}_{n}"', bend right] %
	\arrow[rr, "\TypeF \big( \cone^{\trans'}_{n} \circ \cmor \big)"', bend right=65] %
	& \TypeF (\Car') %
		\arrow[r, "\TypeF \cone^{\trans'}_{n}"'] %
		& \TypeF^{n+1} (\1) %
			\arrow[u, "\TypeF^{n} !"']
\end{tikzcd}
\end{equation*}
that commutes.
\end{itemize}
Thus, what is written above establishes the result.
\end{proof}

\begin{proof}[Proof of \Mref{lem:terminal_transition}]
We let $\zetap : \Terminal \to \Projective_\TypeF$ be defined as $\pi^{\TypeF}_n \zetap = \pi_{n+1}$, for every $n \in \bN$. We prove that $\pi^{\TypeF}_n \zetap \trans = \pi^{\TypeF}_n \cone^{\trans}_{\TypeF} \trans$. Hence, we have that
\begin{align*}
\pi^{\TypeF}_n \circ \cone^{\trans}_{\TypeF} \circ \trans & = \TypeF \cone^{\trans}_n \circ \trans \\
	& = \cone^{\trans}_{n+1} \\
	& = \pi_{n+1} \circ \cone^{\trans} \\
	& = \pi^{\TypeF}_{n} \circ \zetap \circ \cone^{\trans}.
\end{align*}
Thus, this establishes that $\zetap \cone^{\trans} = \cone^{\trans}_{\TypeF} \trans$  along with the fact that $\zetap (\Terminal) \subseteq \Terminal_{\TypeF}$. Concerning the measurability of $\zetap$, it is enough to prove that $(\zetap)^{-1} (E) \in \sAlg_{\Terminal}$ for an arbitrary $E := (\pi^{\TypeF}_n)^{-1} (E_n)$, where $E_n \in \sAlg_{\TypeF^{n+1} (\1)}$, since such sets generate the $\sigma$-algebra on $\Terminal_{\TypeF}$. Thus, this  is immediate, since 
\begin{align*}
(\zetap)^{-1} (E) & = (\zetap)^{-1} \Big( (\pi^{\TypeF}_n)^{-1} (E_n) \Big) \\
	& = (\pi^{\TypeF}_n \circ \zetap)^{-1} (E_n) \\
	& = (\pi_{n+1})^{-1} (E_n) ,
\end{align*} 
where the set in the last equation is measurable, since $E_n \in \sAlg_{\TypeF^{n+1} (\1)}$.
\end{proof}

To establish the following lemma, we need three new pieces of notation. First of all, given that 
\begin{equation*}
\pi^{\TypeF}_{n} : \Terminal_\TypeF \to \TypeF^{n+1} (\1) := \CPSM (M \times \TypeF^{n} (\1)) ,
\end{equation*}
we let $\pi^{\TypeF}_{n, C}$ denote the corresponding projection of the function on $[\Delta (M \times \TypeF^{n} (\1))]^{\{C\}}$, for every $C \in \CEventsM$. In second place, we let $\cps_C := \cps (\cdot | C )$. Finally, for every $\ell < n$ with $\ell, n \in \bN$, let $\rho_{n\ell} : \ProdFM \TypeF^{n} (\1) \to \ProdFM \TypeF^{\ell} (\1)$ be defined as  
\begin{equation*}
\rho_{n\ell} := \ProdFM \TypeF^{\ell} ! \circ \ProdFM \TypeF^{\ell+1} ! \circ \dots \circ \ProdFM \TypeF^{n-1} !  ,
\end{equation*}
i.e., the following diagram
\begin{equation*}
\begin{tikzcd}[column sep =large]
\ProdFM \TypeF^{\ell} (\1) %
	& \ProdFM \TypeF^{\ell+1} (\1) %
		\arrow[l, "\ProdFM \TypeF^{\ell} !"'] %
		& \dots %
			\arrow[l, "\ProdFM \TypeF^{\ell+1} !"'] %
			& \ProdFM \TypeF^{n} (\1) %
				\arrow[l, "\ProdFM \TypeF^{n-1} !"'] %
				\arrow[lll, "\rho_{n\ell}"', bend left]
\end{tikzcd}
\end{equation*}
commutes, with $\rho_{n n} = \id_{\ProdFM \TypeF^{n} (\1)}$, for every $n \in \bN$.

\begin{applemma}
\label{lem:cps}
There exists a measurable morphism 
\begin{equation}
\label{eq:belt_CPSProd}
\belt_{\CPSFM \ProdFM} := (\belt_{\DeltaC \ProdFM})_{C \in \CEventsM} : \Terminal_{\CPSFM \ProdFM} \to \CPSM (\Terminal_{\ProdFM})
\end{equation}
such that:
\begin{enumerate}[leftmargin=*, label=\arabic*)]
    \item $\belt_{\CPSFM \ProdFM} \circ \cone^{\trans}_{\CPSFM \ProdFM} = \CPSFM \cone^{\trans}_{\ProdFM}$, for every $\TypeF$-coalgebra $\Coalg := \la \Car, \trans \ra$;
    
    \item $\CPSFM \pi^{\ProdFM}_n \circ  \belt_{\CPSFM} = \pi^{\CPSFM \ProdFM}_{n}$, for every $n \in \bN$.
\end{enumerate}
\end{applemma}

Before proceeding with the proof of \Mref{lem:cps}, due to the fact that it is going to be used in what follows, it is worth stressing the notational convention implicit in \Mref{eq:belt_CPSProd} according to which $\belt_{\DeltaC \ProdFM}$ essentially stands for the notationally heavier $\belt_{\CPSFM \ProdFM, C}$ with $C \in \CEventsM$ arbitrary.

\begin{proof}[Proof of \Mref{lem:cps}]
The proof proceeds as follows: first, we define the morphism to then establish its measurability. It should be recalled that $\TypeF := \CPSFM \ProdFM$, a point which we occasionally exploit to lighten the notation.

\begin{itemize}[leftmargin=*]
\item \emph{Definition:} %
Given that $\sAlg_{\Terminal_{\ProdFM}}$ is generated by the family of sets
\begin{equation*}
\mcal{G} := \Set { E | \exists n \in \bN, E_n \in \sAlg_{M \times \TypeF^n (\1)}: %
E = \big( \pi^{\ProdFM}_n \big)^{-1} (E_n) } ,
\end{equation*}
where $E_n \in \sAlg_{M \times \TypeF^n (\1)}$ is such that $E = \big( \pi^{\ProdFM}_n \big)^{-1} (E_n)$, we define $\belt_{\CPSFM \ProdFM}$ for the elements of $\mcal{G}$ to then extend the definition to the elements of $\sAlg_{\Terminal_{\ProdFM}}$. Thus, let $\terminal \in \Terminal_{\TypeF}$ be arbitrary and  let
\begin{equation*}
\belt_{\CPSFM \ProdFM} (\terminal) ( E ) := \pi^{\TypeF}_n (\terminal) (E_n) ,
\end{equation*}
i.e., by exploiting the notation set  forth in \Mref{eq:belt_CPSProd}, $\belt_{\DeltaC \ProdFM} (\terminal) (E) := \pi^{\TypeF}_{n, C} (\terminal) (E_n)$, for every $C \in \CEventsM$ and for every $E_n \in \mcal{G}$.\footnote{Alternatively, we could have defined $\belt_{\CPSFM \ProdFM} (\terminal) (E)$ as 
\begin{equation*}
\belt_{\CPSFM \ProdFM} (\terminal) \Rounds{ E | C } := \pi^{\TypeF}_n (\terminal) \Rounds { \pi^{\ProdFM}_{n} (E) | C } .
\end{equation*}
We would like to thank an anonymous referee for this this point.}
 To see that the definition does not depend on the choice of an $n \in \bN$, let $\Coalg := \la \Car, \trans \ra$ be an arbitrary $\TypeF$-coalgebra, let $\cps \in \CPSM (M \times \Car)$ be such that $\cone^{\trans}_{\TypeF} (\cps) = \terminal$, and let $C \in \CEventsM$ be arbitrary. Thus, we have that
\begin{align*}
\belt_{\DeltaC \ProdFM} (\terminal) (E) & := \pi^{\TypeF}_{n, C} (\terminal) (E_n) \\
& = \pi^{\TypeF}_{n, C} \big( \cone^{\trans}_{\TypeF, C} (\cps_C) \big) (E_n) \\
& = \big( \TypeF \cone^{\trans}_{n, C} \big) (\cps_C) (E_n) \\
& = \big( \CPSFM \ProdFM \cone^{\trans}_{n, C} \big) (\cps_C) (E_n) \\
& = \cps_C \Big( \big( \ProdFM \cone^{\trans}_{n, C} \big)^{-1} (E_n) \Big) \\
& = \cps_C \Big( \big( \cone^{\trans}_{\ProdFM, C} \big)^{-1} \big(  (\pi^{\ProdFM}_{n})^{-1} (E_n) \big) \Big) \\
& = \cps_C \Big( \big( \cone^{\trans}_{\ProdFM, C} \big)^{-1}  (E)  \Big) \\
& = \big(\CPSFM \cone^{\trans}_{\ProdFM, C} \big) (\cps_C) (E) ,%
\end{align*}
where this derivation shows at once that the definition is independent of the natural number $n \in \bN$ we started with and that  
\begin{equation*}
\belt_{\CPSFM \ProdFM} \cone^{\trans}_{\TypeF, C}  (\terminal) (E) = %
\big(\CPSFM \cone^{\trans}_{\ProdFM, C} \big) (\cps_C) (E) ,
\end{equation*}
for every $E \in \mcal{G}$. Now, we extend the definition to every $F \in \sAlg_{\Terminal_{\ProdFM}}$ by setting
\begin{equation}
\label{eq:definition}
\belt_{\DeltaC \ProdFM} (\terminal) (F) := \big(\CPSFM \cone^{\trans}_{\ProdFM, C} \big) (\cps_C) (F) ,
\end{equation}
for every $C \in \CEventsM$. We now show that \Mref{eq:definition} is well-defined. Hence, letting $\cps', \cps'' \in \CPSM (M \times \Car)$ be arbitrary with $\terminal = \cone^{\trans}_{\TypeF} (\cps') = \cone^{\trans}_{\TypeF} (\cps'')$, it follows that $\belt_{\CPSFM \ProdFM} \big( \cone^{\trans}_{\TypeF} (\cps') \big)$ and $\belt_{\CPSFM \ProdFM} \big( \cone^{\trans}_{\TypeF} (\cps'') \big)$ agree on all the elements of $\mcal{G}$. Thus, for every $\ell \leq n$, we have 
\begin{equation*}
(\pi^{\ProdFM}_{n})^{-1} (E_n) \cap (\pi^{\ProdFM}_{\ell})^{-1} (E_\ell) = %
(\pi^{\ProdFM}_{n})^{-1} \big( \rho^{-1}_{n\ell} ( E_\ell ) \cap E_n \big) .
\end{equation*}
This implies that $\mcal{G}$ is a $\pi$-system, which in turns implies by \Mref{lem:pi_system} that the CPSs have to agree on all measurable subsets.

\item \emph{Measurability:} Recalling \Mref{cd:crucial} as a visual aid, for every $E_n \in \sAlg_{M \times \TypeF^n (\1)}$ and for every $C \in \CEventsM$, we have that
\begin{equation*}
\big( \CPSFM \pi^{\ProdFM}_{n} \circ \belt_{\CPSFM \ProdFM} \big) (\terminal) (E_n) = %
\belt_{\CPSFM \ProdFM} (\terminal) \Big( (\pi^{\ProdFM}_{n})^{-1} (E_n)  \Big) = %
\pi^{\TypeF}_{n} (\terminal) (E_n) .
\end{equation*}
Hence, from \Mref{lem:HS98}, it is enough to establish that 
\begin{equation*}
(\belt_{\DeltaC \ProdFM})^{-1} \Big( \big( \gamma^{p}_C ( \pi^{\ProdFM}_n )^{-1} (E_n) \big) \Big) ,
\end{equation*}
for every $C \in \CEventsM$ and $E_n \in \sAlg_{M \times \TypeF^n (\1)}$. Thus, we have that
\begin{align*}
(\belt_{\DeltaC \ProdFM})^{-1} \Big( \gamma^{p}_C \big( (\pi^{\ProdFM}_{n} )^{-1} (E_n) \big) \Big) & %
:= (\belt_{\DeltaC \ProdFM})^{-1} \Big( \big( \CPSFM \pi^{\ProdFM}_{n} \big)^{-1} \big( \gamma^{p}_C (E_n) \big) \Big) \\
& = \big( \CPSFM \pi^{\ProdFM}_{n} \circ \belt_{\DeltaC \ProdFM} \big)^{-1} \Big( \gamma^{p}_C (E_n) \Big) \\
& = \big( \pi^{\TypeF}_{n} \big)^{-1} \Big( \gamma^{p}_C (E_n) \Big),
\end{align*}
where the first line is motivated by \Mref{rem:measurability} and the last set is measurable (with the equality motivated by the fact that we established Condition (2)).
\end{itemize}
Thus, what is written above establishes the result.
\end{proof}

\begin{applemma}
\label{lem:products}
There exists a measurable morphism 
\begin{equation*}
\belt_{\ProdFM} : \Terminal_{\Cons_{M} \times \Id} \to \Terminal_{\Cons_M} \times \Terminal_\Id 
\end{equation*}
such that:
\begin{enumerate}[leftmargin=*, label=\arabic*)]
    \item $\belt_{\ProdFM}  \circ \cone^{\trans}_{\Cons_{M} \times \Id} = \cone^{\trans}_{\Cons_M} \times \cone^{\trans}_{\Id}$, for every $\TypeF$-coalgebra $\Coalg := \la \Car, \trans \ra$;
    
    \item $\Big( \pi^{\Cons_M}_{n} \times \pi^{\Id}_{n} \Big) \circ \belt_{\ProdFM} = \pi^{\Cons_M \times \Id}_{n},$ for every $n \in \bN$.
\end{enumerate}
\end{applemma}

\begin{proof}
This result is an immediate consequence of \citet[Lemma 6, p.401]{Viglizzo_2005a} with the binary product applied to the functors $\Cons_M$ and $\Id$.
\end{proof}

We can now provide the proof of \Mref{lem:crucial}, which turns out to be a straightforward application of the previous lemmata.

\begin{proof}[Proof of \Mref{lem:crucial}]
In light of the fact that $\TypeF := \CPSFM \ProdFM = \CPSFM (\Cons_M \times \Id)$, for every functor $\Ing \in \{ \TypeF, \ProdFM, \Cons_M, \Id \}$ we define an appropriate measurable morphism $\rhot_\Ing$ satisfying  Conditions (1)--(2), i.e., there exists a measurable morphism $\rhot_\Ing : \Terminal_\Ing \to \Ing (\Terminal)$ such that:
\begin{enumerate}[leftmargin=*, label=\arabic*)]
\item $\rhot_\Ing \circ \cone^{\trans}_{\Ing} = \Ing \cone^{\trans}$, for every $\TypeF$-coalgebra $\Coalg := \la \Car, \trans \ra$;

\item $\Ing \pi_m \circ \rhot_{\Ing} = \pi^{\Ing}_{n}$, for every $n \in \bN$.
\end{enumerate}
The morphisms $\rhot_\Ing$ with $\Ing \in \{ \TypeF, \ProdFM, \Cons_M, \Id \}$ are established in what follows.
\begin{itemize}[leftmargin=*]
\item \emph{Functor $\Id$:} We let $\rhot_{\Id} := \id_\Terminal$, which is measurable and trivially satisfies Conditions (1)--(2), with $\belt_{\Id} := \rhot_{\Id}$.

\item \emph{Functor $\Cons_M$:} We let $\rhot_{\Cons_M} := \pi^{\Cons_M}_0$, which is measurable. 
\begin{itemize}[leftmargin=*]
\item Regarding Condition (1), we let $\Coalg := \la \Car , \trans \ra$ be an arbitrary $\TypeF$-coalgebra and we observe that  
\begin{align*}
\rhot_{\Cons_M} \circ \cone^{\trans}_{\Cons_M} & = \pi^{\Cons_M}_0 \circ \cone^{\trans}_{\Cons_M} \\ 
	& = (\pi^{\Cons_M}_n \circ \cone^{\trans}_{\Cons_M})_{n \in \bN} \\
	& = \Cons_M \cone^{\trans} ,
\end{align*}
thus, establishing the condition. 

\item Regarding Condition (2), we have that
\begin{align*}
\Cons_M \pi_n \circ \rhot_{\Cons_M} & = \Cons_M \pi_n \circ \pi^{\Cons_M}_0 \\ 
	& = \id_M \circ \pi^{\Cons_M}_0 \\
	& = \pi^{\Cons_M}_n.
\end{align*}
for every $n \in \bN$, thus, establishing the condition.
\end{itemize}

Finally, we set $\belt_{\Cons_M} := \rhot_{\Cons_M}$.

\item \emph{Functor $\ProdFM$:} From \Mref{lem:products}, we set
\begin{align*} 
\rhot_{\ProdFM} & := \big(\pi^{\Cons_M}_0, \id_Z \big) \belt_{\ProdFM} , \\
	& \textcolor{white}{:}=   \big(\belt_{\Cons_M}, \belt_\Id \big) \belt_{\ProdFM} ,
\end{align*}
which is a measurable morphism that satisfies Conditions (1)--(2) so that the following diagram
\begin{equation*}
\label{cd:products}
\begin{tikzcd}[row sep = large]
	& \hspace{1.5cm}\ProdFM \Car = M \times \Car%
		\arrow[ld, "\cone^{\trans}_{\ProdFM}"', %
				start anchor={[shift={(0.35cm,0.1cm)}]}]
		\arrow[d, "\cone^{\trans}_{\ProdFM}" description] 
		\arrow[rd, "\ProdFM \cone^{\trans}_{\Id}", %
				start anchor={[shift={(-0.6cm,0.1cm)}]}] & %
\\
\Terminal_{\ProdFM}%
	\arrow[r, "\belt_{\ProdFM}"] %
	\arrow[rd, "\pi^{\ProdFM}_n"'] %
	& \Terminal_{\Cons_M} \times \Terminal_\Id %
		\arrow[r, "\big(\pi^{\Cons_M}_0 {,} \id_\Terminal \big)", pos=0.3] %
		\arrow[d, "\pi^{\ProdFM}_n" description] %
		& M \times \Terminal%
			\arrow[ld, "\ProdFM \pi_n"] %
\\
	& \hspace{2.5cm}\ProdFM \TypeF^{n} (\1)  = M \times \TypeF^{n} (\1) & %
\end{tikzcd}
\end{equation*}
commutes.

\item \emph{Functor $\TypeF$:} Recalling that  $\TypeF := \CPSFM \ProdFM$, from \Mref{lem:cps}, we set
\begin{equation*} 
\rhot_{\CPSFM \ProdFM} := \big(\CPSFM \rhot_{\ProdFM} \big) \belt_{\CPSFM \ProdFM}, \\
\end{equation*}
which is a measurable morphism  that satisfies Conditions (1)--(2) so that the following diagram
\begin{equation*}
\label{cd:cps}
\begin{tikzcd}[column sep =huge, row sep = huge]
	& \CPSFM \ProdFM \Car %
		\arrow[ld, "\cone^{\trans}_{\CPSFM \ProdFM}"', %
				start anchor={[shift={(-0.4cm,0.1cm)}]}] 
		\arrow[d, "\CPSFM \cone^{\trans}_{\ProdFM}" description] 
		\arrow[rd, "\CPSFM \ProdFM \cone^{\trans}_{\Id}", %
				start anchor={[shift={(0.15cm,0.1cm)}]}, %
				end anchor={[shift={(-0.55cm,0cm)}]}]  & %
\\
\Terminal_{\CPSFM \ProdFM} %
	\arrow[r, "\belt_{\CPSFM \ProdFM}"] %
	\arrow[rd, "\pi^{\CPSFM \ProdFM}_n"', %
			end anchor={[shift={(-0.5cm,-0.1cm)}]}] 
	& \CPSM (\Terminal_{\ProdFM}) %
		\arrow[r, "\CPSFM \rhot_{\ProdFM}"] %
		\arrow[d, "\CPSFM \pi^{\ProdFM}_n" description] %
		& \CPSM (M \times \Terminal) %
			\arrow[ld, "\CPSFM \ProdFM \pi_n", %
				start anchor={[shift={(-0.7cm,0cm)}]}, %
				end anchor={[shift={(0.3cm,-0.1cm)}]}] %
\\
	& \CPSFM \ProdFM \TypeF^{n} (\1) & %
\end{tikzcd}
\end{equation*}
commutes as well.
\end{itemize}
As a result, for every functor $\Ing \in \{ \TypeF, \ProdFM, \Cons_M, \Id \}$ both Conditions (1)--(2) are satisfied and, in particular, we have that the following diagram 
\begin{equation*}
\label{cd:cps}
\begin{tikzcd}[column sep =huge, row sep = huge]
\Terminal_{\CPSFM \ProdFM}%
	\arrow[r, "\belt_{\CPSFM \ProdFM}"] %
	\arrow[rr, bend right=15, "\rhot_{\CPSFM \ProdFM}"']
	&  \CPSM (\Terminal_{\ProdFM}) %
		\arrow[r, "\CPSFM \rhot_{\ProdFM}"] 
		& \CPSM (M \times \Terminal) %
\end{tikzcd}
\end{equation*}
commutes, with 
\begin{align*}
\zetat & := \rhot_{\CPSFM \ProdFM} \\%
	& \textcolor{white}{:}= 
\Big( \CPSFM \rhot_{\ProdFM} \Big) %
			\belt_{\CPSFM \ProdFM} \\
	& \textcolor{white}{:}= \bigg( %
	\CPSFM \Big( \big( \pi^{\Cons_M}_0 , \id_\Terminal \big)  %
		\belt_{\ProdFM} \Big) \bigg) %
			\belt_{\CPSFM \ProdFM} \\
	& \textcolor{white}{:}= \bigg( %
	\CPSFM  \Big( \big( \belt_{\Cons_M} , \belt_\Id \big) %
		\belt_{\ProdFM} \Big) \bigg) %
			\belt_{\CPSFM \ProdFM}
\end{align*}
as in \Mref{eq:definition_zetat}, thus, establishing the result.
\end{proof}

\begin{proof}[Proof of \Mref{lem:terminal_identity}]
To establish the result, it is enough to prove that $\cone_n = \pi_n \circ \cone = \pi_n$, for every $n \in \bN$. Thus, we proceed inductively over $n \in \bN$.
\begin{itemize}[leftmargin=*]
\item ($n = 0$) We have $\cone_0 = \pi_0 \circ \cone = ! = \pi_0 \circ \id_{\Terminal}$.

\item ($n \geq 1$) Assume that the result has been established for $n$, i.e., $\cone_n = \pi_n \circ \cone = \pi_n$. Thus, we have
\begin{align*}
\pi_{n+1} \circ \cone & = \cone_{n+1} \\
& = \TypeF \cone_{n} \circ  \belt \\
& = \TypeF \pi_n \circ  \belt \\
& = \TypeF \pi_n \circ \zetat \circ \zetap \\
& = \pi^{\TypeF}_n \circ \zetap \\
& = \pi_{n+1} ,
\end{align*}
where the second to last equality comes from \Mref{lem:crucial}.
\end{itemize}
Hence, what is written above establishes the result.
\end{proof}

\begin{proof}[Proof of \Mref{lem:terminal_existence}]
Let $\Coalg := \la \Car, \trans \ra$ be an arbitrary $\TypeF$-coalgebra. 
\begin{itemize}[leftmargin=*]
\item \emph{Existence:} By \Mref{lem:coalgebra_morphism}, $\cone^{\trans}$ is a $\TypeF$-coalgebra morphism.

\item \emph{Uniqueness:} Let $\cmor : \Car \to \Terminal$ be an arbitrary $\TypeF$-coalgebra morphism. By \Mref{lem:coalgebra_morphism_preservation}, we have $\cone \circ \cmor = \cone^{\trans}$ and, from \Mref{lem:terminal_identity}, $\cone = \id_{\Terminal}$. Thus, we have that $\id_\Terminal \circ \cmor = \cone^{\trans}$, from which it follows that $\cmor = \cone^{\trans}$.
\end{itemize}
Thus, what is written above establishes the result.
\end{proof}

\begin{proof}[Proof of \Mref{lem:type_coalgebra_morphism}]
Fix a conditional measurable space $(\State, \sAlg_\State, \CEventsT)$ and let  %
$\mscr{T} := \la (T_i, \beta_i )_{i \in I} \ra$ and $\mscr{T}' := \la (T'_i, \beta'_i )_{i \in I} \ra$ be two arbitrary type structures appended to it. From \Mref{lem:type_coalgebra}, we can treat them as two $\TypeFp$-coalgebras. 
\begin{itemize}[leftmargin=*]
\item Regarding Condition (1) in \Mref{def:type_morphism}, the functor $\ProdFT$ ensures that, for every $\Type \in \Ob (\Meas^I)$, $\TypeFp (\Type)$ implies that there exists a morphism $\id_\State$ by definition of $\TypeFp$, with $\tmor_0 = \id_\State$.

\item Regarding Condition (2) in \Mref{def:type_morphism}, observe that, from \Mref{def:coalgebra_morphism}, for a $\TypeFp$-coalgebra morphism  $\cmor : T \to T'$, the following diagram
\begin{equation*}
\begin{tikzcd}[column sep = huge]
T_i %
	\arrow[r, "\cmor_i"] %
	\arrow[d, "\beta_i"'] %
	& T'_i %
		\arrow[d, "\beta'_i"] %
\\
\CPST (\Theta \times T_{-i}) %
	\arrow[r, "\widehat{\cmor_{\pm i}}"'] %
	& \CPST (\Theta \times T'_{-i})
\end{tikzcd}
\end{equation*}
commutes, for every $i \in I$. Clearly, this is exactly Condition (2) in \Mref{def:type_morphism} with $(\tmor_i)_{i \in I} = \cmor$. 
\end{itemize}
Thus, what is written above establishes the result.
\end{proof}

\hypersetup{colorlinks=true,linkcolor=green!50!black}
\phantomsection
\addcontentsline{toc}{section}{References}

\bibliographystyle{ampersand_standard_capital_natbib}

\begin{thebibliography}{104}
\providecommand{\natexlab}[1]{#1}
\expandafter\ifx\csname urlstyle\endcsname\relax
  \providecommand{\doi}[1]{doi:\discretionary{}{}{}#1}\else
  \providecommand{\doi}{doi:\discretionary{}{}{}\begingroup
  \urlstyle{rm}\Url}\fi

\bibitem[{Aczel(1988)}]{Aczel_1988}
\textsc{Aczel, P.} (1988).
\newblock \emph{Non-Well-Founded Sets}.
\newblock CSLI.

\bibitem[{Ad{\'a}mek \& Koubek(1979)}]{Adamek_Koubek_1979}
\textsc{Ad{\'a}mek, J.P., Koubek, V.} (1979).
\newblock Least Fixed Point of a Functor.
\newblock \emph{Journal of Computer and System Sciences}, \textbf{19},
  163--178.

\bibitem[{Ad{\'a}mek \& Koubek(1995)}]{Adamek_Koubek_1995}
---{}---{}--- (1995).
\newblock On the Greatest Fixed Point of a Set Functor.
\newblock \emph{Theoretical Computer Science}, \textbf{150}, 57--75.

\bibitem[{Aliprantis \& Border(2006)}]{Aliprantis_Border_2006}
\textsc{Aliprantis, C.D., Border, K.C.} (2006).
\newblock \emph{Infinite Dimensional Analysis}.
\newblock 3rd edition. Springer-Verlag, Berlin.

\bibitem[{Aluffi(2009)}]{Aluffi_2009}
\textsc{Aluffi, P.} (2009).
\newblock \emph{Algebra: Chapter 0}.
\newblock American Mathematical Society, Providence.

\bibitem[{Armbruster \& B\"oge(1979)}]{Armbruster_Boge_1979}
\textsc{Armbruster, W., B\"oge, W.} (1979).
\newblock Bayesian Game Theory.
\newblock In \emph{Game theory and related topics}. North-Holland, Amsterdam,
  (pp. 17--28).

\bibitem[{Aumann(1999b)}]{Aumann_1999b}
\textsc{Aumann, R.J.} (1999b).
\newblock Interactive Epistemology II: Probability.
\newblock \emph{International Journal of Game Theory}, \textbf{28}, 301--314.

\bibitem[{Aumann \& Heifetz(2002)}]{Aumann_Heifetz_2002}
\textsc{Aumann, R.J., Heifetz, A.} (2002).
\newblock ``Incomplete Information''.
\newblock In \emph{Handbook of Game Theory with Economic Applications} (R.J.
  Aumann, S.~Hart, editors), volume III. Elsevier/North-Holland, Amsterdam.

\bibitem[{Barr(1993)}]{Barr_1993}
\textsc{Barr, M.} (1993).
\newblock Terminal Coalgebras in Well-Founded Set Theory.
\newblock \emph{Theoretical Computer Science}, \textbf{114}, 299--315.

\bibitem[{Battigalli(1997)}]{Battigalli_1997}
\textsc{Battigalli, P.} (1997).
\newblock On Rationalizability in Extensive Form Games.
\newblock \emph{Journal of Economic Theory}, \textbf{74}, 40--61.

\bibitem[{Battigalli et~al.(2011)Battigalli, Di~Tillio, Grillo, \&
  Penta}]{Battigalli_et_al_2011}
\textsc{Battigalli, P., Di~Tillio, A., Grillo, E., Penta, A.} (2011).
\newblock Interactive Epistemology and Solution Concepts for Games with
  Asymmetric Information.
\newblock \emph{The B.E. Journal of Theoretical Economics (Advances)},
  \textbf{11}, Article 6.

\bibitem[{Battigalli \& Dufwenberg(2009)}]{Battigalli_Dufwenberg_2009}
\textsc{Battigalli, P., Dufwenberg, M.} (2009).
\newblock Dynamic Psychological Games.
\newblock \emph{Journal of Economic Theory}, \textbf{144}, 1--35.

\bibitem[{Battigalli \& Dufwenberg(2022)}]{Battigalli_Dufwenberg_2022}
---{}---{}--- (2022).
\newblock Belief-Dependent Motivations and Psychological Game Theory.
\newblock \emph{Journal of Economic Surveys}, \textbf{60}, 833--882.

\bibitem[{Battigalli et~al.(Work in Progress)Battigalli, Friedenberg, \&
  Siniscalchi}]{Battigalli_et_al_Forthcoming}
\textsc{Battigalli, P., Friedenberg, A., Siniscalchi, M.} (Work in Progress).
\newblock Epistemic Game Theory: Reasoning about Strategic Uncertainty.

\bibitem[{Battigalli \& Siniscalchi(1999)}]{Battigalli_Siniscalchi_1999}
\textsc{Battigalli, P., Siniscalchi, M.} (1999).
\newblock Hierarchies of Conditional Beliefs and Interactive Epistemology in
  Dynamic Games.
\newblock \emph{Journal of Economic Theory}, \textbf{88}, 188--230.

\bibitem[{Battigalli \& Siniscalchi(2002)}]{Battigalli_Siniscalchi_2002}
---{}---{}--- (2002).
\newblock Strong Belief and Forward Induction Reasoning.
\newblock \emph{Journal of Economic Theory}, \textbf{106}, 356--391.

\bibitem[{Battigalli \& Siniscalchi(2003)}]{Battigalli_Siniscalchi_2003}
---{}---{}--- (2003).
\newblock Rationalization and Incomplete Information.
\newblock \emph{The B.E. Journal of Theoretical Economics (Advances)},
  \textbf{3}, Article 3.

\bibitem[{Battigalli \& Tebaldi(2019)}]{Battigalli_Tebaldi_2019}
\textsc{Battigalli, P., Tebaldi, P.} (2019).
\newblock Interactive Epistemology in Simple Dynamic Games with a Continuum of
  Strategies.
\newblock \emph{Economic Theory}, \textbf{68}, 737--763.

\bibitem[{Ben-Porath(1997)}]{Ben-Porath_1997}
\textsc{Ben-Porath, E.} (1997).
\newblock Rationality, Nash Equilibrium and Backward Induction in Perfect
  Information Games.
\newblock \emph{Review of Economic Studies}, \textbf{64}, 23--46.

\bibitem[{Bergemann \& Morris(2005)}]{Bergemann_Morris_2005}
\textsc{Bergemann, D., Morris, S.} (2005).
\newblock Robust Mechanism Design.
\newblock \emph{Econometrica}, \textbf{73}, 1771--1813.

\bibitem[{Bergemann \& Morris(2009)}]{Bergemann_Morris_2009}
---{}---{}--- (2009).
\newblock Robust Implementation in Direct Mechanisms.
\newblock \emph{Review of Economic Studies}, \textbf{76}, 1175--1204.

\bibitem[{Bergemann \& Morris(2011)}]{Bergemann_Morris_2011}
---{}---{}--- (2011).
\newblock Robust Implementation in General Mechanisms.
\newblock \emph{Games and Economic Behavior}, \textbf{71}, 261--281.

\bibitem[{Bergemann \& Morris(2012)}]{Bergemann_Morris_2012}
---{}---{}--- (2012).
\newblock Robust Mechanism Design: An Introduction.
\newblock In \emph{Robust Mechanism Design. The Role of Private Information and
  Higher Order Beliefs}. World Scientific.

\bibitem[{Blackburn et~al.(2001)Blackburn, de~Rijke, \&
  Venema}]{Blackburn_et_al_2001}
\textsc{Blackburn, P., de~Rijke, M., Venema, Y.} (2001).
\newblock \emph{Modal Logic}.
\newblock Cambridge University Press.

\bibitem[{Bogachev(2007)}]{Bogachev_2007}
\textsc{Bogachev, V.I.} (2007).
\newblock \emph{Measure Theory}.
\newblock Springer-Verlag, Berlin.

\bibitem[{B\"oge \& Eisele(1979)}]{Boge_Eisele_1979}
\textsc{B\"oge, W., Eisele, T.} (1979).
\newblock On Solutions of Bayesian games.
\newblock \emph{International Journal of Game Theory}, \textbf{8}, 193--215.

\bibitem[{Borceux(1994)}]{Borceux_1994}
\textsc{Borceux, F.} (1994).
\newblock \emph{Handbook of Categorical Algebra 1. Basic Category Theory}.
\newblock Cambridge University Press.

\bibitem[{Brandenburger(2003)}]{Brandenburger_2003}
\textsc{Brandenburger, A.} (2003).
\newblock On the Existence of a `Complete' Possibility Structure.
\newblock In \emph{Cognitive Processes and Economic Behavior} (M.~Basili,
  N.~Dimitri, I.~Gilboa, editors). Routledge, (pp. 30--34).

\bibitem[{Brandenburger \& Dekel(1993)}]{Brandenburger_Dekel_1993}
\textsc{Brandenburger, A., Dekel, E.} (1993).
\newblock Hierarchies of Beliefs and Common Knowledge.
\newblock \emph{Journal of Economic Theory}, \textbf{59}, 189--198.

\bibitem[{Brandenburger \& Keisler(2006)}]{Brandenburger_Keisler_2006}
\textsc{Brandenburger, A., Keisler, J.H.} (2006).
\newblock An Impossibility Theorem on Beliefs in Games.
\newblock \emph{Studia Logica: An International Journal for Symbolic Logic},
  \textbf{84}, 211--240.

\bibitem[{de~Oliveira(2018)}]{de_Oliveira_2018}
\textsc{de~Oliveira, H.} (2018).
\newblock Blackwell's Informativeness Theorem Using Diagrams.
\newblock \emph{Games and Economic Behavior}, \textbf{109}, 126--131.

\bibitem[{De~Vito(2023)}]{De_Vito_2023}
\textsc{De~Vito, N.M.} (2023).
\newblock Complete Conditional Type Structures.
\newblock \emph{Technical report}, Mimeo.

\bibitem[{Dekel et~al.(2007)Dekel, Fudenberg, \& Morris}]{Dekel_et_al_2007}
\textsc{Dekel, E., Fudenberg, D., Morris, S.} (2007).
\newblock Interim Correlated Rationalizability.
\newblock \emph{Theoretical Economics}, \textbf{2}, 15--40.

\bibitem[{Dekel \& Siniscalchi(2015)}]{Dekel_Siniscalchi_2015}
\textsc{Dekel, E., Siniscalchi, M.} (2015).
\newblock Epistemic Game Theory.
\newblock In \emph{Handbook of Game Theory} (H.P. Young, S.~Zamir, editors),
  volume~IV. North-Holland, Amsterdam.

\bibitem[{Di~Tillio(2008)}]{Di_Tillio_2008}
\textsc{Di~Tillio, A.} (2008).
\newblock Subjective Expected Utility in Games.
\newblock \emph{Theoretical Economics}, \textbf{3}, 287--323.

\bibitem[{Di~Tillio et~al.(2014)Di~Tillio, Halpern, \&
  Samet}]{DiTillio_et_al_2014}
\textsc{Di~Tillio, A., Halpern, J.Y., Samet, D.} (2014).
\newblock Conditional Type Spaces.
\newblock \emph{Games and Economic Behavior}, \textbf{87}, 253--268.

\bibitem[{Dieudonn{\'e}(1948)}]{Dieudonne_1948}
\textsc{Dieudonn{\'e}, J.A.E.} (1948).
\newblock Sur le th{\'e}or{\`e}me de Lebesgue-Nikodym (III).
\newblock \emph{Annales de l'universit{\'e} de Grenoble}, \textbf{23}, 25--53.

\bibitem[{Dieudonn{\'e}(1989)}]{Dieudonne_1989}
---{}---{}--- (1989).
\newblock \emph{A History of Algebraic and Differential Topology, 1900 - 1960}.
\newblock Birkh{\"a}user, Boston/Basel/Berlin.

\bibitem[{Freyd(2008)}]{Freyd_2008}
\textsc{Freyd, P.J.} (2008).
\newblock Algebraic Real Analysis.
\newblock \emph{Theory and Applications of Categories}, \textbf{20}, 215--306.

\bibitem[{Friedenberg(2010)}]{Friedenberg_2010}
\textsc{Friedenberg, A.} (2010).
\newblock When do Type Structures Contain All Hierarchies of Beliefs?
\newblock \emph{Games and Economic Behavior}, \textbf{68}, 108--129.

\bibitem[{Friedenberg(2019)}]{Friedenberg_2019}
---{}---{}--- (2019).
\newblock Bargaining Under Strategic Uncertainty: The Role of Second-Order
  Optimism.
\newblock \emph{Econometrica}, \textbf{87}, 1835--1865.

\bibitem[{Friedenberg \& Meier(2011)}]{Friedenberg_Meier_2011}
\textsc{Friedenberg, A., Meier, M.} (2011).
\newblock On the Relationship Between Hierarchy and Type Morphisms.
\newblock \emph{Economic Theory}, \textbf{46}, 377--399.

\bibitem[{Fukuda(2024{\natexlab{a}})}]{Fukuda_2024a}
\textsc{Fukuda, S.} (2024{\natexlab{a}}).
\newblock The Existence of Universal Qualitative Belief Spaces.
\newblock \emph{Journal of Economic Theory}, \textbf{207}, 105590.

\bibitem[{Fukuda(2024{\natexlab{b}})}]{Fukuda_2024b}
---{}---{}--- (2024{\natexlab{b}}).
\newblock Topology-Free Constructions of a Universal Type Space as Coherent
  Belief Hierarchies.
\newblock \emph{Mimeo}.

\bibitem[{Galeazzi \& Marti(2023)}]{Galeazzi_Marti_2023}
\textsc{Galeazzi, P., Marti, J.} (2023).
\newblock Choice Structures in Games.
\newblock \emph{Games and Economic Behavior}, \textbf{140}, 431--455.

\bibitem[{Giry(1982)}]{Giry_1982}
\textsc{Giry, M.} (1982).
\newblock A Categorical Approach to Probability Theory.
\newblock In \emph{Categorical Aspects of Topology and Analysis}
  (B.~Banaschewski, editor). Springer Berlin Heidelberg, (pp. 68--85).

\bibitem[{Govindan \& Wilson(2009)}]{Govindan_Wilson_2009}
\textsc{Govindan, S., Wilson, R.A.} (2009).
\newblock On Forward Induction.
\newblock \emph{Econometrica}, \textbf{77}, 1--28.

\bibitem[{Guarino(2017)}]{Guarino_2017}
\textsc{Guarino, P.} (2017).
\newblock A Topology-Free Construction of the Universal Type Structure for
  Conditional Probability Systems.
\newblock \emph{Electronic Proceedings in Theoretical Computer Science}.

\bibitem[{Guo \& Yannelis(2022)}]{Guo_Yannelis_2022}
\textsc{Guo, H., Yannelis, N.C.} (2022).
\newblock Robust Coalitional Implementation.
\newblock \emph{Games and Economic Behavior}, \textbf{132}, 553--575.

\bibitem[{Guo \& Shmaya(2021)}]{Guo_Shmaya_2021}
\textsc{Guo, Y., Shmaya, E.} (2021).
\newblock Costly Miscalibration.
\newblock \emph{Theoretical Economics}, \textbf{16}, 477--506.

\bibitem[{Halmos(1950)}]{Halmos_1950}
\textsc{Halmos, P.R.} (1950).
\newblock \emph{Measure Theory}.
\newblock 1st edition. Litton Educational Publishing, Inc.

\bibitem[{Hammond(1994)}]{Hammond_1994}
\textsc{Hammond, P.J.} (1994).
\newblock Elementary Non-Archimedean Representations of Probability for
  Decision Theory and Games.
\newblock In \emph{Patrick Suppes: Scientific Philosopher} (P.~Humphreys,
  editor), volume 1. Probability and Probabilistic Causality. Springer,
  Dordrecht.

\bibitem[{Harsanyi(1967)}]{Harsanyi_1967}
\textsc{Harsanyi, J.C.} (1967).
\newblock Games with Incomplete Information Played by ``Bayesian'' Players,
  I--III: Part I. The Basic Model.
\newblock \emph{Management Science}, \textbf{14}, 159--182.

\bibitem[{Heifetz(1993)}]{Heifetz_1993}
\textsc{Heifetz, A.} (1993).
\newblock The Bayesian Formulation of Incomplete Information -- The Non-Compact
  Case.
\newblock \emph{International Journal of Game Theory}, \textbf{21}, 329--338.

\bibitem[{Heifetz(1996)}]{Heifetz_1996}
---{}---{}--- (1996).
\newblock Non-Well-Founded Type Spaces.
\newblock \emph{Games and Economic Behavior}, \textbf{16}, 202--217.

\bibitem[{Heifetz(1997)}]{Heifetz_1997}
---{}---{}--- (1997).
\newblock Infinitary S5 Epistemic Logic.
\newblock \emph{Mathematical Logic Quarterly}, \textbf{43}, 333--342.

\bibitem[{Heifetz \& Mongin(2001)}]{Heifetz_Mongin_2001}
\textsc{Heifetz, A., Mongin, P.} (2001).
\newblock Probability Logic for Type Spaces.
\newblock \emph{Games and Economic Behavior}, \textbf{35}, 31--53.

\bibitem[{Heifetz \& Samet(1998)}]{Heifetz_Samet_1998}
\textsc{Heifetz, A., Samet, D.} (1998).
\newblock Topology-Free Typology of Beliefs.
\newblock \emph{Journal of Economic Theory}, \textbf{82}, 324--341.

\bibitem[{Heifetz \& Samet(1999)}]{Heifetz_Samet_1999}
---{}---{}--- (1999).
\newblock Coherent Beliefs are not always Types.
\newblock \emph{Journal of Mathematical Economics}, \textbf{32}, 475--488.

\bibitem[{Heinsalu(2014)}]{Heinsalu_2014}
\textsc{Heinsalu, S.} (2014).
\newblock Universal Type Structures with Unawareness.
\newblock \emph{Games and Economic Behavior}, \textbf{83}, 255--266.

\bibitem[{Jacobs(2001)}]{Jacobs_2001}
\textsc{Jacobs, B.} (2001).
\newblock Many-Sorted Coalgebraic Modal Logic: A Model-Theoretic Study.
\newblock \emph{Theoretical Informatics and Applications}, \textbf{35}, 31--59.

\bibitem[{Jacobs(2017)}]{Jacobs_2017}
---{}---{}--- (2017).
\newblock \emph{Introduction to Coalgebra. Towards Mathematics of States and
  Observation}.
\newblock Cambridge University Press.

\bibitem[{Jacobs \& Rutten(1997)}]{Jacobs_Rutten_1997}
\textsc{Jacobs, B., Rutten, J.J.J.M.} (1997).
\newblock A Tutorial on (Co)Algebras and (Co)Induction.
\newblock \emph{Bulletin of EATCS}, \textbf{62}, 222--259.

\bibitem[{Jech(2006)}]{Jech_2006}
\textsc{Jech, T.} (2006).
\newblock \emph{Set Theory. The Third Millennium Edition, Revised and
  Expanded}.
\newblock 4th edition. Springer-Verlag, Berlin.

\bibitem[{Kechris(1995)}]{Kechris_1995}
\textsc{Kechris, A.S.} (1995).
\newblock \emph{Classical Descriptive Set Theory}.
\newblock Springer-Verlag, Berlin.

\bibitem[{Kurz(2001)}]{Kurz_2001}
\textsc{Kurz, A.} (2001).
\newblock Specifying Coalgebras with Modal Logic.
\newblock \emph{Theoretical Computer Science}, \textbf{260}, 119--138.

\bibitem[{Lambek(1968)}]{Lambek_1968}
\textsc{Lambek, J.} (1968).
\newblock A Fixpoint Theorem for Complete Categories.
\newblock \emph{Mathematische Zeitschrift}, \textbf{103}, 151--161.

\bibitem[{Lawvere(1962)}]{Lawvere_1962}
\textsc{Lawvere, F.W.} (1962).
\newblock The Category of Probabilistic Mappings.
\newblock \emph{Technical report}, Mimeo.

\bibitem[{Leinster(2014)}]{Leinster_2014}
\textsc{Leinster, T.} (2014).
\newblock \emph{Basic Category Theory}.
\newblock Cambridge University Press.

\bibitem[{Liu(2009)}]{Liu_2009}
\textsc{Liu, Q.} (2009).
\newblock On Redundant Types and Bayesian Formulation of Incomplete
  Information.
\newblock \emph{Journal of Economic Theory}, \textbf{144}, 2115--2145.

\bibitem[{Mac~Lane(1998)}]{MacLane_1998}
\textsc{Mac~Lane, S.} (1998).
\newblock \emph{Categories for the Working Mathematician}.
\newblock Springer-Verlag, New York/Berlin/Heidelberg.

\bibitem[{Maschler et~al.(2013)Maschler, Solan, \& Zamir}]{Maschler_et_al_2013}
\textsc{Maschler, M., Solan, E., Zamir, S.} (2013).
\newblock \emph{Game Theory}.
\newblock Cambridge University Press, Cambridge.

\bibitem[{Meier(2012)}]{Meier_2012}
\textsc{Meier, M.} (2012).
\newblock An Infinitary Probability Logic for Type Spaces.
\newblock \emph{Israel Journal of Mathematics}, \textbf{192}, 1--58.

\bibitem[{Meier \& Perea(2023)}]{Meier_Perea_2023}
\textsc{Meier, M., Perea, A.} (2023).
\newblock Forward Induction in a Backward Inductive Manner.
\newblock \emph{Technical report}, EpiCenter Working Paper.

\bibitem[{Mertens et~al.(2015)Mertens, Sorin, \& Zamir}]{Mertens_et_al_2015}
\textsc{Mertens, J.F., Sorin, S., Zamir, S.} (2015).
\newblock \emph{Repeated Games}.
\newblock Cambridge University Press, Cambridge.

\bibitem[{Mertens \& Zamir(1985)}]{Mertens_Zamir_1985}
\textsc{Mertens, J.F., Zamir, S.} (1985).
\newblock Formulation of Bayesian Analysis for Games with Incomplete
  Information.
\newblock \emph{International Journal of Game Theory}, \textbf{14}, 1--29.

\bibitem[{Moss(2011)}]{Moss_2011}
\textsc{Moss, L.} (2011).
\newblock Connections of Coalgebra and Semantic Modeling.
\newblock \emph{Proceedings Thirteenth Conference on Theoretical Aspects of
  Rationality and Knowledge (TARK)}.

\bibitem[{Moss \& Viglizzo(2004)}]{Moss_Viglizzo_2004}
\textsc{Moss, L.S., Viglizzo, I.D.} (2004).
\newblock Harsanyi Type Spaces and Final Coalgebras Constructed from Satisfied
  Theories.
\newblock \emph{Electronic Notes in Theoretical Computer Science},
  \textbf{106}, 279--295.

\bibitem[{Moss \& Viglizzo(2006)}]{Moss_Viglizzo_2006}
---{}---{}--- (2006).
\newblock Final Coalgebras for Functors on Measurable Spaces.
\newblock \emph{Information and Computation}, \textbf{204}, 610--636.

\bibitem[{M{\"u}ller(2016)}]{Mueller_2016}
\textsc{M{\"u}ller, C.} (2016).
\newblock Robust Virtual Implementation Under Common Strong Belief in
  Rationality.
\newblock \emph{Journal of Economic Theory}, \textbf{162}, 407--450.

\bibitem[{Myerson(1986)}]{Myerson_1986}
\textsc{Myerson, R.B.} (1986).
\newblock Multistage Games with Communication.
\newblock \emph{Econometrica}, \textbf{54}, 323--358.

\bibitem[{Osborne \& Rubinstein(1994)}]{Osborne_Rubinstein_1994}
\textsc{Osborne, M.J., Rubinstein, A.} (1994).
\newblock \emph{A Course in Game Theory}.
\newblock MIT Press.

\bibitem[{Pearce(1984)}]{Pearce_1984}
\textsc{Pearce, D.G.} (1984).
\newblock Rationalizable Strategic Behavior and the Problem of Perfection.
\newblock \emph{Econometrica}, \textbf{52}, 1029--1050.

\bibitem[{Penta(2015)}]{Penta_2015}
\textsc{Penta, A.} (2015).
\newblock Robust Dynamic Implementation.
\newblock \emph{Journal of Economic Theory}, \textbf{160}, 280--316.

\bibitem[{Perea(2012)}]{Perea_2012}
\textsc{Perea, A.} (2012).
\newblock \emph{Epistemic Game Theory: Reasoning and Choice}.
\newblock Cambridge University Press.

\bibitem[{Pint{\'e}r(2010)}]{Pinter_2010}
\textsc{Pint{\'e}r, M.} (2010).
\newblock The Non-Existence of a Universal Topological Type Space.
\newblock \emph{Journal of Mathematical Economics}, \textbf{46}, 223--229.

\bibitem[{Pivato(2024{\natexlab{a}})}]{Pivato_2024b}
\textsc{Pivato, M.} (2024{\natexlab{a}}).
\newblock Global Subjective Expected Utility Representations.
\newblock \emph{Mimeo}.

\bibitem[{Pivato(2024{\natexlab{b}})}]{Pivato_2024}
---{}---{}--- (2024{\natexlab{b}}).
\newblock Universal Recursive Preference Structures.
\newblock \emph{Mimeo}.

\bibitem[{Pivato(Work in Progress)}]{Pivato_unpublished}
---{}---{}--- (Work in Progress).
\newblock Categorical Decision Theory.

\bibitem[{R\^enyi(1955)}]{Renyi_1955}
\textsc{R\^enyi, A.} (1955).
\newblock On a New Axiomatic Theory of Probability.
\newblock \emph{Acta Mathematica Academiae Scientiarum Hungarica}, \textbf{6},
  285--335.

\bibitem[{R{\"o}{\upshape{\ss}}iger(2001)}]{Roessiger_2001}
\textsc{R{\"o}{\upshape{\ss}}iger, M.} (2001).
\newblock From Modal Logic to Terminal Coalgebras.
\newblock \emph{Theoretical Computer Science}, \textbf{260}, 209--228.

\bibitem[{Rutten(2000)}]{Rutten_2000}
\textsc{Rutten, J.J.J.M.} (2000).
\newblock Universal Coalgebra: a Theory of Systems.
\newblock \emph{Theoretical Computer Science}, \textbf{249}, 3--80.

\bibitem[{Schubert(2009)}]{Schubert_2009}
\textsc{Schubert, C.} (2009).
\newblock Terminal Coalgebras for Measure-Polynomial Functors.
\newblock In \emph{Theory and Applications of Models of Computation} (J.~Chen,
  S.B. Cooper, editors). Springer, Berlin, Heidelberg, (pp. 325--334).

\bibitem[{Scott(1976)}]{Scott_1976}
\textsc{Scott, D.S.} (1976).
\newblock Data Types as Lattices.
\newblock \emph{SIAM Journal on Computing}, \textbf{5}, 522--587.

\bibitem[{Siniscalchi(2008)}]{Siniscalchi_2008}
\textsc{Siniscalchi, M.} (2008).
\newblock ``Epistemic Game Theory: Beliefs and Types''.
\newblock In \emph{The New Palgrave Dictionary of Economics} (S.N. Durlauf,
  L.E. Blume, editors), 2nd edition. Palgrave Macmillan, New York.

\bibitem[{Smyth \& Plotkin(1982)}]{Smyth_Plotkin_1982}
\textsc{Smyth, M.B., Plotkin, G.D.} (1982).
\newblock The Category-Theoretic Solution of Recursive Domain Equations.
\newblock \emph{SIAM Journal on Computing}, \textbf{11}, 761--783.

\bibitem[{Sparre~Andersen \& Jessen(1948)}]{Sparre-Andersen_Jessen_1948}
\textsc{Sparre~Andersen, E., Jessen, B.} (1948).
\newblock On the Introduction of Measures in Infinite Product Sets.
\newblock \emph{Kongelige Danske Videnskabernes Selskab Matematisk-Fysiske
  Meddelelser}, \textbf{25}, 3--7.

\bibitem[{Srivastava(1998)}]{Srivastava_1998}
\textsc{Srivastava, S.M.} (1998).
\newblock \emph{A Course on Borel Sets}.
\newblock Springer-Verlag, New York.

\bibitem[{Vassilakis(1991)}]{Vassilakis_1991}
\textsc{Vassilakis, S.} (1991).
\newblock Functional Fixed Points.
\newblock \emph{Technical Report~33}, Stanford Institute for Theoretical
  Economics.

\bibitem[{Vassilakis(1992)}]{Vassilakis_1992}
---{}---{}--- (1992).
\newblock Some Economic Applications of Scott Domains.
\newblock \emph{Mathematical Social Sciences}, \textbf{24}, 173--208.

\bibitem[{Viglizzo(2005{\natexlab{a}})}]{Viglizzo_2005}
\textsc{Viglizzo, I.D.} (2005{\natexlab{a}}).
\newblock \emph{Coalgebras on measurable spaces}.
\newblock Ph.D. thesis, Indiana University.

\bibitem[{Viglizzo(2005{\natexlab{b}})}]{Viglizzo_2005a}
---{}---{}--- (2005{\natexlab{b}}).
\newblock Final Sequences and Final Coalgebras for Measurable Spaces.
\newblock In \emph{Algebra and Coalgebra in Computer Science} (J.L. Fiadeiro,
  N.~Harman, M.~Roggenbach, J.~Rutten, editors). Springer Berlin Heidelberg,
  Berlin, Heidelberg, (pp. 395--407).

\bibitem[{Wilson(1987)}]{Wilson_1987}
\textsc{Wilson, R.B.} (1987).
\newblock Game-Theoretic Analyses of Trading Processes.
\newblock In \emph{Advances in Economics and Econometrics, 5th World Congress
  of the Econometric Society} (T.~Bewley, editor). Cambridge University Press,
  Cambridge.

\bibitem[{Worrell(2005)}]{Worrell_2005}
\textsc{Worrell, J.} (2005).
\newblock On the Final Sequence of a Finitary Set Functor.
\newblock \emph{Theoretical Computer Science}, \textbf{338}, 184--199.

\end{thebibliography}

\end{document}